\documentclass[
 reprint,
 aps,prl,
 groupedaddress,
 notitlepage,
 superscriptaddress
]{revtex4-1}

\usepackage[textsize = scriptsize, color = pink]{todonotes}
\usepackage{color}
\usepackage{float}
\usepackage{graphicx}
\usepackage{amsmath,amstext,amssymb,amsthm}
\usepackage{siunitx}
\DeclareSIUnit{\nothing}{\relax}
\PassOptionsToPackage{bookmarks={false}}{hyperref}
\usepackage[hidelinks]{hyperref}

\usepackage[normalem]{ulem}
\usepackage{upgreek}
\usepackage{threeparttable}
\usepackage{xspace}
\usepackage{comment}
\usepackage{setspace}
\usepackage{xcolor}
\definecolor{orange}{rgb}{0.59, 0.29, 0.1}
\definecolor{yellow}{rgb}{1.0, 0.7, 0.0}

\newcommand{\overbar}[1]{\mkern 1.5mu\overline{\mkern-5mu#1\mkern-1.5mu}\mkern 1.5mu}

\newcommand{\Gbar}{\ensuremath{\,\overbar{\mathrm{\Gamma}}\xspace}}
\newcommand{\Kbar}{\ensuremath{\,\overbar{K}}\xspace}

\begin{document}

\title{Ultrafast low-energy photoelectron diffraction for the study of surface-adsorbate interactions with 100 femtosecond temporal resolution}

%%%%%%%%%%%%%%%%%%%%%%%%%%%%%%%%%%%%%%%%%%%%%%
\author{H.~Erk}
\email{erk@physik.uni-kiel.de}
\affiliation
{Institute of Experimental and Applied Physics, Kiel University, 24098 Kiel, Germany
}
\author{C. E.~Jensen}
\affiliation{Institute of Experimental and Applied Physics, Kiel University, 24098 Kiel, Germany
}
%\author{N.~N}
%\affiliation{Institute of Experimental and Applied Physics, Kiel University, 24098 Kiel, Germany
%}
\author{S.~Jauernik}
\affiliation{Institute of Experimental and Applied Physics, Kiel University, 24098 Kiel, Germany
}
\author{M.~Bauer}
\homepage{https://www.physik.uni-kiel.de/de/institute/ag-bauer}
\affiliation{Institute of Experimental and Applied Physics, Kiel University, 24098 Kiel, Germany
}
\affiliation{
Kiel Nano, Surface and Interface Science KiNSIS, Kiel University, 24118 Kiel, Germany
}

%%%%%%%%%%%%%%%%%%%%%%%%%%%%%%%%%%%%%%%%%%%%%%

\date{\today}

\begin{abstract}
An ultrafast photoemission-based low-energy electron diffraction experiment with monolayer surface sensitivity is presented. %To circumvent the problem of space charge interaction between the probing electrons, which normally limits the time-resolution of ultrafast electron diffraction experiments, the electron pulse is generated by photoexcitation in the sample substrate with ultrashort vacuum ultraviolet laser pulses. In this way, the effective propagation distance from the electron source to the sample surface is reduced from the micrometer range, as realized in state-of-the-art ultrafast low energy electron diffraction  concepts, to a few Angstrom range. 
In a first experiment on tin-phthalocyanine adsorbed on graphite, we demonstrate a time resolution of $\approx \SI{100}{\femto s}$. Analysis of the transient photoelectron diffraction signal indicates a heating of the adsorbate layer on a time scale of a few \SI{}{\pico s}, suggesting coupling to phononic degrees of freedom of the substrate as the primary energy transfer channel for the vibrational excitation of the adsorbate layer. Remarkably, the transient photoelectron diffraction signal not only provides direct information about the structural dynamics of the adsorbate, but also about the charge carrier dynamics of the substrate. The presented concept combined with momentum microscopy could become a versatile tool for the comprehensive investigation of the coupled charge and vibrational dynamics of relevance for ultrafast surface processes.
\end{abstract}

\maketitle
Surface-sensitive time-domain techniques such as time-resolved desorption \cite{Prybyla1992}, two-photon photoemission \cite{Petek2000}, second-harmonic generation \cite{Stepan2005} and sum-frequency generation \cite{Backus2005}, or, more recently, time-resolved X-ray absorption spectroscopy \cite{DellAngela2013, Diesen2021} and orbital tomography \cite{Wallauer2021} have, in the past, provided detailed and comprehensive insights into ultrafast surface-adsorbate interaction processes involving charge and vibrational degrees of freedom of relevance for surface chemical reactions. Time-domain surface electron diffraction techniques \cite{Aeschlimann1995, Baum2014, Curcio2022} hold the potential to greatly enrich this research, as they can provide quantitative and direct information on how structural orders in molecular adsorbate layers are transiently affected by such interactions \cite{Hanisch-Blicharski2013, Gulde2014, Frigge2017}. In this context, ultrafast low-energy electron diffraction (ULEED) would be the first choice technique due to its exceptional surface sensitivity and the available and established methods for the quantitative analysis of the data \cite{vanHove1986}. However, electron dispersion and Coulomb interaction broadens the probing electron pulse and considerably limits the time resolution of ultrafast electron diffraction techniques in general, and it is particularly critical for the low electron energies typically used in ULEED \cite{Karrer2001, Muller2014}. To at least partially compensate for this problem, attempts have been made to minimize the propagation distance of the electron pulses from the source to the sample surfaces \cite{Storeck2017}. The most sophisticated ULEED design so far uses a micrometer-sized electron gun and a gun-sample distance of a few hundred micrometers yielding a time resolution of 
%\SI{1.4}{\pico s} at \SI{50}{eV} electron energy \cite{Vogelgesang2017} and 
$\approx \SI{1}{\pico s}$ at several \SI{10}{eV} electron energy \cite{Vogelgesang2017, Storeck2021}. However, even with this design, the time domain below a picosecond is not yet accessible.

\begin{figure}
        \includegraphics[width=0.9\linewidth]{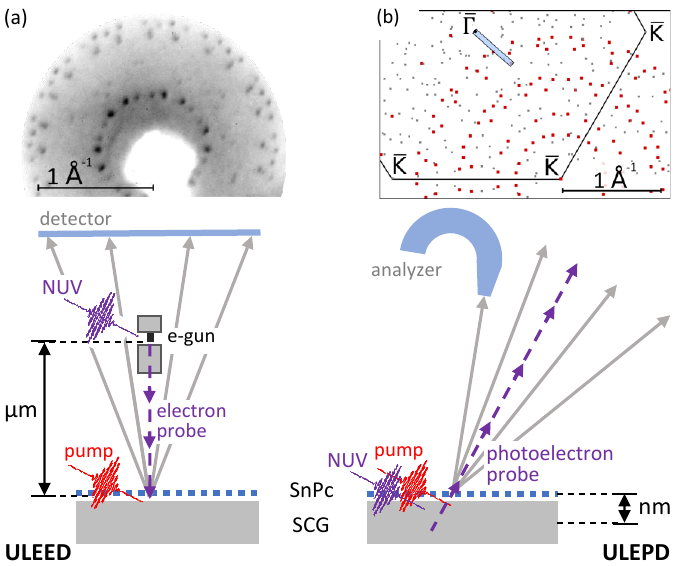}
        \caption{Comparison of ULEED and ULEPD. (a) ULEED: Few-\SI{10}{eV} electron pulses are generated in a photoelectron gun using NUV laser pulses. An imaging detector records the diffraction signal in reflection as illustrated by the static LEED pattern of SnPC/SCG in the top panel. Electron pulse broadening due to Coulomb interaction is limited by placing the electron gun close to the sample surface. (b) LEPD: Electron pulses are generated by photoexcitation in the sample substrate. Emitted photoelectrons are diffracted as they pass an ordered adsorbate layer. The top panel shows a calculated SnPc diffraction pattern for photoelectrons emitted from the Fermi surface of SCG located at the \Kbar-points. Red dots indicate diffraction spots due to photoemission from the red-marked \Kbar-point. Diffraction orders up to the fourth order were considered. The blue bar near \Gbar~indicates the approximate momentum region probed in the experiment.}
    	\label{fig:Fig1}
\end{figure}

In this work, we present a surface-sensitive and ultrafast electron diffraction experiment capable of probing structural dynamics in adsorbate layers with a temporal resolution of $\approx \SI{100}{\femto s}$. The basic concept of the technique is illustrated in Fig.~\ref{fig:Fig1} in comparison to a schematic of a ULEED experiment: We analyze the energy-momentum distribution of low-energy photoelectrons excited by a near ultraviolet (NUV) ultrafast laser pulse in a substrate that are diffracted as they pass through an ordered adsorbate layer. The propagation distance of the (photo-) electron pulse prior diffraction is limited by the inelastic mean free path of the electrons in the substrate to typical values of a few nanometers \cite{Seah1979}, so that a significant temporal broadening is omitted. In our study, which we performed on an ordered monolayer of tin phthalocynanine (SnPc) adsorbed on graphite, we experimentally demonstrate a time resolution of this ultrafast low-energy photoelectron diffraction (ULEPD) technique of $\approx \SI{100}{\femto s}$, currently limited by the pulse width of the NUV laser pulse. The analysis of the transient changes in the photoelectron diffraction intensity indicates a heating of the adsorbate layer on a characteristic time scale of several ps. 
%Remarkably, the ULEPD signal contains also direct information on the hot photocarrier dynamics in the substrate.
%, which shows a characteristic two-component decay as reported in other time-domain studies of graphitic materials before \cite{Kampfrath2005, Wang2010, Johannsen2013}. 
We associate the changes and the related time scale with the formation of vibrational disorder in the adsorbate layer due to coupling to the phonon bath in the graphite, which is excited by the cooling of the photoexcited hot carrier distribution.

 For the experiments, a monolayer of SnPc molecules was thermally evaporated onto a single-crystalline graphite (SCG) surface, prepared in-situ by cleaving and subsequent annealing at \SI{700}{K} for 24 hours. LEED patterns [see top panel of Fig. \ref{fig:Fig1}(a)] indicated an ordered SnPc overlayer formed from six different rotational domains, in good agreement with past studies on the adsorption of phthalocyanines on graphite \cite{Gopakumar2004, Kawakita2017}. The ULEPD experiments were carried out with a setup for time- and angle-resolved photoelectron spectroscopy (trARPES), which is described in detail in Ref. \cite{Jauernik2018}. We employed \SI{827}{\nano m} (\SI{1.5}{eV}), \SI{30}{\femto s} near infrared (NIR) pump pulses to excite the SnPc/SCG surface at an absorbed fluence of $\approx\SI{120}{\micro J/{\centi m}^2}$. The probing photoelectron pulse was generated in the SCG substrate using \SI{210}{\nano m} (\SI{5.9}{eV}), \SI{95}{\femto s} NUV pulses. With this photon energy, about one third of the Brillouin zone of graphite is accessible in a photoemission experiment. The diffracted photoelectrons were detected and analyzed using the hemispherical photoelectron spectrometer of the trARPES setup.

\begin{figure}
        \includegraphics[width=1\linewidth]{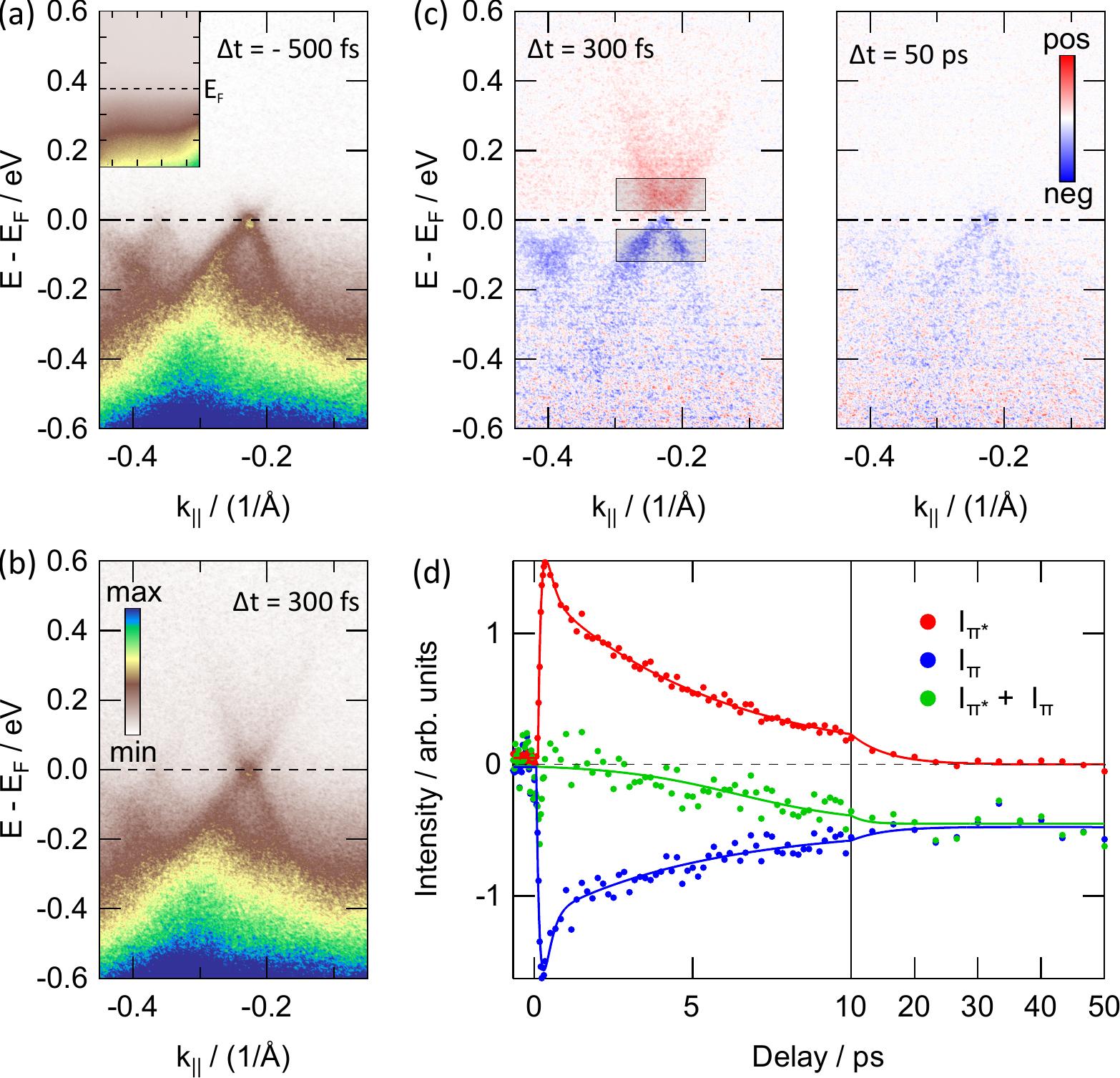}
        \caption{ULEPD of SnPc/SCG. (a) ULEPD intensity map of SnPc/SCG before excitation with \SI{827}{\nano m} pump pulses. The momentum cut is indicated by the blue bar in the top panel of Fig. \ref{fig:Fig1}(b). The cone shaped signal below $E_\text{F}$ is the diffraction signal of the photoelectron pulse due to interaction with the ordered SnPc layer. The inset shows the same energy-momentum region recorded for the pristine SCG substrate. (b) ULEPD intensity map of SnPc/SCG at $\Delta t = \SI{300}{\femto s}$. The additional signal above $E_\text{F}$ indicates the photoexcitation of hot carriers at \Kbar, which is diffracted into the energy-momentum field of view of the experiment. (c) Difference ULEPD intesity maps at $\Delta t = \SI{300}{\femto s}$ and $\Delta t = \SI{50}{\pico s}$ revealing the $\pi$ band dynamics below $E_\text{F}$. The gray shaded areas in the left panel mark the integration regions used to create the ULEPD transients shown in (d). (d) ULEPD transients of $\pi$ band (blue) and $\pi^*$ band (red) including fits to the data (solid lines) as described in the text. The green data points result from the sum of the two transients and represent the transient associated with structural excitations in the adsorbate layer. The green solid line is a fit of a sigmoidal function to the data.}
    	\label{fig:Fig2}
\end{figure}

%In a previous study, we investigated the energy-momentum distribution of low-energy photoelectrons, which were photoexcited with NUV laser pulses in the SCG substrate and diffracted by the ordered SnPc overlayer \cite{Erk2023}. 

Figure \ref{fig:Fig2}(a) shows ULEPD data of SnPc/SCG %the corresponding data of the present study, here 
recorded  near the \Gbar-point at a time delay $\Delta t= -\SI{500}{\femto s}$ before NIR photoexcitation.
%, i.e., at $\Delta t = -\SI{}{\femto s}$. 
In agreement with a previous study \cite{Erk2023}, we observe a cone-shaped structure in the spectrum that is completely absent for pristine SCG [see inset of Fig. \ref{fig:Fig2}(a)]. We assign the cone to the $\pi$ band of SCG at the \Kbar-point \cite{Ahuja1997} that is mapped into the probed energy-momentum region by diffraction of the photoelectrons as they pass through the SnPc layer. A comparison of the data with simulations that take the SnPc superstructure into account shows that the detected photoelectrons experienced a fourth-order momentum transfer during diffraction at the adsorbate layer [see Ref. \cite{Erk2023, Supplemental} and top panel of Fig. \ref{fig:Fig1}(b)]. 

%In the same energy-momentum region, the photoemission spectrum of pristine SCG shows no signal [see inset of Fig. \ref{fig:Fig2}(a)], 

Figure \ref{fig:Fig2}(b) shows ULEPD data recorded at $\Delta t=\SI{300}{\femto s}$ after the arrival of the NIR pump pulse. In comparison to the spectrum in Fig. \ref{fig:Fig2}(a) the data show a noticeable increase of spectral weight above the Fermi energy $E_\text{F}$. We associate this signal with a transient electron population of the cone-shaped $\pi^*$ band at the \Kbar-point of the SCG substrate, generated due to the absorption of the pump pulse. Like the signal from the $\pi$ band, also the photoemitted electrons from the $\pi^*$ band are diffracted by the SnPc layer into the energy-momentum region probed in the experiment. A difference intensity map calculated from the data in Fig. \ref{fig:Fig2}(a) and Fig. \ref{fig:Fig2}(b) and shown in the left panel of Fig. \ref{fig:Fig2}(c) emphasizes the photoinduced changes more distinctly. This representation of the data additionally reveals a decrease in spectral weight below $E_\text{F}$. This decrease results partly from the transient generation of photo-holes in the SCG substrate. Interestingly, we observe residuals of these changes at time delays $\Delta t$ as large as $\SI{50}{\pico s}$ [right panel of Fig. \ref{fig:Fig2}(c)]. A corresponding signal is absent in the $\pi^*$ band data. 
%We interpret this asymmetry as indicator for vibrational disorder in the adsorbate layer due to transient heating, as will be elaborated in more detail below.

%For the quantitative evaluation of the temporal evolution of the ULEPD signal, we extracted momentum distribution curves (MDC) from difference LEPD spectra at selected energies. Figure \ref{fig:Fig2}(d) shows MDCs from the two regions marked by the gray bars in Fig. \ref{fig:Fig2}(c) for $\Delta t = \SI{}{fs}$ illustratung the momentum dependence of the spectral weight changes above and below $E_F$. Each data set was fitted with two Gaussians to account for the two peaks in the MDCs representing the two branches of the $\pi$ and $\pi^*$ cone, respectively. The integral of the Gaussians were taken as a measure for the increase (decrease) of spectral weight above (below) $E_F$. Figure \ref{fig:Fig2}(e) shows difference intensity transients of $\pi$ and $\pi^*$ band evaluated from the sum of the integrals of four MDCs each in the energy windows of $-\SI{}{eV} < E-E_F < -\SI{}{eV}$ and $\SI{}{eV} < E-E_F < \SI{}{eV}$ \textcolor{red}{\cite{Supplemental}}. 
For the quantitative evaluation of the temporal evolution of the ULEPD signal, we separately integrated the difference signal along the two arms of $\pi$ and $\pi^*$ band in the energy windows between $\SI{117}{\milli eV}$ and $\SI{23}{\milli eV}$ below and above $E_{\text{F}}$. The blue and red data points in Fig. \ref{fig:Fig2}(d) are the respective signal intensities $I_{\pi}$ and $I_{\pi^*}$ as a function of $\Delta t$ resulting from this analysis. Both transients show a sudden onset, indicating the initial photoexcitation of the system by the NIR pump pulses at $\Delta t = \SI{0}{\femto s}$. Fitting the signal rise yields a characteristic rise time of the onset of $\approx\SI{115}{\femto s}$. We consider this value as an upper limit for the time resolution of the ULEPD experiment, i.e., the NIR pump-photoelectron diffraction probe cross-correlation signal. The value agrees reasonably well with the full width at half maximum of a reference NIR pump-NUV probe cross-correlation of $\approx \SI{100}{\femto s}$ \cite{Hein2020}. Therefore, we conclude that the photoelectron pulse is not significantly broadened on the way to the surface and suspect that in the current setup of the experiment, the limiting factor for the time resolution is the  temporal width of \SI{95}{\femto s} of the NUV pulse. We expect that this value can be further improved in the future with reasonable efforts \cite{Gauthier2020, Yan2021}.  

The further evolutions of the transient diffraction signals from $\pi$ band and $\pi^*$ band follow a two-phase recovery.
%for both, the signal above and the signal below $E_F$. 
A fit yields characteristic time constants  $\tau_1=\SI{280\pm100}{\femto s}$ and $\tau_2=\SI{5.3\pm0.7}{\pico s}$ for the $\pi$ band and $\tau_1=\SI{360\pm100}{\femto s}$ and $\tau_2=\SI{5.6\pm0.2}{\pico s}$ for the $\pi^*$ band.
%with the longer time constant being slightly but significantly larger for the transient from below $E_F$.  The two-component decay and the values of the associated time constants are in good agreement with findings of previous studies on the decay of hot carrier populations in graphitic materials at \Kbar \cite{Kampfrath2005, Wang2010, Johannsen2013}. 
This two-component decay and the associated time constants $\tau_1$ and $\tau_2$ are in good quantitative agreement with findings of previous studies on the decay of hot carrier populations in graphitic materials \cite{Kampfrath2005, Wang2010, Johannsen2013}.
It has been interpreted as signature for the interaction of the photoexcited carrier distribution with the optical and acoustic phonon bath, respectively \cite{Yan2009, Malic2011}. The diffraction signal obviously carries information about the decay of the hot carrier population in the substrate and the time scales on which the excess energy of the photoexcited electrons is dissipated into the SCG lattice. While the transient ULEPD signal above $E_\text{F}$ has essentially regained its original value at around \SI{20}{\pico s}, the recovery of the signal below $E_\text{F}$ is not completed even after \SI{50}{\pico s}, i.e., the maximum time delay probed in this study. Instead, the signal remains at an intensity level that is reduced by $\approx \SI{9}{\%}$ compared to the photoelectron diffraction signal before time zero \cite{Supplemental}. The value was determined by averaging the ULEPD signal in the time window between \SI{20}{\pico s} and \SI{50}{\pico s}. This observation suggests that the signal changes below $E_\text{F}$ contain additional information beyond pure substrate carrier dynamics, as will be explained in the following.  

\begin{figure}
        \includegraphics[width=1\linewidth]{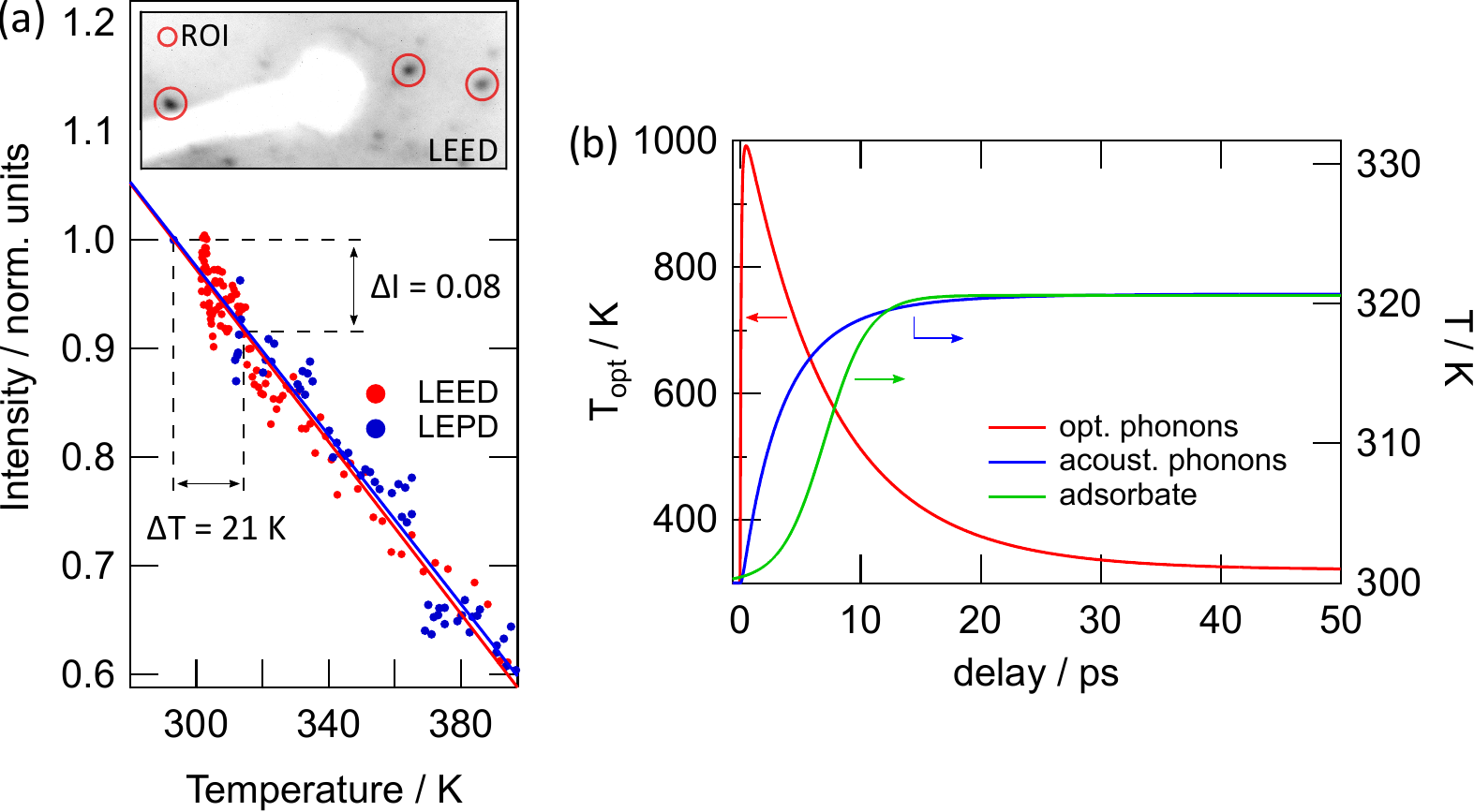}
        \setlength{\abovecaptionskip}{-10pt}
        \caption{Temperature of the adsorbate layer and TTM model for SCG. (a) Comparison of the temperature dependence of LEED and LEPD diffraction intensities from SnPc/SCG. LEPD intensities were determined from the same energy-momentum field of views as used for the analysis of the time-resolved data. LEED intensities were determined from the diffraction spots marked in the section of a LEED image in the inset by the region of interest ROI. The diffraction order of the selected spots correspond to the diffraction order of the LEPD signal. A linear fit to the temperature dependent LEPD data was used to relate the transient changes in the diffraction intensity of the time-resolved data to a temperature rise $\Delta T$. The relative intensity change $\Delta I=0.08$ marked in the figure corresponds to the change in the ULEPD signal at large time delays. (b) Temperature transients simulated for the optical and acoustic phonon bath of SCG within a TTM based on experimental trARPES data of graphite \cite{Stange2015} in comparison with the adsorbate temperature transient derived from the ULEPD data under consideration of the temperature dependence of the LEPD signal shown in (a).}
    	\label{fig:Fig3}
\end{figure}

The integral intensities of electron diffraction peaks show a characteristic temperature dependence, i.e., a decrease with increasing temperature. This behavior is due to vibrational disorder of the diffracting lattice and can be described quantitatively by the Debye-Waller factor $DFW=\exp(-1/3\cdot|\Vec{g}|^2\cdot\langle u^2 \rangle)$ \cite{vanHove1986}. Here, $\Vec{g}$ is a reciprocal lattice vector and $\langle u^2\rangle$ the mean square displacement of the scattering center of the lattice due to vibrational excitation. In thermal equilibrium, also the LEPD signal resulting from diffraction by an ordered adsorbate layer is attenuated by the Debye-Waller factor and quantitatively follows the temperature dependence of the LEED signal \cite{Jauernik2018}. The question arises whether the reduction of the ULEPD signal that we observe for SnPc/SCG for long time delays can be similarly linked to a \textit{transient} change in the adsorbate vibrational disorder. In order to relate our result to a change in the temperature of the adsorbate layer, we performed static, temperature-dependent LEPD and LEED reference measurements. Figure \ref{fig:Fig3}(a) compares relative changes in the LEPD and LEED peak intensities 
%$\Delta I/I = [I(T)-I(T=\SI{300}{K})]/I(\SI{300}{K})$ 
in the temperature range between $T= \SI{300}{K}$ and $T= \SI{400}{K}$. To correctly account for the $\Vec{g}$-dependence of the DWF, we selected  for the quantitative analysis LEED spots of a diffraction order corresponding to that of the LEPD signal [see inset of Fig. \ref{fig:Fig3}(a)]. The very good agreement in the temperature dependence of LEED and LEPD data confirms that both techniques provide equivalent information on the thermal excitation of the ordered adsorbate layer. Note that thermal broadening of the Fermi edge in this temperature range could account only for an intensity change in the LEPD signal of $\approx \SI{4.6}{\%}$ in comparison to the observed \SI{40}{\%} change \cite{Supplemental}. The dashed horizontal lines in Fig. \ref{fig:Fig3}(a) indicate the relative change in the ULEPD intensity of the $\pi$ band at large time delays, $\Delta I = \SI{8}{\%}$. The static diffraction data enable us to relate $\Delta I$ to a temperature rise of the SnPc layer of $\approx\SI{21}{K}$. This value agrees very well with an increase in the sample equilibrium temperature $\Delta T = \SI{16}{K}$ calculated under consideration of the heat capacity of SCG and the absorbed pump laser fluence \cite{Supplemental}. 
%For an absorption depth of \SI{50}{\nano m} at the pump wavelength of \SI{840}{\nano m} \textcolor{red}{[Ref.]} and a room temperature heat capacity of graphite of $\SI{706}{J kg^{-1} K^{-1}}$ \cite{Picard2007} we calculate a value of  . 
%Here, we assumed a rectangular absorption depth profile and neglected losses due to diffusion. 
This is a further confirmation of the interpretation of the long-term changes in the ULEPD signal in terms of vibrational disorder of the adsorbate layer due to a transient heating following the pulsed photoexcitation.     

To quantify the temporal evolution of the vibrational disorder in the adsorbate layer, the substrate electronic contribution to the ULEPD $\pi$-band signal must be separated from the lattice contribution of the adsorbate. The high symmetry of the electronic band structure of graphite with respect to $E_\text{F}$ \cite{Ahuja1997, Fermi} and the observation of almost identical electron and hole relaxation times suggests to use the ULEPD transient of the $\pi^*$ band as a background-free reference for the electronic contribution.
%This is achieved by using the ULEPD transient for $E-E_F>\SI{0}{eV}$ as a background-free reference for the electronic response of the substrate in the $\pi$ band. Due to the high symmetry of the electronic band structure with respect to $E_F$ \cite{Fermi}, this is a very reasonable assumption, which is supported by our observation of essentially identical electron and hole relaxation times $\tau_1$ and $\tau_2$.
%\textcolor{red}{Here, we make the assumption that the decay dynamics of the excited electron and hole distribution in SCG follow the same time dependence. This is justified by the high symmetry of the electronic band structure regarding the occupied $\pi$ band and the unoccupied $\pi^*$ band. 
%, as well as the fact that the Fermi energy is at most a few \SI{10}{\milli eV} below the contact point (Dirac point) of both bands at the $H$-point of SCG \cite{Mikitik2006, Orlita2008, Na2019}. The observation that the tip of the cone structure in the ARPES spectra of SnPc/SCG before photoexcitation touches $E_F$ [Fig. \ref{fig:Fig1}(a)] implies that, if at all,  this situation is not significantly adsorption by the adsorption of the SnPc molecules.}
To account for the difference in the matrix elements for photoemission from $\pi$ and $\pi^*$ band \cite{Gruneis2008, Stange2015}, we normalized the ULEPD transients below and above $E_\text{F}$ to the average signal between \SI{1}{\pico s} and \SI{2}{\pico s}. The direct comparison of the two transients shows that the signal decay/recovery in this time window is governed in both cases by the carrier dynamics in the SCG substrate \cite{Supplemental}. The contribution from the formation of a vibrational disorder of the adsorbate layer becomes relevant only on longer time scales. 
%The transients shown in Fig. \ref{fig:Fig2}(e) have already been normalized in this way.
In the sum of the two transients normalized in this way, the respective electronic contributions to the ULEPD signal compensate each other \cite{Supplemental}.
%The sum of the two transients normalized in this way cancels out any electronic contribution to the ULEPD signal. 
The resulting signal is shown as green data points in Fig. \ref{fig:Fig2}(d) and represents the lattice response of the adsorbate layer due to the build-up of vibrational disorder as interrogated by the diffraction signal below  $E_\text{F}$. 
%The result is  
%They represent a transient whose changes are determined exclusively by the diffraction at the adsorbate layer, i.e., by the time evolution of the diffraction signal  as the adsobate layer heats up. 
%Right after time zero we observe a pronounced negative peaking in the signal. We associate this peculiarity with the nascient carrier momentum anisotropy in graphite that is created in the photoexcitation process, but decays within $\approx$\SI{100}{\femto s} \cite{Aeschlimann2017, Beyer2023}. More specifically, we suspect that the experiment probes non equivalent momentum regions in the electronic structure for electrons and holes due to a tilt and bending of the energy-momentum cut selected by the analyzer. 
The first few ps after photoexcitation, the transient remains close to zero indicating that the adsorbate layer is still hardly excited. At about $\approx \SI{3}{\pico s}$ the signal starts to decrease and finally levels off between \SI{10}{\pico s} and \SI{20}{\pico s} to the final value that is still observed at \SI{50}{\pico s}. Fitting a sigmoidal function to the data [solid green line in Fig. \ref{fig:Fig2}(d)] yields a characteristic delay of the signal decay of $t_{\text{half}}=\SI{6.4\pm0.4}{\pico s}$ with respect to time zero, a time constant that is defined by the time at which half of the final value at \SI{50}{\pico s} is reached. For the characteristic rise time of the signal the fit yields $\tau_{\text{sig}}=\SI{2\pm0.3}{\pico s}$.

In order to identify the relevant excitation channel for the heating of the adsorbate layer we simulated temperature transients for optical phonon gas and acoustic phonon gas in the SCG substrate using a three-temperature model (TTM) \cite{Perfetti2007}. The coupling parameters describing the interaction between initial electronic excitation and the lattice degrees of freedom of SCG were deduced from fits of the TTM to electron temperature transients of graphite evaluated from previous trARPES data \cite{Stange2015}. The pump fluence was adjusted to yield a final temperature rise in the substrate of $\Delta T =\SI{21}{K}$, corresponding to the temperature rise determined experimentally from the ULEPD data for the SnPc overlayer.    
%provides on a qualitative level insight into the relevant coupling pathways between substrate and adsorbate. 
Simulation results together with the fit to the ULEPD data representing the adsorbate lattice dynamics are shown in Fig. \ref{fig:Fig3}(b). 
%The exceptionally short heating and cooling times of electrons and optical phonons are in the order of a few \SI{10}{\femto s} to a few \SI{100}{\femto s}. The mismatch between these time scales and the characteristic heating time of the SnPc layer  preclude any appreciable direct coupling of these subsystems with the vibrational degrees of freedom of the adsorbate. 
The direct comparison of the transients and the related characteristic time scales for heating and cooling hints to an energy transfer between substrate and adsorbate layer mainly governed by an interaction between the acoustic phonon bath of graphite an the vibrational degrees of freedom of the adsorbate.

In summary we presented a low-energy photoelectron diffraction experiment with exceptional surface sensitivity and particularly suitable for the study of long-range structural order dynamics in molecular adsorbate layers. The experimentally proven time resolution of $\approx \SI{100}{\femto s}$ outperforms reported values of state-of-the-art ULEED schemes by a factor of 10 \cite{Storeck2017}. As a trARPES-based technique the experimental data intrinsically contain also information on carrier and electronic band structure dynamics providing unique capabilities to directly correlate ultrafast processes associated with the electronic \textit{and} structural degrees of freedom at surfaces. Limited by the capabilities of the electron analyzer used, the data presented in this paper was collected along a one-dimensional cut in momentum space, which significantly limits the potential depth of information of the technique. A momentum microscope can overcome this restriction \cite{Kromker2008} so that similar to ULEED the full two-dimensional ULEPD signal of an adsorbate layer becomes accessible, including ultrafast diffuse scattering information \cite{Stern2017, Seiler2021}. The use of XUV-pulses from high-harmonic generation sources or free electron lasers for the generation of the probe photoelectron pulse \cite{Rohwer2011, Hellmann2010} will here considerably extend the accessible momentum field of view. It will also significantly increase the bandwidth of the detected photoelecton spectra, which will make it easier to separate the excited carrier and adsorbate lattice contributions of the ULEPD signal.              

\section*{\label{sec:Acknowledgements}Acknowledgements}
This work was supported by the German Research Foundation (DFG), Project ID 433458487.

\bibliography{bibliography}

%merlin.mbs apsrev4-1.bst 2010-07-25 4.21a (PWD, AO, DPC) hacked
%Control: key (0)
%Control: author (8) initials jnrlst
%Control: editor formatted (1) identically to author
%Control: production of article title (-1) disabled
%Control: page (0) single
%Control: year (1) truncated
%Control: production of eprint (0) enabled
\begin{thebibliography}{6}%
\makeatletter
\providecommand \@ifxundefined [1]{%
 \@ifx{#1\undefined}
}%
\providecommand \@ifnum [1]{%
 \ifnum #1\expandafter \@firstoftwo
 \else \expandafter \@secondoftwo
 \fi
}%
\providecommand \@ifx [1]{%
 \ifx #1\expandafter \@firstoftwo
 \else \expandafter \@secondoftwo
 \fi
}%
\providecommand \natexlab [1]{#1}%
\providecommand \enquote  [1]{``#1''}%
\providecommand \bibnamefont  [1]{#1}%
\providecommand \bibfnamefont [1]{#1}%
\providecommand \citenamefont [1]{#1}%
\providecommand \href@noop [0]{\@secondoftwo}%
\providecommand \href [0]{\begingroup \@sanitize@url \@href}%
\providecommand \@href[1]{\@@startlink{#1}\@@href}%
\providecommand \@@href[1]{\endgroup#1\@@endlink}%
\providecommand \@sanitize@url [0]{\catcode `\\12\catcode `\$12\catcode `\&12\catcode `\#12\catcode `\^12\catcode `\_12\catcode `\%12\relax}%
\providecommand \@@startlink[1]{}%
\providecommand \@@endlink[0]{}%
\providecommand \url  [0]{\begingroup\@sanitize@url \@url }%
\providecommand \@url [1]{\endgroup\@href {#1}{\urlprefix }}%
\providecommand \urlprefix  [0]{URL }%
\providecommand \Eprint [0]{\href }%
\providecommand \doibase [0]{http://dx.doi.org/}%
\providecommand \selectlanguage [0]{\@gobble}%
\providecommand \bibinfo  [0]{\@secondoftwo}%
\providecommand \bibfield  [0]{\@secondoftwo}%
\providecommand \translation [1]{[#1]}%
\providecommand \BibitemOpen [0]{}%
\providecommand \bibitemStop [0]{}%
\providecommand \bibitemNoStop [0]{.\EOS\space}%
\providecommand \EOS [0]{\spacefactor3000\relax}%
\providecommand \BibitemShut  [1]{\csname bibitem#1\endcsname}%
\let\auto@bib@innerbib\@empty
%</preamble>
\bibitem [{\citenamefont {Erk}\ \emph {et~al.}(2023)\citenamefont {Erk}, \citenamefont {Opitz}, \citenamefont {Hein}, \citenamefont {Jauernik},\ and\ \citenamefont {Bauer}}]{Erk2023}%
  \BibitemOpen
  \bibfield  {author} {\bibinfo {author} {\bibfnamefont {H.}~\bibnamefont {Erk}}, \bibinfo {author} {\bibfnamefont {K.}~\bibnamefont {Opitz}}, \bibinfo {author} {\bibfnamefont {P.}~\bibnamefont {Hein}}, \bibinfo {author} {\bibfnamefont {S.}~\bibnamefont {Jauernik}}, \ and\ \bibinfo {author} {\bibfnamefont {M.}~\bibnamefont {Bauer}},\ }\href@noop {} {\bibfield  {journal} {\bibinfo  {journal} {J. Phys. Condens. Matter}\ }\textbf {\bibinfo {volume} {35}},\ \bibinfo {pages} {095501} (\bibinfo {year} {2023})}\BibitemShut {NoStop}%
\bibitem [{\citenamefont {Ooi}\ \emph {et~al.}(2006)\citenamefont {Ooi}, \citenamefont {Rairkar},\ and\ \citenamefont {Adams}}]{Ooi2006}%
  \BibitemOpen
  \bibfield  {author} {\bibinfo {author} {\bibfnamefont {N.}~\bibnamefont {Ooi}}, \bibinfo {author} {\bibfnamefont {A.}~\bibnamefont {Rairkar}}, \ and\ \bibinfo {author} {\bibfnamefont {J.~B.}\ \bibnamefont {Adams}},\ }\href {\doibase 10.1016/j.carbon.2005.07.036} {\bibfield  {journal} {\bibinfo  {journal} {Carbon}\ }\textbf {\bibinfo {volume} {44}},\ \bibinfo {pages} {231} (\bibinfo {year} {2006})}\BibitemShut {NoStop}%
\bibitem [{\citenamefont {Picard}\ \emph {et~al.}(2007)\citenamefont {Picard}, \citenamefont {Burns},\ and\ \citenamefont {Roger}}]{Picard2007}%
  \BibitemOpen
  \bibfield  {author} {\bibinfo {author} {\bibfnamefont {S.}~\bibnamefont {Picard}}, \bibinfo {author} {\bibfnamefont {D.~T.}\ \bibnamefont {Burns}}, \ and\ \bibinfo {author} {\bibfnamefont {P.}~\bibnamefont {Roger}},\ }\href {\doibase 10.1088/0026-1394/44/5/005} {\bibfield  {journal} {\bibinfo  {journal} {Metrologia}\ }\textbf {\bibinfo {volume} {44}},\ \bibinfo {pages} {294} (\bibinfo {year} {2007})}\BibitemShut {NoStop}%
\bibitem [{\citenamefont {Djurišić}\ and\ \citenamefont {Li}(1999)}]{Djurišić1999}%
  \BibitemOpen
  \bibfield  {author} {\bibinfo {author} {\bibfnamefont {A.~B.}\ \bibnamefont {Djurišić}}\ and\ \bibinfo {author} {\bibfnamefont {E.~H.}\ \bibnamefont {Li}},\ }\href {\doibase 10.1063/1.369370} {\bibfield  {journal} {\bibinfo  {journal} {Journal of Applied Physics}\ }\textbf {\bibinfo {volume} {85}},\ \bibinfo {pages} {7404} (\bibinfo {year} {1999})}\BibitemShut {NoStop}%
\bibitem [{\citenamefont {Gr{\"{u}}neis}\ \emph {et~al.}(2008)\citenamefont {Gr{\"{u}}neis}, \citenamefont {Attaccalite}, \citenamefont {Rubio}, \citenamefont {Molodtsov}, \citenamefont {Vyalikh}, \citenamefont {Fink}, \citenamefont {Follath},\ and\ \citenamefont {Pichler}}]{Gruneis2008}%
  \BibitemOpen
  \bibfield  {author} {\bibinfo {author} {\bibfnamefont {A.}~\bibnamefont {Gr{\"{u}}neis}}, \bibinfo {author} {\bibfnamefont {C.}~\bibnamefont {Attaccalite}}, \bibinfo {author} {\bibfnamefont {A.}~\bibnamefont {Rubio}}, \bibinfo {author} {\bibfnamefont {S.~L.}\ \bibnamefont {Molodtsov}}, \bibinfo {author} {\bibfnamefont {D.~V.}\ \bibnamefont {Vyalikh}}, \bibinfo {author} {\bibfnamefont {J.}~\bibnamefont {Fink}}, \bibinfo {author} {\bibfnamefont {R.}~\bibnamefont {Follath}}, \ and\ \bibinfo {author} {\bibfnamefont {T.}~\bibnamefont {Pichler}},\ }\href {\doibase 10.1002/pssb.200879662} {\bibfield  {journal} {\bibinfo  {journal} {Phys. Status Solidi B Basic Res.}\ }\textbf {\bibinfo {volume} {245}},\ \bibinfo {pages} {2072} (\bibinfo {year} {2008})}\BibitemShut {NoStop}%
\bibitem [{\citenamefont {Stange}\ \emph {et~al.}(2015)\citenamefont {Stange}, \citenamefont {Sohrt}, \citenamefont {Yang}, \citenamefont {Rohde}, \citenamefont {Janssen}, \citenamefont {Hein}, \citenamefont {Oloff}, \citenamefont {Hanff}, \citenamefont {Rossnagel},\ and\ \citenamefont {Bauer}}]{Stange2015}%
  \BibitemOpen
  \bibfield  {author} {\bibinfo {author} {\bibfnamefont {A.}~\bibnamefont {Stange}}, \bibinfo {author} {\bibfnamefont {C.}~\bibnamefont {Sohrt}}, \bibinfo {author} {\bibfnamefont {L.~X.}\ \bibnamefont {Yang}}, \bibinfo {author} {\bibfnamefont {G.}~\bibnamefont {Rohde}}, \bibinfo {author} {\bibfnamefont {K.}~\bibnamefont {Janssen}}, \bibinfo {author} {\bibfnamefont {P.}~\bibnamefont {Hein}}, \bibinfo {author} {\bibfnamefont {L.-P.}\ \bibnamefont {Oloff}}, \bibinfo {author} {\bibfnamefont {K.}~\bibnamefont {Hanff}}, \bibinfo {author} {\bibfnamefont {K.}~\bibnamefont {Rossnagel}}, \ and\ \bibinfo {author} {\bibfnamefont {M.}~\bibnamefont {Bauer}},\ }\href {\doibase 10.1103/PhysRevB.92.184303} {\bibfield  {journal} {\bibinfo  {journal} {Phys. Rev. B}\ }\textbf {\bibinfo {volume} {92}},\ \bibinfo {pages} {184303} (\bibinfo {year} {2015})}\BibitemShut {NoStop}%
\end{thebibliography}%


@book{vanHove1986,
  added-at = {},
  address = {Berlin},
  author = {Van Hove, M. A.  and Weinberg, W. H. and Chan, C.-M. },
  biburl = {},
  description = {Low-Energy Electron Diffraction},
  interhash = {},
  intrahash = {},
  isbn = {},
  keywords = {},
  publisher = {Springer},
  refid = {},
  timestamp = {},
  title = {Low-Energy Electron Diffraction},
  url = {},
  year = 1986
}


@article{Reich2002,
  title = {Tight-binding description of graphene},
  author = {Reich, S. and Maultzsch, J. and Thomsen, C. and Ordej\'on, P.},
  journal = {Phys. Rev. B},
  volume = {66},
  issue = {3},
  pages = {035412},
  numpages = {5},
  year = {2002},
  month = {Jul},
  publisher = {American Physical Society},
  doi = {10.1103/PhysRevB.66.035412},
  url = {https://link.aps.org/doi/10.1103/PhysRevB.66.035412}
}

@article{Prybyla1992,
  title = {Femtosecond Time-Resolved Surface Reaction: Desorption of Co from Cu(111) in < 325 fsec},
  author = {Prybyla, J. A. and  Tom, H. W. K. and Aumiller, G. D.},
  journal = {Phys. Rev. Lett.},
  volume = {68},
  issue = {4},
  pages = {503},
  numpages = {},
  year = {1992},
  month = {Jan},
  publisher = {American Physical Society},
  doi = {10.1103/PhysRevLett.68.503},
  url = {https://journals.aps.org/prl/pdf/10.1103/PhysRevLett.68.503}
}


@article{Petek2000,
abstract = {The dynamics of cesium atom motion above the copper(111) surface following electronic excitation with light was studied with femtosecond (10(-15) seconds) time resolution. Unusual changes in the surface electronic structure within 160 femtoseconds after excitation, observed by time-resolved two-photon photoemission spectroscopy, are attributed to atomic motion in a copper-cesium bond-breaking process. Describing the change in energy of the cesium antibonding state with a simple classical model provides information on the mechanical forces acting on cesium atoms that are "turned on" by photoexcitation. Within 160 femtoseconds, the copper-cesium bond extends by 0.35 angstrom from its equilibrium value.},
author = {Petek, H and Weida, Mj and Nagano, H and Ogawa, S},
file = {:C\:/Users/mbau/AppData/Local/Mendeley Ltd./Mendeley Desktop/Downloaded/Petek et al. - 2000 - Real-time observation of adsorbate atom motion above a metal surface.svn-base:svn-base},
issn = {1095-9203},
journal = {Science},
month = {may},
number = {5470},
pages = {1402--4},
pmid = {10827946},
title = {{Real-time observation of adsorbate atom motion above a metal surface}},
url = {http://www.ncbi.nlm.nih.gov/pubmed/10827946},
volume = {288},
year = {2000}
}

@article{Backus2005,
abstract = {The laser-induced movement of CO molecules over a platinum surface was followed in real time by means of ultrafast vibrational spectroscopy. Because the CO molecules bound on different surface sites exhibit different C-O stretch vibrational frequencies, the site-to-site hopping, triggered by excitation with a laser pulse, can be determined from subpicosecond changes in the vibrational spectra. The unexpectedly fast motion-characterized by a 500-femtosecond time constant-reveals that a rotational motion of the CO molecules, rather than pure translation, is required for this diffusion process. This conclusion is corroborated by density functional theory calculations.},
author = {Backus, Ellen H.G. and Eichler, Andreas and Kleyn, Aart W. and Bonn, Mischa},
doi = {10.1126/science.1120693},
file = {:C\:/Users/mbau/Documents/work/Kiel/Paper f{\"{u}}r Bibliothek/neuer/science.1120693.pdf:pdf},
issn = {00368075},
journal = {Science},
number = {5755},
pages = {1790--1793},
title = {{Chemistry: Real-time observation of molecular motion on a surface}},
volume = {310},
year = {2005}
}

@article{Wallauer2021,
abstract = {Frontier orbitals determine fundamental molecular properties such as chemical reactivities. Although electron distributions of occupied orbitals can be imaged in momentum space by photoemission tomography, it has so far been impossible to follow the momentum-space dynamics of a molecular orbital in time, for example, through an excitation or a chemical reaction. Here, we combined timeresolved photoemission using high laser harmonics and a momentum microscope to establish a tomographic, femtosecond pump-probe experiment of unoccupied molecular orbitals. We measured the full momentum-space distribution of transiently excited electrons, connecting their excited-state dynamics to real-space excitation pathways. Because in molecules this distribution is closely linked to orbital shapes, our experiment may, in the future, offer the possibility of observing ultrafast electron motion in time and space.},
author = {Wallauer, R. and Raths, M. and Stallberg, K. and M{\"{u}}nster, L. and Brandstetter, D. and Yang, X. and G{\"{u}}dde, J. and Puschnig, P. and Soubatch, S. and Kumpf, C. and Bocquet, F. C. and Tautz, F. S. and H{\"{o}}fer, U.},
doi = {10.1126/science.abf3286},
eprint = {},
file = {:C\:/Users/mbau/Documents/work/Kiel/Paper f{\"{u}}r Bibliothek/neuer/pap.pdf:pdf},
issn = {10959203},
journal = {Science},
number = {6533},
pages = {1056--1059},
pmid = {33602865},
title = {{Tracing orbital images on ultrafast time scales}},
volume = {371},
year = {2021}
}

@article{Diesen2021,
abstract = {We use a pump-probe scheme to measure the time evolution of the C K-edge x-ray absorption spectrum from CO/Ru(0001) after excitation by an ultrashort high-intensity optical laser pulse. Because of the short duration of the x-ray probe pulse and precise control of the pulse delay, the excitation-induced dynamics during the first picosecond after the pump can be resolved with unprecedented time resolution. By comparing with density functional theory spectrum calculations, we find high excitation of the internal stretch and frustrated rotation modes occurring within 200 fs of laser excitation, as well as thermalization of the system in the picosecond regime. The ∼100 fs initial excitation of these CO vibrational modes is not readily rationalized by traditional theories of nonadiabatic coupling of adsorbates to metal surfaces, e.g., electronic frictions based on first order electron-phonon coupling or transient population of adsorbate resonances. We suggest that coupling of the adsorbate to nonthermalized electron-hole pairs is responsible for the ultrafast initial excitation of the modes.},
author = {Diesen, Elias and Wang, Hsin Yi and Schreck, Simon and Weston, Matthew and Ogasawara, Hirohito and Larue, Jerry and Perakis, Fivos and Dell'Angela, Martina and Capotondi, Flavio and Giannessi, Luca and Pedersoli, Emanuele and Naumenko, Denys and Nikolov, Ivaylo and Raimondi, Lorenzo and Spezzani, Carlo and Beye, Martin and Cavalca, Filippo and Liu, Boyang and Gladh, J{\"{o}}rgen and Koroidov, Sergey and Miedema, Piter S. and Costantini, Roberto and Heinz, Tony F. and Abild-Pedersen, Frank and Voss, Johannes and Luntz, Alan C. and Nilsson, Anders},
doi = {10.1103/PhysRevLett.127.016802},
file = {:C\:/Users/mbau/Documents/work/Kiel/Paper f{\"{u}}r Bibliothek/neuer/PhysRevLett.127.016802.pdf:pdf},
issn = {10797114},
journal = {Phys. Rev. Lett.},
keywords = {doi:10.1103/PhysRevLett.127.016802 url:https://doi},
number = {1},
pages = {16802},
pmid = {34270277},
publisher = {American Physical Society},
title = {{Ultrafast Adsorbate Excitation Probed with Subpicosecond-Resolution X-Ray Absorption Spectroscopy}},
url = {https://doi.org/10.1103/PhysRevLett.127.016802},
volume = {127},
year = {2021}
}

@article{Gulde2014,
abstract = {Two-dimensional systems such as surfaces and molecular monolayers exhibit a multitude of intriguing phases and complex transitions. Ultrafast structural probing of such systems offers direct time-domain information on internal interactions and couplings to a substrate or bulk support.We have developed ultrafast low-energy electron diffraction and investigate in transmission the structural relaxation in a polymer/graphene bilayer system excited out of equilibrium. The laser-pump/electron-probe scheme resolves the ultrafast melting of a polymer superstructure consisting of folded-chain crystals registered to a free-standing graphene substrate. We extract the time scales of energy transfer across the bilayer interface, the loss of superstructure order, and the appearance of an amorphous phase with short-range correlations. The high surface sensitivity makes this experimental approach suitable for numerous problems in ultrafast surface science.},
author = {Gulde, Max and Schweda, Simon and Storeck, Gero and Maiti, Manisankar and Yu, Hak Ki and Wodtke, Alec M. and Sch\"afer, Sascha and Ropers, Claus},
doi = {10.1126/science.1250658},
file = {:C\:/Users/mbau/Documents/work/Kiel/Paper f{\"{u}}r Bibliothek/neuer/gulde.sm.pdf:pdf},
issn = {10959203},
journal = {Science},
number = {6193},
pages = {200--204},
title = {{Ultrafast low-energy electron diffraction in transmission resolves polymer/graphene superstructure dynamics}},
volume = {345},
year = {2014}
}

@article{DellAngela2013,
author = {Dell'Angela, M and Anniyev, T and Beye, M and Coffee, R and F{\"{o}}hlisch, A and Gladh, J and Katayama, T and Kaya, S and Krupin, O and Larue, J and M{\o}gelh{\o}j, A and Nordlund, D and N{\o}rskov, J K and Oberg, H and Ogasawara, H and Ostr{\"{o}}m, H and Pettersson, L G M and Schlotter, W F and Sellberg, J A and Sorgenfrei, F and Turner, J J and Wolf, M and Wurth, W and Nilsson, A},
file = {:C\:/Users/mbau/Documents/work/Kiel/Paper f{\"{u}}r Bibliothek/neuer/science.1231711.pdf:pdf},
journal = {Science},
number = {March},
pages = {1302--1306},
title = {{Real-Time Observation of Surface Bond Breaking with an X-ray Laser}},
volume = {339},
year = {2013}
}

@article{Stepan2005,
abstract = {Diffusion of atomic oxygen on a vicinal Pt(111) surface induced by femtosecond laser pulses has been studied using optical second-harmonic generation as a sensitive in situ probe of the step coverage. Time-resolved studies of the hopping rate for step-terrace diffusion with a two-pulse correlation scheme reveal a time constant of 1.5 ps for the energy transfer from the electronic excitation of the substrate to the frustrated translations of the adsorbate. {\textcopyright} 2005 The American Physical Society.},
author = {St{\'{e}}p{\'{a}}n, K. and G{\"{u}}dde, J. and H{\"{o}}fer, U.},
doi = {10.1103/PhysRevLett.94.236103},
file = {:C\:/Users/mbau/Documents/work/Kiel/Paper f{\"{u}}r Bibliothek/neuer/PhysRevLett.94.236103.pdf:pdf},
issn = {00319007},
journal = {Phys. Rev. Lett.},
number = {23},
pages = {236103},
title = {{Time-resolved measurement of surface diffusion induced by femtosecond laser pulses}},
volume = {94},
year = {2005}
}

@article{Frigge2017,
abstract = {Ultrafast diffraction techniques enable us to observe laser-induced structural changes at the atomic scale and with high temporal resolution. A decade of such experiments has indicated that structural changes on surfaces are several orders of magnitude slower than changes in bulk materials, raising the question of whether there is a fundamental limit for low-dimensional systems. Tim Frigge et al. apply laser excitation to a one-dimensional wire of indium atoms on a silicon surface and find that structural changes take place on femtosecond timescales. This short timescale is made possible by electronic coupling to the underlying surface and indicates that structural changes at the surface can, in principle, be as fast as in the bulk material. The findings point to a new method for controlling the dynamic structural responses of solids to laser excitation.},
author = {Frigge, T. and Hafke, B. and Witte, T. and Krenzer, B. and Streub{\"{u}}hr, C. and {Samad Syed}, A. and {Mik{\v{s}}i{\'{c}} Trontl}, V. and Avigo, I. and Zhou, P. and Ligges, M. and {Von Der Linde}, D. and Bovensiepen, U. and {Horn-Von Hoegen}, M. and Wippermann, S. and L{\"{u}}cke, A. and Sanna, S. and Gerstmann, U. and Schmidt, W. G.},
doi = {10.1038/nature21432},
file = {:C\:/Users/mbau/AppData/Local/Mendeley Ltd./Mendeley Desktop/Downloaded/Frigge et al. - 2017 - Optically excited structural transition in atomic wires on surfaces at the quantum limit.pdf:pdf},
isbn = {1476-4687 (Electronic)\r0028-0836 (Linking)},
issn = {14764687},
journal = {Nature},
number = {7649},
pages = {207--211},
pmid = {28355177},
publisher = {Nature Publishing Group},
title = {{Optically excited structural transition in atomic wires on surfaces at the quantum limit}},
url = {http://dx.doi.org/10.1038/nature21432},
volume = {544},
year = {2017}
}

@article{Vogelgesang2017,
abstract = {We introduce ultrafast low-energy electron diffraction (ULEED) in backscattering for the study of structural dynamics at surfaces. Using a tip-based source of ultrashort electron pulses, we investigate the optically-driven transition between charge-density wave phases at the surface of 1T-TaS2. Employing spot-profile analysis enabled by the large transfer width of the instrument, we resolve the phase-ordering kinetics in the nascent incommensurate charge-density wave phase. We attribute the observed power-law scaling of the correlation length to the annihilation of topological defects resembling edge-dislocations of the charge-ordered lattice. Our work opens up the study of a wide class of structural transitions at surfaces and in low-dimensional systems.},
archivePrefix = {},
arxivId = {},
author = {Vogelgesang, S. and Storeck, G. and Horstmann, J. G. and Diekmann, T. and Sivis, M. and Schramm, S. and Rossnagel, K. and Sch{\"{a}}fer, S. and Ropers, C.},
doi = {},
eprint = {},
file = {:C\:/Users/mbau/AppData/Local/Mendeley Ltd./Mendeley Desktop/Downloaded/Vogelgesang et al. - 2017 - Phase ordering of charge density waves traced by ultrafast low-energy electron diffraction.pdf:pdf},
issn = {1745-2473},
journal = {Nat. Phys.},
number = {November},
pages = {184},
title = {{Phase ordering of charge density waves traced by ultrafast low-energy electron diffraction}},
url = {http://www.nature.com/doifinder/10.1038/nphys4309},
volume = {14},
year = {2017}
}

@article{Hanisch-Blicharski2013,
abstract = {Many fundamental processes of structural changes at surfaces occur on a pico- or femtosecond time scale. In order to study such ultra-fast processes, we have combined modern surface science techniques with fs-laser pulses in a pump-probe scheme. Reflection high energy electron diffraction (RHEED) with grazing incident electrons ensures surface sensitivity for the probing electron pulses. Utilizing the Debye-Waller effect, we studied the cooling of vibrational excitations in monolayer adsorbate systems or the nanoscale heat transport from an ultra-thin film through a hetero-interface on the lower ps-time scale. The relaxation dynamics of a driven phase transition far away from thermal equilibrium is demonstrated with the In-induced (8×2) reconstruction on Si(111). This surface exhibits a Peierls-like phase transition at 100. K from a (8×2) ground state to (4×1) excited state. Upon excitation by a fs-laser pulse, this structural phase transition is driven into an excited (4×1) state at a sample temperature of 20. K. Relaxation into the (8×2) ground state occurs after more than 150. ps. {\textcopyright} 2012 Elsevier B.V.},
author = {Hanisch-Blicharski, A. and Janzen, A. and Krenzer, B. and Wall, S. and Klasing, F. and Kalus, A. and Frigge, T. and Kammler, M. and {Horn-von Hoegen}, M.},
doi = {10.1016/j.ultramic.2012.07.017},
file = {:C\:/Users/mbau/Documents/work/Kiel/Paper f{\"{u}}r Bibliothek/neuer/1-s2.0-S0304399112001878-main.pdf:pdf},
issn = {03043991},
journal = {Ultramicroscopy},
pages = {2--8},
publisher = {Elsevier},
title = {{Ultra-fast electron diffraction at surfaces: From nanoscale heat transport to driven phase transitions}},
url = {http://dx.doi.org/10.1016/j.ultramic.2012.07.017},
volume = {127},
year = {2013}
}

@article{Karrer2001,
abstract = {We present the design and performance tests of a miniaturized pulsed low-energy electron gun. Electrons photoemitted from a gold cathode are accelerated over a distance of 75 $\mu$m and then collimated by a microchannel plate. According to calculations, this novel concept will allow the time spread of the electron pulses to be kept below 5 ps for kinetic energies as low as 100 eV. The achievement of a minimum angular beam divergence (≈1°) along with an energy resolution of 1.1 eV has to be paid for by low signal intensities. We demonstrate the performance of the gun and the high electron-beam coherence by presenting low-energy-electron diffraction images taken from a submonolayer of lead adsorbed on the germanium (111) surface. We anticipate that this electron gun will open up new possibilities for following structural changes on solid surfaces in real time. {\textcopyright} 2001 American Institute of Physics.},
author = {Karrer, R. and Neff, H. J. and Hengsberger, M. and Greber, T. and Osterwalder, J.},
doi = {10.1063/1.1419219},
file = {:C\:/Users/mbau/Documents/work/Kiel/Paper f{\"{u}}r Bibliothek/neuer/4404_1_online.pdf:pdf},
issn = {00346748},
journal = {Rev. Sci. Instr.},
number = {12},
pages = {4404--4407},
title = {{Design of a miniature picosecond low-energy electron gun for time-resolved scattering experiments}},
volume = {72},
year = {2001}
}

@article{Storeck2017,
abstract = {We present the design and fabrication of a micrometer-scale electron gun for the implementation of ultrafast low-energy electron diffraction from surfaces. A multi-step process involving photolithography and focused-ion-beam nanostructuring is used to assemble and electrically contact the photoelectron gun, which consists of a nanotip photocathode in a Schottky geometry and an einzel lens for beam collimation. We characterize the low-energy electron pulses by a transient electric field effect and achieve pulse durations of 1.3 ps at an electron energy of 80 eV. First diffraction images in a backscattering geometry (at 50 eV electron energy) are shown.},
author = {Storeck, Gero and Vogelgesang, Simon and Sivis, Murat and Sch{\"{a}}fer, Sascha and Ropers, Claus},
doi = {},
file = {:C\:/Users/mbau/Documents/work/Kiel/Paper f{\"{u}}r Bibliothek/neuer/044024_1_online.pdf:pdf},
issn = {23297778},
journal = {Struct. Dyn.},
pages = {044042},
number = {4},
title = {{Nanotip-based photoelectron microgun for ultrafast LEED}},
url = {http://dx.doi.org/10.1063/1.4982947},
volume = {4},
year = {2017}
}

@article{Storeck2021,
abstract = {We investigate the optically driven phase transition between two charge-density wave (CDW) states at the surface of tantalum disulfide (1T-TaS2). Specifically, we employ a recently improved ultrafast low-energy electron diffraction setup to study the transition from the nearly commensurate to the incommensurate (IC) CDW state. The experimental setup allows us to follow transient changes in the diffraction pattern with high momentum resolution and 1-ps electron pulse duration. In particular, we trace the diffraction intensities and spot profiles of the crystal lattice, including main and CDW superstructure peaks, as well as the diffuse background. Harnessing the enhanced data quality of the instrumental upgrade, we follow the laser-induced transient disorder in the system and perform a spot-profile analysis that yields a substantial IC-peak broadening for very short time scales followed by a prolonged spot narrowing.},
author = {Storeck, G. and Rossnagel, K. and Ropers, C.},
doi = {},
file = {:C\:/Users/mbau/Documents/work/Kiel/Paper f{\"{u}}r Bibliothek/neuer/221603_1_online.pdf:pdf},
issn = {00036951},
journal = {Appl. Phys. Lett.},
pages = {221603},
number = {22},
publisher = {AIP Publishing LLC},
title = {{Ultrafast spot-profile LEED of a charge-density wave phase transition}},
volume = {118},
year = {2021}
}

@article{Muller2014,
abstract = {One- and two-dimensional crystalline materials have emerged as fundamental building blocks for nanoscale devices. Compared to the respective bulk materials, the reduced dimensionality of the translational symmetry has profound effects on the ground state properties of nanomaterials as well as on the coupling between electronic, nuclear and spin degrees of freedom, dictating the dynamical behavior. As all devices operate in states out of equilibrium, and as the dwell time of excited electrons in nanostructures is comparable to the time scale of typical relaxation processes, electron-lattice-spin interactions crucially determine the functionality of future nanodevices. Aspiring towards an ultrafast technique providing real-time access to electronic and structural properties in one- and two-dimensional systems, we developed a hybrid approach for femtosecond low-energy electron diffraction (fsLEED) and femtosecond point projection imaging (fsPPI) with electron energies in the range 20 to 1000 eV. A laser-triggered metal nanotip provides a compact point-like source of coherent femtosecond electron wave packets, optionally collimated for diffraction or spatially diverging for imaging. We demonstrate the feasibility of probing electric currents in nanowires with sub-100 femtosecond temporal and few 10 nm spatial resolution and exemplify accessing femtosecond structural dynamics in crystalline single-layer materials by diffraction with femtosecond electron pulses.},
archivePrefix = {},
arxivId = {},
author = {M{\"{u}}ller, Melanie and Paarmann, Alexander and Ernstorfer, Ralph},
doi = {10.1038/ncomms6292},
eprint = {},
file = {:C\:/Users/mbau/AppData/Local/Mendeley Ltd./Mendeley Desktop/Downloaded/M{\"{u}}ller, Paarmann, Ernstorfer - 2014 - Femtosecond electrons probing currents and atomic structure in nanomaterials.pdf:pdf},
issn = {2041-1723},
journal = {Nat. Commun.},
volume = {5},
pages = {5292},
pmid = {25358554},
title = {{Femtosecond electrons probing currents and atomic structure in nanomaterials}},
url = {http://arxiv.org/abs/1405.4992},
year = {2014}
}

@article{Kampfrath2005,
author = {Kampfrath, Tobias and Perfetti, Luca and Schapper, Florian and Frischkorn, Christian and Wolf, Martin},
doi = {10.1103/PhysRevLett.95.187403},
file = {:C\:/Users/mbau/AppData/Local/Mendeley Ltd./Mendeley Desktop/Downloaded/Kampfrath et al. - 2005 - Strongly Coupled Optical Phonons in the Ultrafast Dynamics of the Electronic Energy and Current Relaxation in.pdf:pdf},
issn = {0031-9007},
journal = {Phys. Rev. Lett.},
month = {oct},
number = {18},
pages = {187403},
title = {{Strongly Coupled Optical Phonons in the Ultrafast Dynamics of the Electronic Energy and Current Relaxation in Graphite}},
url = {http://link.aps.org/doi/10.1103/PhysRevLett.95.187403},
volume = {95},
year = {2005}
}

@article{Wang2010,
author = {Wang, Haining and Strait, Jared H and George, Paul A and Shivaraman, Shriram and Shields, Virgil B and Chandrashekhar, Mvs and Hwang, Jeonghyun and Rana, Farhan and Spencer, Michael G and Ruiz-Vargas, Carlos S and Park, Jiwoong},
doi = {10.1063/1.3291615},
file = {:C\:/Users/mbau/AppData/Local/Mendeley Ltd./Mendeley Desktop/Downloaded/Wang et al. - 2010 - Ultrafast relaxation dynamics of hot optical phonons in graphene Ultrafast relaxation dynamics of hot optical phono.pdf:pdf},
journal = {Appl. Phys. Lett.},
pages = {081917},
title = {{Ultrafast relaxation dynamics of hot optical phonons in graphene}},
url = {https://aip.scitation.org/doi/10.1063/1.3291615},
volume = {96},
year = {2010}
}

@article{Johannsen2013,
author = {Johannsen, Jens Christian and Ulstrup, S{\o}ren and Cilento, Federico and Crepaldi, Alberto and Zacchigna, Michele and Cacho, Cephise and Turcu, I. C. Edmond and Springate, Emma and Fromm, Felix and Raidel, Christian and Seyller, Thomas and Parmigiani, Fulvio and Grioni, Marco and Hofmann, Philip},
doi = {10.1103/PhysRevLett.111.027403},
file = {:C\:/Users/mbau/AppData/Local/Mendeley Ltd./Mendeley Desktop/Downloaded/Johannsen et al. - 2013 - Direct View of Hot Carrier Dynamics in Graphene.pdf:pdf},
issn = {0031-9007},
journal = {Phys. Rev. Lett.},
month = {jul},
number = {2},
pages = {027403},
title = {{Direct View of Hot Carrier Dynamics in Graphene}},
url = {http://link.aps.org/doi/10.1103/PhysRevLett.111.027403},
volume = {111},
year = {2013}
}

@article{Stange2015,
author = {Stange, A. and Sohrt, C. and Yang, L. X. and Rohde, G. and Janssen, K. and Hein, P. and Oloff, L.-P. and Hanff, K. and Rossnagel, K. and Bauer, M.},
doi = {10.1103/PhysRevB.92.184303},
file = {},
issn = {},
journal = {Phys. Rev. B},
month = {},
number = {},
pages = {184303},
title = {{Hot electron cooling in graphite: Supercollision versus hot phonon decay}},
url = {https://journals.aps.org/prb/abstract/10.1103/PhysRevB.92.184303},
volume = {92},
year = {2015}
}

@article{Jauernik2018,
abstract = {Laser-based angle-resolved photoelectron spectroscopy is performed on tin-phthalocyanine (SnPc) adsorbed on silver Ag(111). Upon adsorption of SnPc, strongly dispersing bands are observed which are identified as secondary Mahan cones formed by surface umklapp processes acting on photoelectrons from the silver substrate as they transit through the ordered adsorbate layer. We show that the photoemission data carry quantitative structural information on the adsorbate layer similar to what can be obtained from a conventional low-energy electron diffraction (LEED) study. More specifically, we compare photoemission data and LEED data probing an incommensurate-to-commensurate structural phase transition of the adsorbate layer. Based on our results we propose that Mahan-cone spectroscopy operated in a pump-probe configuration can be used in the future to probe structural dynamics at surfaces with a temporal resolution in the sub-100-fs regime.},
author = {Jauernik, Stephan and Hein, Petra and Gurgel, Max and Falke, Julian and Bauer, Michael},
doi = {10.1103/PhysRevB.97.125413},
file = {:C\:/Users/mbau/Documents/work/Kiel/Paper f{\"{u}}r Bibliothek/neuer/PhysRevB.97.125413.pdf:pdf},
issn = {24699969},
journal = {Phys. Rev. B},
keywords = {doi:10.1103/PhysRevB.97.125413 url:https://doi.org},
number = {12},
pages = {125413},
publisher = {American Physical Society},
title = {{Probing long-range structural order in SnPc/Ag(111) by umklapp process assisted low-energy angle-resolved photoelectron spectroscopy}},
volume = {97},
year = {2018}
}

@article{Gopakumar2004,
abstract = {Adsorption of planar palladium phthalocyanine (PdPc) molecules on the basal plane of graphite is studied using STM (sample at 50 K) and LEED under UHV condition. The monolayer was grown in-situ by organic molecular beam epitaxy (OMBE) on the substrate at room temperature. PdPc molecules diffuse freely on the graphite substrate and form highly ordered, almost defect-free monolayers with a quadratic lattice. One lattice vector of the molecular adlattice is rotated by ±10° with respect to one of the graphite lattice vectors. On the basis of the high-resolution STM images, a model of the adlayer primitive unit cell is proposed, which reveals the adsorption geometry of PdPc molecules on graphite. For a better understanding of the submolecular contrast, the software package Gaussian 98 was used to calculate the electronic structure of isolated PdPc molecules. The STM contrasts at positive sample bias can be assigned to the wave function of the LUMO.},
author = {Gopakumar, T. G. and Lackinger, M. and Hackert, M. and Millier, F. and Hietschold, M.},
doi = {10.1021/jp037751i},
file = {:C\:/Users/mbau/Documents/work/Kiel/Paper f{\"{u}}r Bibliothek/neuer/jp037751i.pdf:pdf},
issn = {15206106},
journal = {J. Phys. Chem. B},
number = {23},
pages = {7839--7843},
title = {{Adsorption of palladium phthalocyanine on graphite: STM and LEED study}},
url = {https://pubs.acs.org/doi/10.1021/jp037751i},
volume = {108},
year = {2004}
}

@article{Kawakita2017,
abstract = {The geometrical and electronic structures of a metastable phase of lead phthalocyanine (PbPc) films on graphite have been studied by combined use of low energy electron diffraction (LEED) and two-photon photoemission (2PPE) spectroscopy. In submonolayer (sub-ML) PbPc films on graphite, islands in a metastable phase are formed just after deposition, as we reported previously by use of photoelectron emission microscopy (PEEM) [I. Yamamoto, N. Matsuura, M. Mikamori, R. Yamamoto, T. Yamada, K. Miyakubo, N. Ueno, and T. Munakata, Surf. Sci. 602, 2232 (2008)10.1016/j.susc.2008.04.037]. On single crystalline graphite substrates, the metastable islands produce clearly discernible LEED spots. By comparing the unit cell with that of annealed 1 ML films, molecules in the metastable islands are standing upright with a molecular density 1.8 times higher than that in the well-ordered 1 ML films. The LEED spots for the sub-ML films disappear after annealing. The islands in the metastable phase are surrounded by areas of a two-dimensional (2D) gaslike phase composed of flat-lying molecules. The metastable islands melt into the 2D gas phase, consistent with the PEEM results. In 2PPE spectroscopy, the lowest unoccupied molecular orbital (LUMO) derived level of the metastable phase is clearly distinguishable from that of flat-lying molecules. By tracking the thermal annealing process of the films by 2PPE spectroscopy, we clarify the decay of the LUMO derived peak intensity, the work function shift, and the energy shifts of molecular states associated with the transition from the metastable phase to the 2D gas phase. With this, we demonstrate the complementary capabilities of LEED and 2PPE spectroscopy to probe phase transitions of organic films in a nondestructive manner.},
author = {Kawakita, N. and Yamada, T. and Meissner, M. and Forker, R. and Fritz, T. and Munakata, T.},
doi = {10.1103/PhysRevB.95.045419},
file = {:C\:/Users/mbau/Documents/work/Kiel/Paper f{\"{u}}r Bibliothek/neuer/PhysRevB.95.045419.pdf:pdf},
issn = {24699969},
journal = {Phys. Rev. B},
number = {4},
pages = {045419},
title = {{Metastable phase of lead phthalocyanine films on graphite: Correlation between geometrical and electronic structures}},
url = {https://journals.aps.org/prb/abstract/10.1103/PhysRevB.95.045419},
volume = {95},
year = {2017}
}

@article{Erk2023,
abstract = {Cone-type bands near the center of the surface Brillouin zone were observed in low-energy angle-resolved photoemission spectra of tin phthalocyanine adsorbed on graphite. Simulations in comparison with the experimental data show that the spectral features represent replicas of the electronic structure of graphite near K ‾ resulting from high-order momentum transfer processes up to the fourth order owing to the interaction of substrate electrons with the long-range structural order of the adsorbate overlayer. The analysis of time-resolved photoemission data from one of the replicas yields a quantitative and very good agreement with previous studies on the excited carrier dynamics in the Dirac cones of graphite and graphene.},
author = {Erk, Hermann and Opitz, Klaas and Hein, Petra and Jauernik, Stephan and Bauer, Michael},
doi = {},
file = {:C\:/Users/mbau/Documents/work/Kiel/Paper f{\"{u}}r Bibliothek/neuer/Erk_2023_J._Phys. _Condens._Matter_35_095501.pdf:pdf},
issn = {1361648X},
journal = {J. Phys. Condens. Matter},
keywords = {adsorbate superlattice,graphite,photoemission spectroscopy,phthalocyanine,ultrafast},
number = {9},
pages = {095501},
pmid = {36535026},
title = {{Observation of electronic structure replicas in photoemission spectra of graphite upon adsorption of tin phthalocyanine}},
volume = {35},
year = {2023}
}

@article{Gauthier2020,
abstract = {Time- and angle-resolved photoemission spectroscopy is a powerful probe of electronic band structures out of equilibrium. Tuning time and energy resolution to suit a particular scientific question has become an increasingly important experimental consideration. Many instruments use cascaded frequency doubling in nonlinear crystals to generate the required ultraviolet probe pulses. We demonstrate how calculations clarify the relationship between laser bandwidth and nonlinear crystal thickness contributing to experimental resolutions and place intrinsic limits on the achievable time-bandwidth product. Experimentally, we tune time and energy resolution by varying the thickness of nonlinear $\beta$-BaB 2O 4 crystals for frequency upconversion, providing a flexible experiment design. We achieve time resolutions of 58-103 fs and corresponding energy resolutions of 55-27 meV. We propose a method to select crystal thickness based on desired experimental resolutions.},
archivePrefix = {},
arxivId = {},
author = {Gauthier, Alexandre and Sobota, Jonathan A. and Gauthier, Nicolas and Xu, Ke Jun and Pfau, Heike and Rotundu, Costel R. and Shen, Zhi Xun and Kirchmann, Patrick S.},
doi = {},
eprint = {},
file = {:C\:/Users/mbau/Documents/work/Kiel/Paper f{\"{u}}r Bibliothek/neuer/5.0018834.pdf:pdf},
issn = {10897550},
journal = {J. Appl. Phys.},
number = {9},
pages = {093101},
publisher = {AIP Publishing LLC},
title = {{Tuning time and energy resolution in time-resolved photoemission spectroscopy with nonlinear crystals}},
url = {https://doi.org/10.1063/5.0018834},
volume = {128},
year = {2020}
}

@article{Yan2021,
abstract = {We present the development of a multi-resolution photoemission spectroscopy (MRPES) setup, which probes quantum materials in energy, momentum, space, and time. This versatile setup integrates three light sources in one photoemission setup and can conveniently switch between traditional angle-resolved photoemission spectroscopy (ARPES), time-resolved ARPES (trARPES), and micrometer-scale spatially resolved ARPES. It provides a first-time all-in-one solution to achieve an energy resolution of <4 meV, a time resolution of <35 fs, and a spatial resolution of ∼10 $\mu$m in photoemission spectroscopy. Remarkably, we obtain the shortest time resolution among the trARPES setups using solid-state nonlinear crystals for frequency upconversion. Furthermore, this MRPES setup is integrated with a shadow-mask assisted molecular beam epitaxy system, which transforms the traditional photoemission spectroscopy into a quantum device characterization instrument. We demonstrate the functionalities of this novel quantum material testbed using FeSe/SrTiO3 thin films and MnBi4Te7 magnetic topological insulators.},
archivePrefix = {},
arxivId = {},
author = {Yan, Chenhui and Green, Emanuel and Fukumori, Riku and Protic, Nikola and Lee, Seng Huat and Fernandez-Mulligan, Sebastian and Raja, Rahim and Erdakos, Robin and Mao, Zhiqiang and Yang, Shuolong},
doi = {},
eprint = {},
file = {:C\:/Users/mbau/Documents/work/Kiel/Paper f{\"{u}}r Bibliothek/neuer/113907_1_online.pdf:pdf},
issn = {10897623},
journal = {Rev. Sci. Instr.},
number = {11},
pages = {113907},
pmid = {34852521},
publisher = {AIP Publishing, LLC},
title = {{An integrated quantum material testbed with multi-resolution photoemission spectroscopy}},
url = {https://doi.org/10.1063/5.0072979},
volume = {92},
year = {2021}
}

@article{Zhou2006,
abstract = {Originating from relativistic quantum field theory, Dirac fermions have been recently applied to study various peculiar phenomena in condensed matter physics, including the novel quantum Hall effect in graphene, magnetic field driven metal-insulator-like transition in graphite, superfluid in 3He, and the exotic pseudogap phase of high temperature superconductors. Although Dirac fermions are proposed to play a key role in these systems, so far direct experimental evidence of Dirac fermions has been limited. Here we report the first direct observation of massless Dirac fermions with linear dispersion near the Brillouin zone (BZ) corner H in graphite, coexisting with quasiparticles with parabolic dispersion near another BZ corner K. In addition, we report a large electron pocket which we attribute to defect-induced localized states. Thus, graphite presents a novel system where massless Dirac fermions, quasiparticles with finite effective mass, and defect states all contribute to the low energy electronic dynamics.},
archivePrefix = {arXiv},
arxivId = {cond-mat/0608069},
author = {Zhou, S. Y. and Gweon, G. H. and Graf, J. and Fedorov, A. V. and Spataru, C. D. and Diehl, R. D. and Kopelevich, Y. and Lee, D. H. and Louie, Steven G. and Lanzara, A.},
doi = {10.1038/nphys393},
eprint = {0608069},
file = {:C\:/Users/mbau/AppData/Local/Mendeley Ltd./Mendeley Desktop/Downloaded/Zhou et al. - 2006 - First direct observation of Dirac fermions in graphite(2).pdf:pdf},
isbn = {1745-2473},
issn = {17452473},
journal = {Nature Physics},
number = {9},
pages = {595--599},
pmid = {25246403},
primaryClass = {cond-mat},
title = {{First direct observation of Dirac fermions in graphite}},
volume = {2},
year = {2006}
}

@article{Gruneis2008,
abstract = {Here we revisited the electronic structure of potassium doped graphite using angle-resolved photoemission spectroscopy (ARPES). We perform an analysis of the charge transfer from potassium to graphite employing a tight-binding model for the band structure. Our ARPES measurements clearly show that the $\pi$* band of graphite becomes occupied upon deposition and intercalation of small amounts of potassium. The observed shift in the Fermi level of ∼ 400 meV is well below the appearance of staged graphite compounds and is described by a rigid band shift model. The $\pi$* band in pristine graphite is unoccupied and can therefore not be measured by ARPES. Thus potassium doping in the dilute limit provides a key for measuring the $\pi$* bands close to the Fermi level by ARPES. {\textcopyright} 2008 Wiley-VCH Verlag GmbH & Co. KGaA.},
author = {Gr{\"{u}}neis, A. and Attaccalite, C. and Rubio, A. and Molodtsov, S. L. and Vyalikh, D. V. and Fink, J. and Follath, R. and Pichler, T.},
doi = {10.1002/pssb.200879662},
file = {:C\:/Users/mbau/Documents/work/Kiel/Paper f{\"{u}}r Bibliothek/neuer/Physica Status Solidi b - 2008 - Gr{\"{u}}neis - Preparation and electronic properties of potassium doped graphite single.pdf:pdf},
issn = {03701972},
journal = {Phys. Status Solidi B Basic Res.},
number = {10},
pages = {2072--2076},
title = {{Preparation and electronic properties of potassium doped graphite single crystals}},
volume = {245},
year = {2008}
}

@article{Aeschlimann2017,
abstract = {The pseudospin of Dirac electrons in graphene manifests itself in a peculiar momentum anisotropy for photo-excited electron-hole pairs. These interband excitations are in fact forbidden along the direction of the light polarization, and are maximum perpendicular to it. Here, we use time- and angle-resolved photoemission spectroscopy to investigate the resulting unconventional hot carrier dynamics, sampling carrier distributions as a function of energy and in-plane momentum. We first show that the rapidly-established quasi-thermal electron distribution initially exhibits an azimuth-dependent temperature, consistent with relaxation through collinear electron-electron scattering. Azimuthal thermalization is found to occur only at longer time delays, at a rate that depends on the substrate and the static doping level. Further, we observe pronounced differences in the electron and hole dynamics in n-doped samples. By simulating the Coulomb- and phonon-mediated carrier dynamics we are able to disentangle the influence of excitation fluence, screening, and doping, and develop a microscopic picture of the carrier dynamics in photo-excited graphene. Our results clarify new aspects of hot carrier dynamics that are unique to Dirac materials, with relevance for photo-control experiments and optoelectronic device applications.},
archivePrefix = {},
arxivId = {},
author = {Aeschlimann, S. and Krause, R. and Ch{\'{a}}vez-Cervantes, M. and Bromberger, H. and Jago, R. and Mali{\'{c}}, E. and Al-Temimy, A. and Coletti, C. and Cavalleri, A. and Gierz, I.},
doi = {},
eprint = {},
file = {:C\:/Users/mbau/AppData/Local/Mendeley Ltd./Mendeley Desktop/Downloaded/Aeschlimann et al. - 2017 - Ultrafast momentum imaging of pseudospin-flip excitations in graphene.pdf:pdf;:C\:/Users/mbau/AppData/Local/Mendeley Ltd./Mendeley Desktop/Downloaded/Aeschlimann et al. - 2017 - Ultrafast momentum imaging of pseudospin-flip excitations in graphene(2).pdf:pdf;:C\:/Users/mbau/AppData/Local/Mendeley Ltd./Mendeley Desktop/Downloaded/Ss, Riet - 2013 -  Supplementary Material.pdf:pdf},
issn = {24699969},
journal = {Phys. Rev. B},
number = {2},
pages = {020301(R)},
title = {{Ultrafast momentum imaging of pseudospin-flip excitations in graphene}},
volume = {96},
year = {2017}
}

@article{Beyer2023,
abstract = {Time- and angle-resolved photoelectron spectroscopy is employed to study the thermalization of a nonequilibrium carrier distribution in the Dirac cone of graphite generated upon absorption of linearly polarized near-infrared laser pulses. The data reveal a characteristic time constant of (20±3)fs for the decay of a photoinduced momentum anisotropy in the nascent carrier population. Additionally, the nascent carrier spectral peak shows a downshift in energy by ∼100meV within the first 30fs after excitation, which is associated with the emission of strongly coupled A1′ optical phonons at K. In agreement with previous findings for graphene, a pronounced momentum anisotropy is also observed for an intermediate quasithermalized state of the carrier population. A fully thermalized distribution (with respect to energy and momentum) is finally established on a characteristic timescale of (40±10)fs. The results underscore the complexity of carrier thermalization on ultrafast timescales resulting from the interplay of carrier-carrier and carrier-phonon interaction.},
author = {Beyer, H. and Hein, P. and Rossnagel, K. and Bauer, M.},
doi = {10.1103/PhysRevB.107.115136},
file = {:C\:/Users/mbau/Documents/work/Kiel/Paper f{\"{u}}r Bibliothek/neuer/PhysRevB.107.115136.pdf:pdf},
issn = {24699969},
journal = {Phys. Rev. B},
keywords = {doi:10.1103/PhysRevB.107.115136 url:https://doi.or},
number = {11},
pages = {115136},
publisher = {American Physical Society},
title = {{Ultrafast decay of carrier momentum anisotropy in graphite}},
volume = {107},
year = {2023}
}

@article{Perfetti2007,
author = {Perfetti, L. and Loukakos, P. and Lisowski, M. and Bovensiepen, U. and Eisaki, H. and Wolf, M.},
doi = {10.1103/PhysRevLett.99.197001},
file = {:C\:/Users/mbau/AppData/Local/Mendeley Ltd./Mendeley Desktop/Downloaded/Perfetti et al. - 2007 - Ultrafast Electron Relaxation in Superconducting Bi_{2}Sr_{2}CaCu_{2}O_{8$\delta$} by Time-Resolved Photoelectron Spec.pdf:pdf},
issn = {0031-9007},
journal = {Phys. Rev. Lett.},
month = {nov},
number = {19},
pages = {197001},
title = {{Ultrafast Electron Relaxation in Superconducting Bi_{2}Sr_{2}CaCu_{2}O_{8+$\delta$} by Time-Resolved Photoelectron Spectroscopy}},
url = {http://link.aps.org/doi/10.1103/PhysRevLett.99.197001},
volume = {99},
year = {2007}
}

@article{Aeschlimann1995,
abstract = {Theoretical and experimental investigations for a new design of an ultrashort pulsed laser activated electron gun for time resolved surface analysis are described. In addition, a novel electron detection and image analysis system, as it applies specifically to time resolved reflection high-energy electron diffraction in the multiple-shot operation, are reviewed. Special attention is directed to minimize the photoelectron transit-time spread from the electron gun, in spite of an unusually long focal length and a small convergence angle of the pulsed electron beam. Both requirements are necessary to use the electron gun for diffraction techniques. The design value for the temporal resolution in the synchroscan operation is 1.3 ps. Based on a thorough theoretical investigation, a new electron gun has been designed, constructed, and tested. {\textcopyright} 1995 American Institute of Physics.},
author = {Aeschlimann, M. and Hull, E. and Cao, J. and Schmuttenmaer, C. A. and Jahn, L. G. and Gao, Y. and Elsayed-Ali, H. E. and Mantell, D. A. and Scheinfein, M. R.},
doi = {10.1063/1.1146036},
file = {:C\:/Users/mbau/Documents/work/Kiel/Paper f{\"{u}}r Bibliothek/neuer/1000_1_online.pdf:pdf},
issn = {00346748},
journal = {Rev. Sci. Instr.},
number = {2},
pages = {1000--1009},
title = {{A picosecond electron gun for surface analysis}},
volume = {66},
year = {1995}
}

@article{Baum2014,
abstract = {Complex systems in condensed phases involve a multidimensional energy landscape, and knowledge of transitional structures and separation of time scales for atomic movements is critical to understanding their dynamical behavior. Here, we report, using four-dimensional (4D) femtosecond electron diffraction, the visualization of transitional structures from the initial monoclinic to the final tetragonal phase in crystalline vanadium dioxide; the change was initiated by a near-infrared excitation. By revealing the spatiotemporal behavior from all observed Bragg diffractions in 3D, the femtosecond primary vanadium-vanadium bond dilation, the displacements of atoms in picoseconds, and the sound wave shear motion on hundreds of picoseconds were resolved, elucidating the nature of the structural pathways and the nonconcerted mechanism of the transformation.},
author = {Baum, Peter and Yang, Ding Shyue and Zewail, Ahmed H.},
doi = {10.1126/science.1147724},
file = {:C\:/Users/mbau/Documents/work/Kiel/Paper f{\"{u}}r Bibliothek/neuer/science.1147724.pdf:pdf},
isbn = {9781783265060},
issn = {00368075},
journal = {Science},
number = {November},
pages = {788},
title = {{4D Visualization of Transitional Structures in Phase Transformations by Electron Diffraction}},
volume = {318},
year = {2007}
}

@article{Ahuja1997,
abstract = {We present theoretical results for the frequency-dependent dielectric response, both for the electric field parallel and perpendicular to the c axis of graphite. The calculations are performed using a full-potential linear muffin-tin orbital method. Our calculations show fair agreement with experimental data and the different features observed are identified from interband transitions in various regions of the Brillouin zone. The anisotropy of the dielectric function is discussed in detail and shown to be due to the difference in the optical matrix elements for the two different polarizations, which is a result of the anisotropic crystallographic and electronic properties of graphite. {\textcopyright} 1997 The American Physical Society.},
author = {Ahuja, R. and Auluck, S.},
doi = {10.1103/PhysRevB.55.4999},
file = {:C\:/Users/mbau/Documents/work/Kiel/Paper f{\"{u}}r Bibliothek/neuer/PhysRevB.55.4999.pdf:pdf},
issn = {1550235X},
journal = {Phys. Rev. B},
number = {8},
pages = {4999--5005},
title = {{Optical properties of graphite from first-principles calculations}},
volume = {55},
year = {1997}
}

@article{Yan2009,
author = {Yan, Hugen and Song, Daohua and Mak, Kin Fai and Chatzakis, Ioannis and Maultzsch, Janina and Heinz, Tony F.},
doi = {10.1103/PhysRevB.80.121403},
file = {:C\:/Users/mbau/AppData/Local/Mendeley Ltd./Mendeley Desktop/Downloaded/Yan et al. - 2009 - Time-resolved Raman spectroscopy of optical phonons in graphite Phonon anharmonic coupling and anomalous stiffening.pdf:pdf},
issn = {1098-0121},
journal = {Phys. Rev. B},
month = {sep},
number = {12},
pages = {121403(R)},
title = {{Time-resolved Raman spectroscopy of optical phonons in graphite: Phonon anharmonic coupling and anomalous stiffening}},
url = {http://link.aps.org/doi/10.1103/PhysRevB.80.121403},
volume = {80},
year = {2009}
}

@article{Malic2011,
author = {Malic, Ermin and Winzer, Torben and Bobkin, Evgeny and Knorr, Andreas},
doi = {10.1103/PhysRevB.84.205406},
file = {:C\:/Users/mbau/AppData/Local/Mendeley Ltd./Mendeley Desktop/Downloaded/Malic et al. - 2011 - Microscopic theory of absorption and ultrafast many-particle kinetics in graphene.pdf:pdf;:C\:/Users/mbau/AppData/Local/Mendeley Ltd./Mendeley Desktop/Downloaded/Malic et al. - 2011 - Microscopic theory of absorption and ultrafast many-particle kinetics in graphene(2).pdf:pdf},
issn = {1098-0121},
journal = {Phys. Rev. B},
month = {nov},
number = {20},
pages = {205406},
title = {{Microscopic theory of absorption and ultrafast many-particle kinetics in graphene}},
url = {http://link.aps.org/doi/10.1103/PhysRevB.84.205406},
volume = {84},
year = {2011}
}

@article{Mikitik2006,
abstract = {We discuss the known experimental data on the phase of the de Haas-van Alphen oscillations in graphite. These data can be understood if one takes into account that four band-contact lines exist near the HKH edge of the Brillouin zone of graphite. {\textcopyright} 2006 The American Physical Society.},
author = {Mikitik, G. P. and Sharlai, Yu V.},
doi = {10.1103/PhysRevB.73.235112},
file = {:C\:/Users/mbau/Documents/work/Kiel/Paper f{\"{u}}r Bibliothek/neuer/PhysRevB.73.235112.pdf:pdf},
issn = {10980121},
journal = {Phys. Rev. B},
number = {23},
pages = {235112},
title = {{Band-contact lines in the electron energy spectrum of graphite}},
volume = {73},
year = {2006}
}

@article{Orlita2008,
abstract = {We report on far infrared magnetotransmission measurements on a thin graphite sample prepared by exfoliation of highly oriented pyrolytic graphite. In a magnetic field, absorption lines exhibiting a blueshift proportional to B are observed. This is a fingerprint for massless Dirac holes at the H point in bulk graphite. The Fermi velocity is found to be c=(1.02±0.02) ×106m/s and the pseudogap |$\Delta$| at the H point is estimated to be below 10meV. Although the holes behave to a first approximation as a strictly 2D gas of Dirac fermions, the full 3D band structure has to be taken into account to explain all the observed spectral features. {\textcopyright} 2008 The American Physical Society.},
archivePrefix = {},
arxivId = {},
author = {Orlita, M. and Faugeras, C. and Martinez, G. and Maude, D. K. and Sadowski, M. L. and Potemski, M.},
doi = {10.1103/PhysRevLett.100.136403},
eprint = {},
file = {:C\:/Users/mbau/Documents/work/Kiel/Paper f{\"{u}}r Bibliothek/neuer/PhysRevLett.100.136403.pdf:pdf},
issn = {00319007},
journal = {Phys. Rev. Lett.},
number = {13},
pages = {4--7},
title = {{Dirac fermions at the H point of graphite: Magnetotransmission studies}},
volume = {100},
year = {2008}
}

@article{Na2019,
abstract = {Ultrafast spectroscopies have become an important tool for elucidating the microscopic description and dynamical properties of quantum materials. In particular, by tracking the dynamics of nonthermal electrons, a material's dominant scattering processes can be revealed. Here, we present a method for extracting the electron-phonon coupling strength in the time domain, using timeand angle-resolved photoemission spectroscopy (TR-ARPES). This method is demonstrated in graphite, where we investigate the dynamics of photoinjected electrons at the K point, detecting quantized energy-loss processes that correspond to the emission of strongly coupled optical phonons. We show that the observed characteristic time scale for spectral weight transfer mediated by phonon-scattering processes allows for the direct quantitative extraction of electron-phonon matrix elements for specific modes.},
archivePrefix = {},
arxivId = {},
author = {Na, M. X. and Mills, A. K. and Boschini, F. and Michiardi, M. and Nosarzewski, B. and Day, R. P. and Razzoli, E. and Sheyerman, A. and Schneider, M. and Levy, G. and Zhdanovich, S. and Devereaux, T. P. and Kemper, A. F. and Jones, D. J. and Damascelli, A.},
doi = {10.1126/science.aaw1662},
eprint = {},
file = {:C\:/Users/mbau/AppData/Local/Mendeley Ltd./Mendeley Desktop/Downloaded/Na et al. - 2019 - Direct determination of mode-projected electron-phonon coupling in the time domain.pdf:pdf;:C\:/Users/mbau/Documents/work/Kiel/Paper f{\"{u}}r Bibliothek/neuer/1231.full.pdf:pdf},
issn = {10959203},
journal = {Science},
month = {dec},
number = {6470},
pages = {1231--1236},
pmid = {31806810},
publisher = {American Association for the Advancement of Science},
title = {{Direct determination of mode-projected electron-phonon coupling in the time domain}},
volume = {366},
year = {2019}
}

@article{Picard2007,
abstract = {An experimental assembly has been constructed to measure the specific heat capacity of macroscopic graphite samples at room temperature. The same batch of graphite constitutes the core of a graphite calorimeter, which is currently being realized to measure the absorbed dose due to ionizing radiation. Two different experimental procedures have been applied. In the first method the specific heat capacity of graphite was measured directly, where its value is corrected for the influence of impurities. The second method, to our knowledge not previously applied to macroscopic samples, is based on a series of differential measurements where no correction for added impurities is needed. By its nature, the second method reduces systematic effects. The specific heat capacity of a particular graphite sample is determined to be 706.9 J K -1 kg-1 with a combined relative standard uncertainty of 9 parts in 104 at 295.15 K. The specific heat capacity of cyanoacrylate has also been determined. {\textcopyright} 2007 BIPM and IOP Publishing Ltd.},
author = {Picard, Susanne and Burns, David T. and Roger, Philippe},
doi = {10.1088/0026-1394/44/5/005},
file = {:C\:/Users/mbau/Documents/work/Kiel/Paper f{\"{u}}r Bibliothek/neuer/Picard_2007_Metrologia_44_294.pdf:pdf},
issn = {00261394},
journal = {Metrologia},
number = {5},
pages = {294--302},
title = {{Determination of the specific heat capacity of a graphite sample using absolute and differential methods}},
volume = {44},
year = {2007}
}

@article{Seah1979,
abstract = {A compilation is presented of all published measurements of electron inelastic mean free path lengths in solids for energies in the range 0-10 000 eV above the Fermi level. For analysis, the materials are grouped under one of the headings: element, inorganic compound, organic compound and adsorbed gas, with the path lengths each time expressed in nanometres, monolayers and milligrams per square metre. The path lengths are very high at low energies, fall to 0.1-0.8 nm for energies in the range 30-100 eV and then rise again as the energy increases further. For elements and inorganic compounds the scatter about a ‘universal curve' is least when the path lengths are expressed in monolayers, A,. Analysis of the inter-element and inter-compound effects shows that A, is related to atom size and the most accurate relations are A, = 538E-2+0.41(aE)”2 for elements and A, = 2170E-2+0.72(aE)”2 for inorganic compounds, where a is the monolayer thickness (nm) and E is the electron energy above the Fermi level in eV. For organic compounds A,, = 49E-” + O.llE1'” mg m-”. Published general theoretical predictions for A, valid above 150 eV, do not show as good correlations with the experimental data as the above relations. INTRODUCTION},
author = {Seah, M.P. and Dench, W.A.},
file = {:C\:/Users/mbau/Documents/work/Kiel/Paper f{\"{u}}r Bibliothek/neuer/Surface Interface Analysis - February 1979 - Seah - Quantitative electron spectroscopy of surfaces A standard data base.pdf:pdf},
journal = {Surf. Interface Anal.},
number = {1},
pages = {2--11},
title = {{Quantitative Electron Spectroscopy of Surfaces}},
volume = {1},
year = {1979}
}

@article{Kromker2008,
abstract = {We demonstrate the use of a novel design of a photoelectron microscope in combination to an imaging energy filter for momentum resolved photoelectron detection. Together with a time resolved imaging detector, it is possible to combine spatial, momentum, energy, and time resolution of photoelectrons within the same instrument. The time resolution of this type of energy analyzer can be reduced to below 100 ps. The complete ARUPS pattern of a Cu(111) sample excited with He I, is imaged in parallel and energy resolved up to the photoelectron emission horizon. Excited with a mercury light source (h =4.9 eV), the Shockley surface state at the energy threshold is clearly imaged in k -space. Electron-electron interactions are observed in momentum space as a correlation hole in two-electron photoemission. With the high transmission and the time resolution of this instrument, possible new measurements are discussed: Time and polarization resolved ARUPS measurements, probing change of bandstructure due to chemical reaction, growth of films, or phase transitions, e.g., melting or martensitic transformations. {\textcopyright} 2008 American Institute of Physics.},
author = {Kr{\"{o}}mker, B. and Escher, M. and Funnemann, D. and Hartung, D. and Engelhard, H. and Kirschner, J.},
doi = {},
file = {:C\:/Users/mbau/Documents/work/Kiel/Paper f{\"{u}}r Bibliothek/neuer/053702_1_online.pdf:pdf},
issn = {00346748},
journal = {Rev. Sci. Instr.},
number = {5},
title = {{Development of a momentum microscope for time resolved band structure imaging}},
volume = {79},
year = {2008}
}

@article{Rohwer2011,
abstract = {Intense femtosecond (10-15s) light pulses can be used to transform electronic, magnetic and structural order in condensed-matter systems on timescales of electronic and atomic motion. This technique is particularly useful in the study and in the control of materials whose physical properties are governed by the interactions between multiple degrees of freedom. Time- and angle-resolved photoemission spectroscopy is in this context a direct and comprehensive, energy- and momentum-selective probe of the ultrafast processes that couple to the electronic degrees of freedom. Previously, the capability of such studies to access electron momentum space away from zero momentum was, however, restricted owing to limitations of the available probing photon energy. Here, using femtosecond extreme-ultraviolet pulses delivered by a high-harmonic-generation source, we use time- and angle-resolved photoemission spectroscopy to measure the photoinduced vaporization of a charge-ordered state in the potential excitonic insulator 1T-TiSe 2 (refs 12, 13). By way of stroboscopic imaging of electronic band dispersions at large momentum, in the vicinity of the edge of the first Brillouin zone, we reveal that the collapse of atomic-scale periodic long-range order happens on a timescale as short as 20femtoseconds. The surprisingly fast response of the system is assigned to screening by the transient generation of free charge carriers. Similar screening scenarios are likely to be relevant in other photoinduced solid-state transitions and may generally determine the response times. Moreover, as electron states with large momenta govern fundamental electronic properties in condensed matter systems, we anticipate that the experimental advance represented by the present study will be useful to study the ultrafast dynamics and microscopic mechanisms of electronic phenomena in a wide range of materials. {\textcopyright} 2011 Macmillan Publishers Limited. All rights reserved.},
author = {Rohwer, Timm and Hellmann, Stefan and Wiesenmayer, Martin and Sohrt, Christian and Stange, Ankatrin and Slomski, Bartosz and Carr, Adra and Liu, Yanwei and Avila, Luis Miaja and Kall{\"{a}}signne, Matthias and Mathias, Stefan and Kipp, Lutz and Rossnagel, Kai and Bauer, Michael},
doi = {10.1038/nature09829},
file = {:C\:/Users/mbau/Documents/work/Kiel/paper/2011/Timm Rohwer TiSe2 Nature 2011/final version/Rohwer et al.pdf:pdf},
issn = {00280836},
journal = {Nature},
number = {7339},
pages = {490--494},
title = {{Collapse of long-range charge order tracked by time-resolved photoemission at high momenta}},
volume = {471},
year = {2011}
}

@article{Hellmann2010,
abstract = {Femtosecond time-resolved core-level photoemission spectroscopy with a free-electron laser is used to measure the atomic-site specific charge-order dynamics of the charge-density wave in the Mott insulator 1T-TaS2. After strong photoexcitation, a prompt loss of charge order and subsequent fast equilibration dynamics of the electron-lattice system are observed. On the time scale of electron-phonon thermalization, about 1 ps, the system is driven across a phase transition from a long-range charge ordered state to a quasiequilibrium state with domainlike short-range charge and lattice order. The experiment opens the way to study the nonequilibrium dynamics of condensed matter systems with full elemental, chemical, and atomic-site selectivity. {\textcopyright} 2010 The American Physical Society.},
archivePrefix = {},
arxivId = {},
author = {Hellmann, S. and Beye, M. and Sohrt, C. and Rohwer, T. and Sorgenfrei, F. and Redlin, H. and Kall{\"{a}}ne, M. and Marczynski-B{\"{u}}hlow, M. and Hennies, F. and Bauer, M. and F{\"{o}}hlisch, A. and Kipp, L. and Wurth, W. and Rossnagel, K.},
doi = {10.1103/PhysRevLett.105.187401},
eprint = {},
file = {:C\:/Users/mbau/Documents/work/Kiel/Paper f{\"{u}}r Bibliothek/neuer/PhysRevLett.105.187401.pdf:pdf},
issn = {00319007},
journal = {Phys. Rev. Lett.},
number = {18},
pages = {187401},
pmid = {21231136},
title = {{Ultrafast melting of a charge-density wave in the mott insulator 1T-TaS2}},
volume = {105},
year = {2010}
}


@misc{Supplemental,
author* ={Hermann Erk},
note ={See Supplemental Material at [URL will be inserted by publisher] for additional information},
year* ={2024}
}

@misc{Fermi,
author* ={Hermann Erk},
note ={The Fermi energy of graphite lies within a few \SI{10}{\milli eV} below the contact point (Dirac point) of both bands at the $H$-point of SCG \cite{Mikitik2006, Orlita2008, Na2019}. The observation that the tip of the cone structure in the ARPES spectra of SnPc/SCG before photoexcitation touches $E_F$ [Fig. \ref{fig:Fig1}(a)] implies that, if at all, this situation is not significantly changed by the adsorption of the SnPc molecules.},
year* ={2024}

}

@article{Stern2017,
abstract = {Despite their fundamental role in determining material properties, detailed momentum-dependent information on the strength of electron-phonon and phonon-phonon coupling (EPC and PPC, respectively) across the entire Brillouin zone (BZ) has proved difficult to obtain. Here we demonstrate that ultrafast electron diffuse scattering (UEDS) directly provides such information. By exploiting symmetry-based selection rules and time-resolution, scattering from different phonon branches can be distinguished even without energy resolution. Using graphite as a model system, we show that UEDS patterns map the relative EPC and PPC strength through their profound sensitivity to photoinduced changes in phonon populations. We measure strong EPC to the $K$-point transverse optical phonon of $A_1'$ symmetry ($K-A_1'$) and along the entire longitudinal optical branch between $\Gamma-K$, not only to the $\Gamma-E_{2g}$ phonon as previously emphasized. We also determine that the subsequent phonon relaxation pathway involves three stages; decay via several identifiable channels to transverse acoustic (TA) and longitudinal acoustic (LA) phonons (1-2 ps), intraband thermalization of the non-equilibrium TA/LA phonon populations (30-40 ps) and interband relaxation of the LA/TA modes (115 ps). Combining UEDS with ultrafast angle-resolved photoelectron spectroscopy will yield a complete picture of the dynamics within and between electron and phonon subsystems, helping to unravel complex phases in which the intertwined nature of these systems have a strong influence on emergent properties.},
archivePrefix = {},
arxivId = {},
author = {Stern, Mark J. and de Cotret, Laurent P. Ren{\'{e}} and Otto, Martin R. and Chatelain, Robert P. and Boisvert, Jean-Philippe and Sutton, Mark and Siwick, Bradley J.},
doi = {10.1103/PhysRevB.97.165416},
eprint = {},
file = {:C\:/Users/mbau/AppData/Local/Mendeley Ltd./Mendeley Desktop/Downloaded/Stern et al. - 2017 - Mapping momentum-dependent electron-phonon coupling and non-equilibrium phonon dynamics with ultrafast electron di.pdf:pdf},
issn = {2469-9950},
keywords = {doi:10.1103/PhysRevB.97.165416 url:https://doi.org},
pages = {165416},
publisher = {American Physical Society},
journal = {Phys. Rev. B},
title = {{Mapping momentum-dependent electron-phonon coupling and non-equilibrium phonon dynamics with ultrafast electron diffuse scattering}},
url = {http://arxiv.org/abs/1708.01251},
volume = {79},
year = {2017}
}

@article{Seiler2021,
abstract = {We combine ultrafast electron diffuse scattering experiments and first-principles calculations of the coupled electron-phonon dynamics to provide a detailed momentum-resolved picture of lattice thermalization in black phosphorus. The measurements reveal the emergence of highly anisotropic nonthermal phonon populations persisting for several picoseconds after exciting the electrons with a light pulse. Ultrafast dynamics simulations based on the time-dependent Boltzmann formalism are supplemented by calculations of the structure factor, defining an approach to reproduce the experimental signatures of nonequilibrium structural dynamics. The combination of experiments and theory enables us to identify highly anisotropic electron-phonon scattering processes as the primary driving force of the nonequilibrium lattice dynamics in black phosphorus. Our approach paves the way toward unravelling and controlling microscopic energy flows in two-dimensional materials and van der Waals heterostructures, and may be extended to other nonequilibrium phenomena involving coupled electron-phonon dynamics such as superconductivity, phase transitions, or polaron physics.},
archivePrefix = {},
arxivId = {},
author = {Seiler, H{\'{e}}l{\`{e}}ne and Zahn, Daniela and Zacharias, Marios and Hildebrandt, Patrick Nigel and Vasileiadis, Thomas and Windsor, Yoav William and Qi, Yingpeng and Carbogno, Christian and Draxl, Claudia and Ernstorfer, Ralph and Caruso, Fabio},
doi = {10.1021/acs.nanolett.1c01786},
eprint = {},
file = {:C\:/Users/mbau/Documents/work/Kiel/Paper f{\"{u}}r Bibliothek/neuer/acs.nanolett.1c01786.pdf:pdf},
issn = {15306992},
journal = {Nano Lett.},
keywords = {black phosphorus,electron-phonon coupling,first-principles calculations,layered materials,ultrafast electron diffraction},
number = {14},
pages = {6171--6178},
pmid = {34279103},
title = {{Accessing the Anisotropic Nonthermal Phonon Populations in Black Phosphorus}},
volume = {21},
year = {2021}
}

@article{Hein2020,
abstract = {The excitation of coherent phonons provides unique capabilities to control fundamental properties of quantum materials on ultrafast time scales. Recently, it was predicted that a topologically protected Weyl semimetal phase in the transition metal dichalcogenide Td-WTe2 can be controlled and, ultimately, be destroyed upon the coherent excitation of an interlayer shear mode. By monitoring electronic structure changes with femtosecond resolution, we provide here direct experimental evidence that the shear mode acts on the electronic states near the phase-defining Weyl points. Furthermore, we observe a periodic reduction in the spin splitting of bands, a distinct electronic signature of the Weyl phase-stabilizing non-centrosymmetric Td ground state of WTe2. The comparison with higher-frequency coherent phonon modes finally proves the shear mode-selectivity of the observed changes in the electronic structure. Our real-time observations reveal direct experimental insights into electronic processes that are of vital importance for a coherent phonon-induced topological phase transition in Td-WTe2.},
archivePrefix = {},
arxivId = {},
author = {Hein, Petra and Jauernik, Stephan and Erk, Hermann and Yang, Lexian and Qi, Yanpeng and Sun, Yan and Felser, Claudia and Bauer, Michael},
doi = {},
eprint = {},
file = {:C\:/Users/mbau/Documents/work/Kiel/Paper f{\"{u}}r Bibliothek/neuer/s41467-020-16076-0.pdf:pdf},
issn = {20411723},
journal = {Nat. Commun.},
number = {1},
pages ={2613},
pmid = {32457344},
publisher = {Springer US},
title = {{Mode-resolved reciprocal space mapping of electron-phonon interaction in the Weyl semimetal candidate Td-WTe2}},
url = {http://dx.doi.org/10.1038/s41467-020-16076-0},
volume = {11},
year = {2020}
}

@article{Djurišić1999,
    author = {Djurišić, Aleksandra B. and Li, E. Herbert},
    title = "{Optical properties of graphite}",
    journal = {Journal of Applied Physics},
    volume = {85},
    number = {10},
    pages = {7404-7410},
    year = {1999},
    month = {05},
    abstract = "{Optical constants of graphite for ordinary and extraordinary waves are modeled with a modified Lorentz–Drude model with frequency-dependent damping. The model enables the shape of the spectral line to vary over a range of broadening functions with similar kernels and different wings, the broadening type being its adjustable parameter. The model parameters are determined by the acceptance-probability-controlled simulated annealing algorithm. Good agreement with the experimental data is obtained in the entire investigated spectral range (0.12–40 eV for ordinary wave and 2.1–40 eV for extraordinary wave). The significant discrepancies between the experimental data obtained by the reflectance measurements and the electron-energy-loss spectroscopy data are analyzed in details. Inconsistency in terms of unsatisfied Kramers–Kronig relations is discovered in the index of refraction data derived from reflectance measurements, and a method for correcting the data is proposed.}",
    issn = {0021-8979},
    doi = {10.1063/1.369370},
    url = {https://doi.org/10.1063/1.369370},
    eprint = {},
}

@article{Ooi2006,
abstract = {The structural and electronic properties of bulk graphite were compared using density functional theory calculations with the local density (LDA) and generalized gradient (GGA) approximations to determine the relative ability of each to model this material. The GGA fails to generate interplanar bonding, but the LDA does, even though the band structures obtained from both approximations were essentially identical. The atomic geometry, electronic structure, and enthalpy of the graphite (0 0 0 1) surface were then obtained using the LDA. The calculated surface energy was ∼0.075 J/m2 and the calculated work function was in 4.4-5.2 eV range, both of which correspond well to published, measured and calculated values. The surface is semi-metallic, just like in the bulk, with the conduction band minimum and valence band maximum just touching with minimal overlap in the H-K region in the Brillouin Zone, so the electron affinity was identical to the work function. The (0 0 0 1) surface undergoes no noticeable relaxation and no reconstruction, as the strong covalent bonding prevents any corrugation of the basal planes. {\textcopyright} 2005 Elsevier Ltd. All rights reserved.},
author = {Ooi, Newton and Rairkar, Asit and Adams, James B.},
doi = {10.1016/j.carbon.2005.07.036},
file = {:C\:/Users/mbau/Documents/work/Kiel/Paper f{\"{u}}r Bibliothek/neuer/1-s2.0-S0008622305004719-main.pdf:pdf},
issn = {00086223},
journal = {Carbon},
keywords = {Computational chemistry,Electrical (electronic) properties,Electronic structure,Graphite,Surface properties},
number = {2},
pages = {231--242},
title = {{Density functional study of graphite bulk and surface properties}},
volume = {44},
year = {2006}
}

@article{Curcio2022,
abstract = {X-ray photoelectron diffraction is a powerful tool for determining the structure of clean and adsorbate-covered surfaces. Extending the technique into the ultrafast time domain will open the door to studies as diverse as the direct determination of the electron-phonon coupling strength in solids and the mapping of atomic motion in surface chemical reactions. Here we demonstrate time-resolved photoelectron diffraction using ultrashort soft X-ray pulses from the free electron laser FLASH. We collect Se 3d photoelectron diffraction patterns over a wide angular range from optically excited Bi$_2$Se$_3$ with a time resolution of 140 fs. Combining these with multiple scattering simulations allows us to track the motion of near-surface atoms within the first 3 ps after triggering a coherent vibration of the A$_{1g}$ optical phonons. Using a fluence of 4.2 mJ/cm$^2$ from a 1.55 eV pump laser, we find the resulting coherent vibrational amplitude in the first two interlayer spacings to be on the order of 1 pm.},
archivePrefix = {},
arxivId = {},
author = {Curcio, Davide and Volckaert, Klara and Kutnyakhov, Dmytro and Agustsson, Steinn Ymir and B{\"{u}}hlmann, Kevin and Pressacco, Federico and Heber, Michael and Dziarzhytski, Siarhei and Acremann, Yves and Demsar, Jure and Wurth, Wilfried and Sanders, Charlotte E. and Hofmann, Philip},
doi = {10.1103/PhysRevB.106.L201409},
eprint = {},
file = {:C\:/Users/mbau/Documents/work/Kiel/Paper f{\"{u}}r Bibliothek/neuer/PhysRevB.106.L201409.pdf:pdf},
keywords = {doi:10.1103/PhysRevB.106.L201409 url:https://doi.o},
journal = {Phys. Rev. B},
pages = {L201409},
publisher = {American Physical Society},
title = {{Tracking the surface atomic motion in a coherent phonon oscillation}},
url = {http://arxiv.org/abs/2205.13212},
volume = {106},
year = {2022}
}

%merlin.mbs apsrev4-1.bst 2010-07-25 4.21a (PWD, AO, DPC) hacked
%Control: key (0)
%Control: author (8) initials jnrlst
%Control: editor formatted (1) identically to author
%Control: production of article title (-1) disabled
%Control: page (0) single
%Control: year (1) truncated
%Control: production of eprint (0) enabled
\begin{thebibliography}{46}%
\makeatletter
\providecommand \@ifxundefined [1]{%
 \@ifx{#1\undefined}
}%
\providecommand \@ifnum [1]{%
 \ifnum #1\expandafter \@firstoftwo
 \else \expandafter \@secondoftwo
 \fi
}%
\providecommand \@ifx [1]{%
 \ifx #1\expandafter \@firstoftwo
 \else \expandafter \@secondoftwo
 \fi
}%
\providecommand \natexlab [1]{#1}%
\providecommand \enquote  [1]{``#1''}%
\providecommand \bibnamefont  [1]{#1}%
\providecommand \bibfnamefont [1]{#1}%
\providecommand \citenamefont [1]{#1}%
\providecommand \href@noop [0]{\@secondoftwo}%
\providecommand \href [0]{\begingroup \@sanitize@url \@href}%
\providecommand \@href[1]{\@@startlink{#1}\@@href}%
\providecommand \@@href[1]{\endgroup#1\@@endlink}%
\providecommand \@sanitize@url [0]{\catcode `\\12\catcode `\$12\catcode `\&12\catcode `\#12\catcode `\^12\catcode `\_12\catcode `\%12\relax}%
\providecommand \@@startlink[1]{}%
\providecommand \@@endlink[0]{}%
\providecommand \url  [0]{\begingroup\@sanitize@url \@url }%
\providecommand \@url [1]{\endgroup\@href {#1}{\urlprefix }}%
\providecommand \urlprefix  [0]{URL }%
\providecommand \Eprint [0]{\href }%
\providecommand \doibase [0]{http://dx.doi.org/}%
\providecommand \selectlanguage [0]{\@gobble}%
\providecommand \bibinfo  [0]{\@secondoftwo}%
\providecommand \bibfield  [0]{\@secondoftwo}%
\providecommand \translation [1]{[#1]}%
\providecommand \BibitemOpen [0]{}%
\providecommand \bibitemStop [0]{}%
\providecommand \bibitemNoStop [0]{.\EOS\space}%
\providecommand \EOS [0]{\spacefactor3000\relax}%
\providecommand \BibitemShut  [1]{\csname bibitem#1\endcsname}%
\let\auto@bib@innerbib\@empty
%</preamble>
\bibitem [{\citenamefont {Prybyla}\ \emph {et~al.}(1992)\citenamefont {Prybyla}, \citenamefont {Tom},\ and\ \citenamefont {Aumiller}}]{Prybyla1992}%
  \BibitemOpen
  \bibfield  {author} {\bibinfo {author} {\bibfnamefont {J.~A.}\ \bibnamefont {Prybyla}}, \bibinfo {author} {\bibfnamefont {H.~W.~K.}\ \bibnamefont {Tom}}, \ and\ \bibinfo {author} {\bibfnamefont {G.~D.}\ \bibnamefont {Aumiller}},\ }\href {\doibase 10.1103/PhysRevLett.68.503} {\bibfield  {journal} {\bibinfo  {journal} {Phys. Rev. Lett.}\ }\textbf {\bibinfo {volume} {68}},\ \bibinfo {pages} {503} (\bibinfo {year} {1992})}\BibitemShut {NoStop}%
\bibitem [{\citenamefont {Petek}\ \emph {et~al.}(2000)\citenamefont {Petek}, \citenamefont {Weida}, \citenamefont {Nagano},\ and\ \citenamefont {Ogawa}}]{Petek2000}%
  \BibitemOpen
  \bibfield  {author} {\bibinfo {author} {\bibfnamefont {H.}~\bibnamefont {Petek}}, \bibinfo {author} {\bibfnamefont {M.}~\bibnamefont {Weida}}, \bibinfo {author} {\bibfnamefont {H.}~\bibnamefont {Nagano}}, \ and\ \bibinfo {author} {\bibfnamefont {S.}~\bibnamefont {Ogawa}},\ }\href {http://www.ncbi.nlm.nih.gov/pubmed/10827946} {\bibfield  {journal} {\bibinfo  {journal} {Science}\ }\textbf {\bibinfo {volume} {288}},\ \bibinfo {pages} {1402} (\bibinfo {year} {2000})}\BibitemShut {NoStop}%
\bibitem [{\citenamefont {St{\'{e}}p{\'{a}}n}\ \emph {et~al.}(2005)\citenamefont {St{\'{e}}p{\'{a}}n}, \citenamefont {G{\"{u}}dde},\ and\ \citenamefont {H{\"{o}}fer}}]{Stepan2005}%
  \BibitemOpen
  \bibfield  {author} {\bibinfo {author} {\bibfnamefont {K.}~\bibnamefont {St{\'{e}}p{\'{a}}n}}, \bibinfo {author} {\bibfnamefont {J.}~\bibnamefont {G{\"{u}}dde}}, \ and\ \bibinfo {author} {\bibfnamefont {U.}~\bibnamefont {H{\"{o}}fer}},\ }\href {\doibase 10.1103/PhysRevLett.94.236103} {\bibfield  {journal} {\bibinfo  {journal} {Phys. Rev. Lett.}\ }\textbf {\bibinfo {volume} {94}},\ \bibinfo {pages} {236103} (\bibinfo {year} {2005})}\BibitemShut {NoStop}%
\bibitem [{\citenamefont {Backus}\ \emph {et~al.}(2005)\citenamefont {Backus}, \citenamefont {Eichler}, \citenamefont {Kleyn},\ and\ \citenamefont {Bonn}}]{Backus2005}%
  \BibitemOpen
  \bibfield  {author} {\bibinfo {author} {\bibfnamefont {E.~H.}\ \bibnamefont {Backus}}, \bibinfo {author} {\bibfnamefont {A.}~\bibnamefont {Eichler}}, \bibinfo {author} {\bibfnamefont {A.~W.}\ \bibnamefont {Kleyn}}, \ and\ \bibinfo {author} {\bibfnamefont {M.}~\bibnamefont {Bonn}},\ }\href {\doibase 10.1126/science.1120693} {\bibfield  {journal} {\bibinfo  {journal} {Science}\ }\textbf {\bibinfo {volume} {310}},\ \bibinfo {pages} {1790} (\bibinfo {year} {2005})}\BibitemShut {NoStop}%
\bibitem [{\citenamefont {Dell'Angela}\ \emph {et~al.}(2013)\citenamefont {Dell'Angela}, \citenamefont {Anniyev}, \citenamefont {Beye}, \citenamefont {Coffee}, \citenamefont {F{\"{o}}hlisch}, \citenamefont {Gladh}, \citenamefont {Katayama}, \citenamefont {Kaya}, \citenamefont {Krupin}, \citenamefont {Larue}, \citenamefont {M{\o}gelh{\o}j}, \citenamefont {Nordlund}, \citenamefont {N{\o}rskov}, \citenamefont {Oberg}, \citenamefont {Ogasawara}, \citenamefont {Ostr{\"{o}}m}, \citenamefont {Pettersson}, \citenamefont {Schlotter}, \citenamefont {Sellberg}, \citenamefont {Sorgenfrei}, \citenamefont {Turner}, \citenamefont {Wolf}, \citenamefont {Wurth},\ and\ \citenamefont {Nilsson}}]{DellAngela2013}%
  \BibitemOpen
  \bibfield  {author} {\bibinfo {author} {\bibfnamefont {M.}~\bibnamefont {Dell'Angela}}, \bibinfo {author} {\bibfnamefont {T.}~\bibnamefont {Anniyev}}, \bibinfo {author} {\bibfnamefont {M.}~\bibnamefont {Beye}}, \bibinfo {author} {\bibfnamefont {R.}~\bibnamefont {Coffee}}, \bibinfo {author} {\bibfnamefont {A.}~\bibnamefont {F{\"{o}}hlisch}}, \bibinfo {author} {\bibfnamefont {J.}~\bibnamefont {Gladh}}, \bibinfo {author} {\bibfnamefont {T.}~\bibnamefont {Katayama}}, \bibinfo {author} {\bibfnamefont {S.}~\bibnamefont {Kaya}}, \bibinfo {author} {\bibfnamefont {O.}~\bibnamefont {Krupin}}, \bibinfo {author} {\bibfnamefont {J.}~\bibnamefont {Larue}}, \bibinfo {author} {\bibfnamefont {A.}~\bibnamefont {M{\o}gelh{\o}j}}, \bibinfo {author} {\bibfnamefont {D.}~\bibnamefont {Nordlund}}, \bibinfo {author} {\bibfnamefont {J.~K.}\ \bibnamefont {N{\o}rskov}}, \bibinfo {author} {\bibfnamefont {H.}~\bibnamefont {Oberg}}, \bibinfo {author} {\bibfnamefont {H.}~\bibnamefont {Ogasawara}}, \bibinfo {author} {\bibfnamefont
  {H.}~\bibnamefont {Ostr{\"{o}}m}}, \bibinfo {author} {\bibfnamefont {L.~G.~M.}\ \bibnamefont {Pettersson}}, \bibinfo {author} {\bibfnamefont {W.~F.}\ \bibnamefont {Schlotter}}, \bibinfo {author} {\bibfnamefont {J.~A.}\ \bibnamefont {Sellberg}}, \bibinfo {author} {\bibfnamefont {F.}~\bibnamefont {Sorgenfrei}}, \bibinfo {author} {\bibfnamefont {J.~J.}\ \bibnamefont {Turner}}, \bibinfo {author} {\bibfnamefont {M.}~\bibnamefont {Wolf}}, \bibinfo {author} {\bibfnamefont {W.}~\bibnamefont {Wurth}}, \ and\ \bibinfo {author} {\bibfnamefont {A.}~\bibnamefont {Nilsson}},\ }\href@noop {} {\bibfield  {journal} {\bibinfo  {journal} {Science}\ }\textbf {\bibinfo {volume} {339}},\ \bibinfo {pages} {1302} (\bibinfo {year} {2013})}\BibitemShut {NoStop}%
\bibitem [{\citenamefont {Diesen}\ \emph {et~al.}(2021)\citenamefont {Diesen}, \citenamefont {Wang}, \citenamefont {Schreck}, \citenamefont {Weston}, \citenamefont {Ogasawara}, \citenamefont {Larue}, \citenamefont {Perakis}, \citenamefont {Dell'Angela}, \citenamefont {Capotondi}, \citenamefont {Giannessi}, \citenamefont {Pedersoli}, \citenamefont {Naumenko}, \citenamefont {Nikolov}, \citenamefont {Raimondi}, \citenamefont {Spezzani}, \citenamefont {Beye}, \citenamefont {Cavalca}, \citenamefont {Liu}, \citenamefont {Gladh}, \citenamefont {Koroidov}, \citenamefont {Miedema}, \citenamefont {Costantini}, \citenamefont {Heinz}, \citenamefont {Abild-Pedersen}, \citenamefont {Voss}, \citenamefont {Luntz},\ and\ \citenamefont {Nilsson}}]{Diesen2021}%
  \BibitemOpen
  \bibfield  {author} {\bibinfo {author} {\bibfnamefont {E.}~\bibnamefont {Diesen}}, \bibinfo {author} {\bibfnamefont {H.~Y.}\ \bibnamefont {Wang}}, \bibinfo {author} {\bibfnamefont {S.}~\bibnamefont {Schreck}}, \bibinfo {author} {\bibfnamefont {M.}~\bibnamefont {Weston}}, \bibinfo {author} {\bibfnamefont {H.}~\bibnamefont {Ogasawara}}, \bibinfo {author} {\bibfnamefont {J.}~\bibnamefont {Larue}}, \bibinfo {author} {\bibfnamefont {F.}~\bibnamefont {Perakis}}, \bibinfo {author} {\bibfnamefont {M.}~\bibnamefont {Dell'Angela}}, \bibinfo {author} {\bibfnamefont {F.}~\bibnamefont {Capotondi}}, \bibinfo {author} {\bibfnamefont {L.}~\bibnamefont {Giannessi}}, \bibinfo {author} {\bibfnamefont {E.}~\bibnamefont {Pedersoli}}, \bibinfo {author} {\bibfnamefont {D.}~\bibnamefont {Naumenko}}, \bibinfo {author} {\bibfnamefont {I.}~\bibnamefont {Nikolov}}, \bibinfo {author} {\bibfnamefont {L.}~\bibnamefont {Raimondi}}, \bibinfo {author} {\bibfnamefont {C.}~\bibnamefont {Spezzani}}, \bibinfo {author} {\bibfnamefont
  {M.}~\bibnamefont {Beye}}, \bibinfo {author} {\bibfnamefont {F.}~\bibnamefont {Cavalca}}, \bibinfo {author} {\bibfnamefont {B.}~\bibnamefont {Liu}}, \bibinfo {author} {\bibfnamefont {J.}~\bibnamefont {Gladh}}, \bibinfo {author} {\bibfnamefont {S.}~\bibnamefont {Koroidov}}, \bibinfo {author} {\bibfnamefont {P.~S.}\ \bibnamefont {Miedema}}, \bibinfo {author} {\bibfnamefont {R.}~\bibnamefont {Costantini}}, \bibinfo {author} {\bibfnamefont {T.~F.}\ \bibnamefont {Heinz}}, \bibinfo {author} {\bibfnamefont {F.}~\bibnamefont {Abild-Pedersen}}, \bibinfo {author} {\bibfnamefont {J.}~\bibnamefont {Voss}}, \bibinfo {author} {\bibfnamefont {A.~C.}\ \bibnamefont {Luntz}}, \ and\ \bibinfo {author} {\bibfnamefont {A.}~\bibnamefont {Nilsson}},\ }\href {\doibase 10.1103/PhysRevLett.127.016802} {\bibfield  {journal} {\bibinfo  {journal} {Phys. Rev. Lett.}\ }\textbf {\bibinfo {volume} {127}},\ \bibinfo {pages} {16802} (\bibinfo {year} {2021})}\BibitemShut {NoStop}%
\bibitem [{\citenamefont {Wallauer}\ \emph {et~al.}(2021)\citenamefont {Wallauer}, \citenamefont {Raths}, \citenamefont {Stallberg}, \citenamefont {M{\"{u}}nster}, \citenamefont {Brandstetter}, \citenamefont {Yang}, \citenamefont {G{\"{u}}dde}, \citenamefont {Puschnig}, \citenamefont {Soubatch}, \citenamefont {Kumpf}, \citenamefont {Bocquet}, \citenamefont {Tautz},\ and\ \citenamefont {H{\"{o}}fer}}]{Wallauer2021}%
  \BibitemOpen
  \bibfield  {author} {\bibinfo {author} {\bibfnamefont {R.}~\bibnamefont {Wallauer}}, \bibinfo {author} {\bibfnamefont {M.}~\bibnamefont {Raths}}, \bibinfo {author} {\bibfnamefont {K.}~\bibnamefont {Stallberg}}, \bibinfo {author} {\bibfnamefont {L.}~\bibnamefont {M{\"{u}}nster}}, \bibinfo {author} {\bibfnamefont {D.}~\bibnamefont {Brandstetter}}, \bibinfo {author} {\bibfnamefont {X.}~\bibnamefont {Yang}}, \bibinfo {author} {\bibfnamefont {J.}~\bibnamefont {G{\"{u}}dde}}, \bibinfo {author} {\bibfnamefont {P.}~\bibnamefont {Puschnig}}, \bibinfo {author} {\bibfnamefont {S.}~\bibnamefont {Soubatch}}, \bibinfo {author} {\bibfnamefont {C.}~\bibnamefont {Kumpf}}, \bibinfo {author} {\bibfnamefont {F.~C.}\ \bibnamefont {Bocquet}}, \bibinfo {author} {\bibfnamefont {F.~S.}\ \bibnamefont {Tautz}}, \ and\ \bibinfo {author} {\bibfnamefont {U.}~\bibnamefont {H{\"{o}}fer}},\ }\href {\doibase 10.1126/science.abf3286} {\bibfield  {journal} {\bibinfo  {journal} {Science}\ }\textbf {\bibinfo {volume} {371}},\ \bibinfo {pages}
  {1056} (\bibinfo {year} {2021})}\BibitemShut {NoStop}%
\bibitem [{\citenamefont {Aeschlimann}\ \emph {et~al.}(1995)\citenamefont {Aeschlimann}, \citenamefont {Hull}, \citenamefont {Cao}, \citenamefont {Schmuttenmaer}, \citenamefont {Jahn}, \citenamefont {Gao}, \citenamefont {Elsayed-Ali}, \citenamefont {Mantell},\ and\ \citenamefont {Scheinfein}}]{Aeschlimann1995}%
  \BibitemOpen
  \bibfield  {author} {\bibinfo {author} {\bibfnamefont {M.}~\bibnamefont {Aeschlimann}}, \bibinfo {author} {\bibfnamefont {E.}~\bibnamefont {Hull}}, \bibinfo {author} {\bibfnamefont {J.}~\bibnamefont {Cao}}, \bibinfo {author} {\bibfnamefont {C.~A.}\ \bibnamefont {Schmuttenmaer}}, \bibinfo {author} {\bibfnamefont {L.~G.}\ \bibnamefont {Jahn}}, \bibinfo {author} {\bibfnamefont {Y.}~\bibnamefont {Gao}}, \bibinfo {author} {\bibfnamefont {H.~E.}\ \bibnamefont {Elsayed-Ali}}, \bibinfo {author} {\bibfnamefont {D.~A.}\ \bibnamefont {Mantell}}, \ and\ \bibinfo {author} {\bibfnamefont {M.~R.}\ \bibnamefont {Scheinfein}},\ }\href {\doibase 10.1063/1.1146036} {\bibfield  {journal} {\bibinfo  {journal} {Rev. Sci. Instr.}\ }\textbf {\bibinfo {volume} {66}},\ \bibinfo {pages} {1000} (\bibinfo {year} {1995})}\BibitemShut {NoStop}%
\bibitem [{\citenamefont {Baum}\ \emph {et~al.}(2007)\citenamefont {Baum}, \citenamefont {Yang},\ and\ \citenamefont {Zewail}}]{Baum2014}%
  \BibitemOpen
  \bibfield  {author} {\bibinfo {author} {\bibfnamefont {P.}~\bibnamefont {Baum}}, \bibinfo {author} {\bibfnamefont {D.~S.}\ \bibnamefont {Yang}}, \ and\ \bibinfo {author} {\bibfnamefont {A.~H.}\ \bibnamefont {Zewail}},\ }\href {\doibase 10.1126/science.1147724} {\bibfield  {journal} {\bibinfo  {journal} {Science}\ }\textbf {\bibinfo {volume} {318}},\ \bibinfo {pages} {788} (\bibinfo {year} {2007})}\BibitemShut {NoStop}%
\bibitem [{\citenamefont {Curcio}\ \emph {et~al.}(2022)\citenamefont {Curcio}, \citenamefont {Volckaert}, \citenamefont {Kutnyakhov}, \citenamefont {Agustsson}, \citenamefont {B{\"{u}}hlmann}, \citenamefont {Pressacco}, \citenamefont {Heber}, \citenamefont {Dziarzhytski}, \citenamefont {Acremann}, \citenamefont {Demsar}, \citenamefont {Wurth}, \citenamefont {Sanders},\ and\ \citenamefont {Hofmann}}]{Curcio2022}%
  \BibitemOpen
  \bibfield  {author} {\bibinfo {author} {\bibfnamefont {D.}~\bibnamefont {Curcio}}, \bibinfo {author} {\bibfnamefont {K.}~\bibnamefont {Volckaert}}, \bibinfo {author} {\bibfnamefont {D.}~\bibnamefont {Kutnyakhov}}, \bibinfo {author} {\bibfnamefont {S.~Y.}\ \bibnamefont {Agustsson}}, \bibinfo {author} {\bibfnamefont {K.}~\bibnamefont {B{\"{u}}hlmann}}, \bibinfo {author} {\bibfnamefont {F.}~\bibnamefont {Pressacco}}, \bibinfo {author} {\bibfnamefont {M.}~\bibnamefont {Heber}}, \bibinfo {author} {\bibfnamefont {S.}~\bibnamefont {Dziarzhytski}}, \bibinfo {author} {\bibfnamefont {Y.}~\bibnamefont {Acremann}}, \bibinfo {author} {\bibfnamefont {J.}~\bibnamefont {Demsar}}, \bibinfo {author} {\bibfnamefont {W.}~\bibnamefont {Wurth}}, \bibinfo {author} {\bibfnamefont {C.~E.}\ \bibnamefont {Sanders}}, \ and\ \bibinfo {author} {\bibfnamefont {P.}~\bibnamefont {Hofmann}},\ }\href {\doibase 10.1103/PhysRevB.106.L201409} {\bibfield  {journal} {\bibinfo  {journal} {Phys. Rev. B}\ }\textbf {\bibinfo {volume} {106}},\
  \bibinfo {pages} {L201409} (\bibinfo {year} {2022})}\BibitemShut {NoStop}%
\bibitem [{\citenamefont {Hanisch-Blicharski}\ \emph {et~al.}(2013)\citenamefont {Hanisch-Blicharski}, \citenamefont {Janzen}, \citenamefont {Krenzer}, \citenamefont {Wall}, \citenamefont {Klasing}, \citenamefont {Kalus}, \citenamefont {Frigge}, \citenamefont {Kammler},\ and\ \citenamefont {{Horn-von Hoegen}}}]{Hanisch-Blicharski2013}%
  \BibitemOpen
  \bibfield  {author} {\bibinfo {author} {\bibfnamefont {A.}~\bibnamefont {Hanisch-Blicharski}}, \bibinfo {author} {\bibfnamefont {A.}~\bibnamefont {Janzen}}, \bibinfo {author} {\bibfnamefont {B.}~\bibnamefont {Krenzer}}, \bibinfo {author} {\bibfnamefont {S.}~\bibnamefont {Wall}}, \bibinfo {author} {\bibfnamefont {F.}~\bibnamefont {Klasing}}, \bibinfo {author} {\bibfnamefont {A.}~\bibnamefont {Kalus}}, \bibinfo {author} {\bibfnamefont {T.}~\bibnamefont {Frigge}}, \bibinfo {author} {\bibfnamefont {M.}~\bibnamefont {Kammler}}, \ and\ \bibinfo {author} {\bibfnamefont {M.}~\bibnamefont {{Horn-von Hoegen}}},\ }\href {\doibase 10.1016/j.ultramic.2012.07.017} {\bibfield  {journal} {\bibinfo  {journal} {Ultramicroscopy}\ }\textbf {\bibinfo {volume} {127}},\ \bibinfo {pages} {2} (\bibinfo {year} {2013})}\BibitemShut {NoStop}%
\bibitem [{\citenamefont {Gulde}\ \emph {et~al.}(2014)\citenamefont {Gulde}, \citenamefont {Schweda}, \citenamefont {Storeck}, \citenamefont {Maiti}, \citenamefont {Yu}, \citenamefont {Wodtke}, \citenamefont {Sch\"afer},\ and\ \citenamefont {Ropers}}]{Gulde2014}%
  \BibitemOpen
  \bibfield  {author} {\bibinfo {author} {\bibfnamefont {M.}~\bibnamefont {Gulde}}, \bibinfo {author} {\bibfnamefont {S.}~\bibnamefont {Schweda}}, \bibinfo {author} {\bibfnamefont {G.}~\bibnamefont {Storeck}}, \bibinfo {author} {\bibfnamefont {M.}~\bibnamefont {Maiti}}, \bibinfo {author} {\bibfnamefont {H.~K.}\ \bibnamefont {Yu}}, \bibinfo {author} {\bibfnamefont {A.~M.}\ \bibnamefont {Wodtke}}, \bibinfo {author} {\bibfnamefont {S.}~\bibnamefont {Sch\"afer}}, \ and\ \bibinfo {author} {\bibfnamefont {C.}~\bibnamefont {Ropers}},\ }\href {\doibase 10.1126/science.1250658} {\bibfield  {journal} {\bibinfo  {journal} {Science}\ }\textbf {\bibinfo {volume} {345}},\ \bibinfo {pages} {200} (\bibinfo {year} {2014})}\BibitemShut {NoStop}%
\bibitem [{\citenamefont {Frigge}\ \emph {et~al.}(2017)\citenamefont {Frigge}, \citenamefont {Hafke}, \citenamefont {Witte}, \citenamefont {Krenzer}, \citenamefont {Streub{\"{u}}hr}, \citenamefont {{Samad Syed}}, \citenamefont {{Mik{\v{s}}i{\'{c}} Trontl}}, \citenamefont {Avigo}, \citenamefont {Zhou}, \citenamefont {Ligges}, \citenamefont {{Von Der Linde}}, \citenamefont {Bovensiepen}, \citenamefont {{Horn-Von Hoegen}}, \citenamefont {Wippermann}, \citenamefont {L{\"{u}}cke}, \citenamefont {Sanna}, \citenamefont {Gerstmann},\ and\ \citenamefont {Schmidt}}]{Frigge2017}%
  \BibitemOpen
  \bibfield  {author} {\bibinfo {author} {\bibfnamefont {T.}~\bibnamefont {Frigge}}, \bibinfo {author} {\bibfnamefont {B.}~\bibnamefont {Hafke}}, \bibinfo {author} {\bibfnamefont {T.}~\bibnamefont {Witte}}, \bibinfo {author} {\bibfnamefont {B.}~\bibnamefont {Krenzer}}, \bibinfo {author} {\bibfnamefont {C.}~\bibnamefont {Streub{\"{u}}hr}}, \bibinfo {author} {\bibfnamefont {A.}~\bibnamefont {{Samad Syed}}}, \bibinfo {author} {\bibfnamefont {V.}~\bibnamefont {{Mik{\v{s}}i{\'{c}} Trontl}}}, \bibinfo {author} {\bibfnamefont {I.}~\bibnamefont {Avigo}}, \bibinfo {author} {\bibfnamefont {P.}~\bibnamefont {Zhou}}, \bibinfo {author} {\bibfnamefont {M.}~\bibnamefont {Ligges}}, \bibinfo {author} {\bibfnamefont {D.}~\bibnamefont {{Von Der Linde}}}, \bibinfo {author} {\bibfnamefont {U.}~\bibnamefont {Bovensiepen}}, \bibinfo {author} {\bibfnamefont {M.}~\bibnamefont {{Horn-Von Hoegen}}}, \bibinfo {author} {\bibfnamefont {S.}~\bibnamefont {Wippermann}}, \bibinfo {author} {\bibfnamefont {A.}~\bibnamefont {L{\"{u}}cke}},
  \bibinfo {author} {\bibfnamefont {S.}~\bibnamefont {Sanna}}, \bibinfo {author} {\bibfnamefont {U.}~\bibnamefont {Gerstmann}}, \ and\ \bibinfo {author} {\bibfnamefont {W.~G.}\ \bibnamefont {Schmidt}},\ }\href {\doibase 10.1038/nature21432} {\bibfield  {journal} {\bibinfo  {journal} {Nature}\ }\textbf {\bibinfo {volume} {544}},\ \bibinfo {pages} {207} (\bibinfo {year} {2017})}\BibitemShut {NoStop}%
\bibitem [{\citenamefont {Van~Hove}\ \emph {et~al.}(1986)\citenamefont {Van~Hove}, \citenamefont {Weinberg},\ and\ \citenamefont {Chan}}]{vanHove1986}%
  \BibitemOpen
  \bibfield  {author} {\bibinfo {author} {\bibfnamefont {M.~A.}\ \bibnamefont {Van~Hove}}, \bibinfo {author} {\bibfnamefont {W.~H.}\ \bibnamefont {Weinberg}}, \ and\ \bibinfo {author} {\bibfnamefont {C.-M.}\ \bibnamefont {Chan}},\ }\href@noop {} {\emph {\bibinfo {title} {Low-Energy Electron Diffraction}}}\ (\bibinfo  {publisher} {Springer},\ \bibinfo {address} {Berlin},\ \bibinfo {year} {1986})\BibitemShut {NoStop}%
\bibitem [{\citenamefont {Karrer}\ \emph {et~al.}(2001)\citenamefont {Karrer}, \citenamefont {Neff}, \citenamefont {Hengsberger}, \citenamefont {Greber},\ and\ \citenamefont {Osterwalder}}]{Karrer2001}%
  \BibitemOpen
  \bibfield  {author} {\bibinfo {author} {\bibfnamefont {R.}~\bibnamefont {Karrer}}, \bibinfo {author} {\bibfnamefont {H.~J.}\ \bibnamefont {Neff}}, \bibinfo {author} {\bibfnamefont {M.}~\bibnamefont {Hengsberger}}, \bibinfo {author} {\bibfnamefont {T.}~\bibnamefont {Greber}}, \ and\ \bibinfo {author} {\bibfnamefont {J.}~\bibnamefont {Osterwalder}},\ }\href {\doibase 10.1063/1.1419219} {\bibfield  {journal} {\bibinfo  {journal} {Rev. Sci. Instr.}\ }\textbf {\bibinfo {volume} {72}},\ \bibinfo {pages} {4404} (\bibinfo {year} {2001})}\BibitemShut {NoStop}%
\bibitem [{\citenamefont {M{\"{u}}ller}\ \emph {et~al.}(2014)\citenamefont {M{\"{u}}ller}, \citenamefont {Paarmann},\ and\ \citenamefont {Ernstorfer}}]{Muller2014}%
  \BibitemOpen
  \bibfield  {author} {\bibinfo {author} {\bibfnamefont {M.}~\bibnamefont {M{\"{u}}ller}}, \bibinfo {author} {\bibfnamefont {A.}~\bibnamefont {Paarmann}}, \ and\ \bibinfo {author} {\bibfnamefont {R.}~\bibnamefont {Ernstorfer}},\ }\href {\doibase 10.1038/ncomms6292} {\bibfield  {journal} {\bibinfo  {journal} {Nat. Commun.}\ }\textbf {\bibinfo {volume} {5}},\ \bibinfo {pages} {5292} (\bibinfo {year} {2014})}\BibitemShut {NoStop}%
\bibitem [{\citenamefont {Storeck}\ \emph {et~al.}(2017)\citenamefont {Storeck}, \citenamefont {Vogelgesang}, \citenamefont {Sivis}, \citenamefont {Sch{\"{a}}fer},\ and\ \citenamefont {Ropers}}]{Storeck2017}%
  \BibitemOpen
  \bibfield  {author} {\bibinfo {author} {\bibfnamefont {G.}~\bibnamefont {Storeck}}, \bibinfo {author} {\bibfnamefont {S.}~\bibnamefont {Vogelgesang}}, \bibinfo {author} {\bibfnamefont {M.}~\bibnamefont {Sivis}}, \bibinfo {author} {\bibfnamefont {S.}~\bibnamefont {Sch{\"{a}}fer}}, \ and\ \bibinfo {author} {\bibfnamefont {C.}~\bibnamefont {Ropers}},\ }\href {http://dx.doi.org/10.1063/1.4982947} {\bibfield  {journal} {\bibinfo  {journal} {Struct. Dyn.}\ }\textbf {\bibinfo {volume} {4}},\ \bibinfo {pages} {044042} (\bibinfo {year} {2017})}\BibitemShut {NoStop}%
\bibitem [{\citenamefont {Vogelgesang}\ \emph {et~al.}(2017)\citenamefont {Vogelgesang}, \citenamefont {Storeck}, \citenamefont {Horstmann}, \citenamefont {Diekmann}, \citenamefont {Sivis}, \citenamefont {Schramm}, \citenamefont {Rossnagel}, \citenamefont {Sch{\"{a}}fer},\ and\ \citenamefont {Ropers}}]{Vogelgesang2017}%
  \BibitemOpen
  \bibfield  {author} {\bibinfo {author} {\bibfnamefont {S.}~\bibnamefont {Vogelgesang}}, \bibinfo {author} {\bibfnamefont {G.}~\bibnamefont {Storeck}}, \bibinfo {author} {\bibfnamefont {J.~G.}\ \bibnamefont {Horstmann}}, \bibinfo {author} {\bibfnamefont {T.}~\bibnamefont {Diekmann}}, \bibinfo {author} {\bibfnamefont {M.}~\bibnamefont {Sivis}}, \bibinfo {author} {\bibfnamefont {S.}~\bibnamefont {Schramm}}, \bibinfo {author} {\bibfnamefont {K.}~\bibnamefont {Rossnagel}}, \bibinfo {author} {\bibfnamefont {S.}~\bibnamefont {Sch{\"{a}}fer}}, \ and\ \bibinfo {author} {\bibfnamefont {C.}~\bibnamefont {Ropers}},\ }\href {http://www.nature.com/doifinder/10.1038/nphys4309} {\bibfield  {journal} {\bibinfo  {journal} {Nat. Phys.}\ }\textbf {\bibinfo {volume} {14}},\ \bibinfo {pages} {184} (\bibinfo {year} {2017})}\BibitemShut {NoStop}%
\bibitem [{\citenamefont {Storeck}\ \emph {et~al.}(2021)\citenamefont {Storeck}, \citenamefont {Rossnagel},\ and\ \citenamefont {Ropers}}]{Storeck2021}%
  \BibitemOpen
  \bibfield  {author} {\bibinfo {author} {\bibfnamefont {G.}~\bibnamefont {Storeck}}, \bibinfo {author} {\bibfnamefont {K.}~\bibnamefont {Rossnagel}}, \ and\ \bibinfo {author} {\bibfnamefont {C.}~\bibnamefont {Ropers}},\ }\href@noop {} {\bibfield  {journal} {\bibinfo  {journal} {Appl. Phys. Lett.}\ }\textbf {\bibinfo {volume} {118}},\ \bibinfo {pages} {221603} (\bibinfo {year} {2021})}\BibitemShut {NoStop}%
\bibitem [{\citenamefont {Seah}\ and\ \citenamefont {Dench}(1979)}]{Seah1979}%
  \BibitemOpen
  \bibfield  {author} {\bibinfo {author} {\bibfnamefont {M.}~\bibnamefont {Seah}}\ and\ \bibinfo {author} {\bibfnamefont {W.}~\bibnamefont {Dench}},\ }\href@noop {} {\bibfield  {journal} {\bibinfo  {journal} {Surf. Interface Anal.}\ }\textbf {\bibinfo {volume} {1}},\ \bibinfo {pages} {2} (\bibinfo {year} {1979})}\BibitemShut {NoStop}%
\bibitem [{\citenamefont {Gopakumar}\ \emph {et~al.}(2004)\citenamefont {Gopakumar}, \citenamefont {Lackinger}, \citenamefont {Hackert}, \citenamefont {Millier},\ and\ \citenamefont {Hietschold}}]{Gopakumar2004}%
  \BibitemOpen
  \bibfield  {author} {\bibinfo {author} {\bibfnamefont {T.~G.}\ \bibnamefont {Gopakumar}}, \bibinfo {author} {\bibfnamefont {M.}~\bibnamefont {Lackinger}}, \bibinfo {author} {\bibfnamefont {M.}~\bibnamefont {Hackert}}, \bibinfo {author} {\bibfnamefont {F.}~\bibnamefont {Millier}}, \ and\ \bibinfo {author} {\bibfnamefont {M.}~\bibnamefont {Hietschold}},\ }\href {\doibase 10.1021/jp037751i} {\bibfield  {journal} {\bibinfo  {journal} {J. Phys. Chem. B}\ }\textbf {\bibinfo {volume} {108}},\ \bibinfo {pages} {7839} (\bibinfo {year} {2004})}\BibitemShut {NoStop}%
\bibitem [{\citenamefont {Kawakita}\ \emph {et~al.}(2017)\citenamefont {Kawakita}, \citenamefont {Yamada}, \citenamefont {Meissner}, \citenamefont {Forker}, \citenamefont {Fritz},\ and\ \citenamefont {Munakata}}]{Kawakita2017}%
  \BibitemOpen
  \bibfield  {author} {\bibinfo {author} {\bibfnamefont {N.}~\bibnamefont {Kawakita}}, \bibinfo {author} {\bibfnamefont {T.}~\bibnamefont {Yamada}}, \bibinfo {author} {\bibfnamefont {M.}~\bibnamefont {Meissner}}, \bibinfo {author} {\bibfnamefont {R.}~\bibnamefont {Forker}}, \bibinfo {author} {\bibfnamefont {T.}~\bibnamefont {Fritz}}, \ and\ \bibinfo {author} {\bibfnamefont {T.}~\bibnamefont {Munakata}},\ }\href {\doibase 10.1103/PhysRevB.95.045419} {\bibfield  {journal} {\bibinfo  {journal} {Phys. Rev. B}\ }\textbf {\bibinfo {volume} {95}},\ \bibinfo {pages} {045419} (\bibinfo {year} {2017})}\BibitemShut {NoStop}%
\bibitem [{\citenamefont {Jauernik}\ \emph {et~al.}(2018)\citenamefont {Jauernik}, \citenamefont {Hein}, \citenamefont {Gurgel}, \citenamefont {Falke},\ and\ \citenamefont {Bauer}}]{Jauernik2018}%
  \BibitemOpen
  \bibfield  {author} {\bibinfo {author} {\bibfnamefont {S.}~\bibnamefont {Jauernik}}, \bibinfo {author} {\bibfnamefont {P.}~\bibnamefont {Hein}}, \bibinfo {author} {\bibfnamefont {M.}~\bibnamefont {Gurgel}}, \bibinfo {author} {\bibfnamefont {J.}~\bibnamefont {Falke}}, \ and\ \bibinfo {author} {\bibfnamefont {M.}~\bibnamefont {Bauer}},\ }\href {\doibase 10.1103/PhysRevB.97.125413} {\bibfield  {journal} {\bibinfo  {journal} {Phys. Rev. B}\ }\textbf {\bibinfo {volume} {97}},\ \bibinfo {pages} {125413} (\bibinfo {year} {2018})}\BibitemShut {NoStop}%
\bibitem [{\citenamefont {Erk}\ \emph {et~al.}(2023)\citenamefont {Erk}, \citenamefont {Opitz}, \citenamefont {Hein}, \citenamefont {Jauernik},\ and\ \citenamefont {Bauer}}]{Erk2023}%
  \BibitemOpen
  \bibfield  {author} {\bibinfo {author} {\bibfnamefont {H.}~\bibnamefont {Erk}}, \bibinfo {author} {\bibfnamefont {K.}~\bibnamefont {Opitz}}, \bibinfo {author} {\bibfnamefont {P.}~\bibnamefont {Hein}}, \bibinfo {author} {\bibfnamefont {S.}~\bibnamefont {Jauernik}}, \ and\ \bibinfo {author} {\bibfnamefont {M.}~\bibnamefont {Bauer}},\ }\href@noop {} {\bibfield  {journal} {\bibinfo  {journal} {J. Phys. Condens. Matter}\ }\textbf {\bibinfo {volume} {35}},\ \bibinfo {pages} {095501} (\bibinfo {year} {2023})}\BibitemShut {NoStop}%
\bibitem [{\citenamefont {Ahuja}\ and\ \citenamefont {Auluck}(1997)}]{Ahuja1997}%
  \BibitemOpen
  \bibfield  {author} {\bibinfo {author} {\bibfnamefont {R.}~\bibnamefont {Ahuja}}\ and\ \bibinfo {author} {\bibfnamefont {S.}~\bibnamefont {Auluck}},\ }\href {\doibase 10.1103/PhysRevB.55.4999} {\bibfield  {journal} {\bibinfo  {journal} {Phys. Rev. B}\ }\textbf {\bibinfo {volume} {55}},\ \bibinfo {pages} {4999} (\bibinfo {year} {1997})}\BibitemShut {NoStop}%
\bibitem [{Sup()}]{Supplemental}%
  \BibitemOpen
  \href@noop {} {}\bibinfo {note} {See Supplemental Material at [URL will be inserted by publisher] for additional information}\BibitemShut {NoStop}%
\bibitem [{\citenamefont {Hein}\ \emph {et~al.}(2020)\citenamefont {Hein}, \citenamefont {Jauernik}, \citenamefont {Erk}, \citenamefont {Yang}, \citenamefont {Qi}, \citenamefont {Sun}, \citenamefont {Felser},\ and\ \citenamefont {Bauer}}]{Hein2020}%
  \BibitemOpen
  \bibfield  {author} {\bibinfo {author} {\bibfnamefont {P.}~\bibnamefont {Hein}}, \bibinfo {author} {\bibfnamefont {S.}~\bibnamefont {Jauernik}}, \bibinfo {author} {\bibfnamefont {H.}~\bibnamefont {Erk}}, \bibinfo {author} {\bibfnamefont {L.}~\bibnamefont {Yang}}, \bibinfo {author} {\bibfnamefont {Y.}~\bibnamefont {Qi}}, \bibinfo {author} {\bibfnamefont {Y.}~\bibnamefont {Sun}}, \bibinfo {author} {\bibfnamefont {C.}~\bibnamefont {Felser}}, \ and\ \bibinfo {author} {\bibfnamefont {M.}~\bibnamefont {Bauer}},\ }\href {http://dx.doi.org/10.1038/s41467-020-16076-0} {\bibfield  {journal} {\bibinfo  {journal} {Nat. Commun.}\ }\textbf {\bibinfo {volume} {11}},\ \bibinfo {pages} {2613} (\bibinfo {year} {2020})}\BibitemShut {NoStop}%
\bibitem [{\citenamefont {Gauthier}\ \emph {et~al.}(2020)\citenamefont {Gauthier}, \citenamefont {Sobota}, \citenamefont {Gauthier}, \citenamefont {Xu}, \citenamefont {Pfau}, \citenamefont {Rotundu}, \citenamefont {Shen},\ and\ \citenamefont {Kirchmann}}]{Gauthier2020}%
  \BibitemOpen
  \bibfield  {author} {\bibinfo {author} {\bibfnamefont {A.}~\bibnamefont {Gauthier}}, \bibinfo {author} {\bibfnamefont {J.~A.}\ \bibnamefont {Sobota}}, \bibinfo {author} {\bibfnamefont {N.}~\bibnamefont {Gauthier}}, \bibinfo {author} {\bibfnamefont {K.~J.}\ \bibnamefont {Xu}}, \bibinfo {author} {\bibfnamefont {H.}~\bibnamefont {Pfau}}, \bibinfo {author} {\bibfnamefont {C.~R.}\ \bibnamefont {Rotundu}}, \bibinfo {author} {\bibfnamefont {Z.~X.}\ \bibnamefont {Shen}}, \ and\ \bibinfo {author} {\bibfnamefont {P.~S.}\ \bibnamefont {Kirchmann}},\ }\href {https://doi.org/10.1063/5.0018834} {\bibfield  {journal} {\bibinfo  {journal} {J. Appl. Phys.}\ }\textbf {\bibinfo {volume} {128}},\ \bibinfo {pages} {093101} (\bibinfo {year} {2020})}\BibitemShut {NoStop}%
\bibitem [{\citenamefont {Yan}\ \emph {et~al.}(2021)\citenamefont {Yan}, \citenamefont {Green}, \citenamefont {Fukumori}, \citenamefont {Protic}, \citenamefont {Lee}, \citenamefont {Fernandez-Mulligan}, \citenamefont {Raja}, \citenamefont {Erdakos}, \citenamefont {Mao},\ and\ \citenamefont {Yang}}]{Yan2021}%
  \BibitemOpen
  \bibfield  {author} {\bibinfo {author} {\bibfnamefont {C.}~\bibnamefont {Yan}}, \bibinfo {author} {\bibfnamefont {E.}~\bibnamefont {Green}}, \bibinfo {author} {\bibfnamefont {R.}~\bibnamefont {Fukumori}}, \bibinfo {author} {\bibfnamefont {N.}~\bibnamefont {Protic}}, \bibinfo {author} {\bibfnamefont {S.~H.}\ \bibnamefont {Lee}}, \bibinfo {author} {\bibfnamefont {S.}~\bibnamefont {Fernandez-Mulligan}}, \bibinfo {author} {\bibfnamefont {R.}~\bibnamefont {Raja}}, \bibinfo {author} {\bibfnamefont {R.}~\bibnamefont {Erdakos}}, \bibinfo {author} {\bibfnamefont {Z.}~\bibnamefont {Mao}}, \ and\ \bibinfo {author} {\bibfnamefont {S.}~\bibnamefont {Yang}},\ }\href {https://doi.org/10.1063/5.0072979} {\bibfield  {journal} {\bibinfo  {journal} {Rev. Sci. Instr.}\ }\textbf {\bibinfo {volume} {92}},\ \bibinfo {pages} {113907} (\bibinfo {year} {2021})}\BibitemShut {NoStop}%
\bibitem [{\citenamefont {Kampfrath}\ \emph {et~al.}(2005)\citenamefont {Kampfrath}, \citenamefont {Perfetti}, \citenamefont {Schapper}, \citenamefont {Frischkorn},\ and\ \citenamefont {Wolf}}]{Kampfrath2005}%
  \BibitemOpen
  \bibfield  {author} {\bibinfo {author} {\bibfnamefont {T.}~\bibnamefont {Kampfrath}}, \bibinfo {author} {\bibfnamefont {L.}~\bibnamefont {Perfetti}}, \bibinfo {author} {\bibfnamefont {F.}~\bibnamefont {Schapper}}, \bibinfo {author} {\bibfnamefont {C.}~\bibnamefont {Frischkorn}}, \ and\ \bibinfo {author} {\bibfnamefont {M.}~\bibnamefont {Wolf}},\ }\href {\doibase 10.1103/PhysRevLett.95.187403} {\bibfield  {journal} {\bibinfo  {journal} {Phys. Rev. Lett.}\ }\textbf {\bibinfo {volume} {95}},\ \bibinfo {pages} {187403} (\bibinfo {year} {2005})}\BibitemShut {NoStop}%
\bibitem [{\citenamefont {Wang}\ \emph {et~al.}(2010)\citenamefont {Wang}, \citenamefont {Strait}, \citenamefont {George}, \citenamefont {Shivaraman}, \citenamefont {Shields}, \citenamefont {Chandrashekhar}, \citenamefont {Hwang}, \citenamefont {Rana}, \citenamefont {Spencer}, \citenamefont {Ruiz-Vargas},\ and\ \citenamefont {Park}}]{Wang2010}%
  \BibitemOpen
  \bibfield  {author} {\bibinfo {author} {\bibfnamefont {H.}~\bibnamefont {Wang}}, \bibinfo {author} {\bibfnamefont {J.~H.}\ \bibnamefont {Strait}}, \bibinfo {author} {\bibfnamefont {P.~A.}\ \bibnamefont {George}}, \bibinfo {author} {\bibfnamefont {S.}~\bibnamefont {Shivaraman}}, \bibinfo {author} {\bibfnamefont {V.~B.}\ \bibnamefont {Shields}}, \bibinfo {author} {\bibfnamefont {M.}~\bibnamefont {Chandrashekhar}}, \bibinfo {author} {\bibfnamefont {J.}~\bibnamefont {Hwang}}, \bibinfo {author} {\bibfnamefont {F.}~\bibnamefont {Rana}}, \bibinfo {author} {\bibfnamefont {M.~G.}\ \bibnamefont {Spencer}}, \bibinfo {author} {\bibfnamefont {C.~S.}\ \bibnamefont {Ruiz-Vargas}}, \ and\ \bibinfo {author} {\bibfnamefont {J.}~\bibnamefont {Park}},\ }\href {\doibase 10.1063/1.3291615} {\bibfield  {journal} {\bibinfo  {journal} {Appl. Phys. Lett.}\ }\textbf {\bibinfo {volume} {96}},\ \bibinfo {pages} {081917} (\bibinfo {year} {2010})}\BibitemShut {NoStop}%
\bibitem [{\citenamefont {Johannsen}\ \emph {et~al.}(2013)\citenamefont {Johannsen}, \citenamefont {Ulstrup}, \citenamefont {Cilento}, \citenamefont {Crepaldi}, \citenamefont {Zacchigna}, \citenamefont {Cacho}, \citenamefont {Turcu}, \citenamefont {Springate}, \citenamefont {Fromm}, \citenamefont {Raidel}, \citenamefont {Seyller}, \citenamefont {Parmigiani}, \citenamefont {Grioni},\ and\ \citenamefont {Hofmann}}]{Johannsen2013}%
  \BibitemOpen
  \bibfield  {author} {\bibinfo {author} {\bibfnamefont {J.~C.}\ \bibnamefont {Johannsen}}, \bibinfo {author} {\bibfnamefont {S.}~\bibnamefont {Ulstrup}}, \bibinfo {author} {\bibfnamefont {F.}~\bibnamefont {Cilento}}, \bibinfo {author} {\bibfnamefont {A.}~\bibnamefont {Crepaldi}}, \bibinfo {author} {\bibfnamefont {M.}~\bibnamefont {Zacchigna}}, \bibinfo {author} {\bibfnamefont {C.}~\bibnamefont {Cacho}}, \bibinfo {author} {\bibfnamefont {I.~C.~E.}\ \bibnamefont {Turcu}}, \bibinfo {author} {\bibfnamefont {E.}~\bibnamefont {Springate}}, \bibinfo {author} {\bibfnamefont {F.}~\bibnamefont {Fromm}}, \bibinfo {author} {\bibfnamefont {C.}~\bibnamefont {Raidel}}, \bibinfo {author} {\bibfnamefont {T.}~\bibnamefont {Seyller}}, \bibinfo {author} {\bibfnamefont {F.}~\bibnamefont {Parmigiani}}, \bibinfo {author} {\bibfnamefont {M.}~\bibnamefont {Grioni}}, \ and\ \bibinfo {author} {\bibfnamefont {P.}~\bibnamefont {Hofmann}},\ }\href {\doibase 10.1103/PhysRevLett.111.027403} {\bibfield  {journal} {\bibinfo  {journal} {Phys.
  Rev. Lett.}\ }\textbf {\bibinfo {volume} {111}},\ \bibinfo {pages} {027403} (\bibinfo {year} {2013})}\BibitemShut {NoStop}%
\bibitem [{\citenamefont {Yan}\ \emph {et~al.}(2009)\citenamefont {Yan}, \citenamefont {Song}, \citenamefont {Mak}, \citenamefont {Chatzakis}, \citenamefont {Maultzsch},\ and\ \citenamefont {Heinz}}]{Yan2009}%
  \BibitemOpen
  \bibfield  {author} {\bibinfo {author} {\bibfnamefont {H.}~\bibnamefont {Yan}}, \bibinfo {author} {\bibfnamefont {D.}~\bibnamefont {Song}}, \bibinfo {author} {\bibfnamefont {K.~F.}\ \bibnamefont {Mak}}, \bibinfo {author} {\bibfnamefont {I.}~\bibnamefont {Chatzakis}}, \bibinfo {author} {\bibfnamefont {J.}~\bibnamefont {Maultzsch}}, \ and\ \bibinfo {author} {\bibfnamefont {T.~F.}\ \bibnamefont {Heinz}},\ }\href {\doibase 10.1103/PhysRevB.80.121403} {\bibfield  {journal} {\bibinfo  {journal} {Phys. Rev. B}\ }\textbf {\bibinfo {volume} {80}},\ \bibinfo {pages} {121403(R)} (\bibinfo {year} {2009})}\BibitemShut {NoStop}%
\bibitem [{\citenamefont {Malic}\ \emph {et~al.}(2011)\citenamefont {Malic}, \citenamefont {Winzer}, \citenamefont {Bobkin},\ and\ \citenamefont {Knorr}}]{Malic2011}%
  \BibitemOpen
  \bibfield  {author} {\bibinfo {author} {\bibfnamefont {E.}~\bibnamefont {Malic}}, \bibinfo {author} {\bibfnamefont {T.}~\bibnamefont {Winzer}}, \bibinfo {author} {\bibfnamefont {E.}~\bibnamefont {Bobkin}}, \ and\ \bibinfo {author} {\bibfnamefont {A.}~\bibnamefont {Knorr}},\ }\href {\doibase 10.1103/PhysRevB.84.205406} {\bibfield  {journal} {\bibinfo  {journal} {Phys. Rev. B}\ }\textbf {\bibinfo {volume} {84}},\ \bibinfo {pages} {205406} (\bibinfo {year} {2011})}\BibitemShut {NoStop}%
\bibitem [{\citenamefont {Stange}\ \emph {et~al.}(2015)\citenamefont {Stange}, \citenamefont {Sohrt}, \citenamefont {Yang}, \citenamefont {Rohde}, \citenamefont {Janssen}, \citenamefont {Hein}, \citenamefont {Oloff}, \citenamefont {Hanff}, \citenamefont {Rossnagel},\ and\ \citenamefont {Bauer}}]{Stange2015}%
  \BibitemOpen
  \bibfield  {author} {\bibinfo {author} {\bibfnamefont {A.}~\bibnamefont {Stange}}, \bibinfo {author} {\bibfnamefont {C.}~\bibnamefont {Sohrt}}, \bibinfo {author} {\bibfnamefont {L.~X.}\ \bibnamefont {Yang}}, \bibinfo {author} {\bibfnamefont {G.}~\bibnamefont {Rohde}}, \bibinfo {author} {\bibfnamefont {K.}~\bibnamefont {Janssen}}, \bibinfo {author} {\bibfnamefont {P.}~\bibnamefont {Hein}}, \bibinfo {author} {\bibfnamefont {L.-P.}\ \bibnamefont {Oloff}}, \bibinfo {author} {\bibfnamefont {K.}~\bibnamefont {Hanff}}, \bibinfo {author} {\bibfnamefont {K.}~\bibnamefont {Rossnagel}}, \ and\ \bibinfo {author} {\bibfnamefont {M.}~\bibnamefont {Bauer}},\ }\href {\doibase 10.1103/PhysRevB.92.184303} {\bibfield  {journal} {\bibinfo  {journal} {Phys. Rev. B}\ }\textbf {\bibinfo {volume} {92}},\ \bibinfo {pages} {184303} (\bibinfo {year} {2015})}\BibitemShut {NoStop}%
\bibitem [{Fer()}]{Fermi}%
  \BibitemOpen
  \href@noop {} {}\bibinfo {note} {The Fermi energy of graphite lies within a few \SI{10}{\milli eV} below the contact point (Dirac point) of both bands at the $H$-point of SCG \cite{Mikitik2006, Orlita2008, Na2019}. The observation that the tip of the cone structure in the ARPES spectra of SnPc/SCG before photoexcitation touches $E_F$ [Fig. \ref{fig:Fig1}(a)] implies that, if at all, this situation is not significantly changed by the adsorption of the SnPc molecules.}\BibitemShut {Stop}%
\bibitem [{\citenamefont {Gr{\"{u}}neis}\ \emph {et~al.}(2008)\citenamefont {Gr{\"{u}}neis}, \citenamefont {Attaccalite}, \citenamefont {Rubio}, \citenamefont {Molodtsov}, \citenamefont {Vyalikh}, \citenamefont {Fink}, \citenamefont {Follath},\ and\ \citenamefont {Pichler}}]{Gruneis2008}%
  \BibitemOpen
  \bibfield  {author} {\bibinfo {author} {\bibfnamefont {A.}~\bibnamefont {Gr{\"{u}}neis}}, \bibinfo {author} {\bibfnamefont {C.}~\bibnamefont {Attaccalite}}, \bibinfo {author} {\bibfnamefont {A.}~\bibnamefont {Rubio}}, \bibinfo {author} {\bibfnamefont {S.~L.}\ \bibnamefont {Molodtsov}}, \bibinfo {author} {\bibfnamefont {D.~V.}\ \bibnamefont {Vyalikh}}, \bibinfo {author} {\bibfnamefont {J.}~\bibnamefont {Fink}}, \bibinfo {author} {\bibfnamefont {R.}~\bibnamefont {Follath}}, \ and\ \bibinfo {author} {\bibfnamefont {T.}~\bibnamefont {Pichler}},\ }\href {\doibase 10.1002/pssb.200879662} {\bibfield  {journal} {\bibinfo  {journal} {Phys. Status Solidi B Basic Res.}\ }\textbf {\bibinfo {volume} {245}},\ \bibinfo {pages} {2072} (\bibinfo {year} {2008})}\BibitemShut {NoStop}%
\bibitem [{\citenamefont {Perfetti}\ \emph {et~al.}(2007)\citenamefont {Perfetti}, \citenamefont {Loukakos}, \citenamefont {Lisowski}, \citenamefont {Bovensiepen}, \citenamefont {Eisaki},\ and\ \citenamefont {Wolf}}]{Perfetti2007}%
  \BibitemOpen
  \bibfield  {author} {\bibinfo {author} {\bibfnamefont {L.}~\bibnamefont {Perfetti}}, \bibinfo {author} {\bibfnamefont {P.}~\bibnamefont {Loukakos}}, \bibinfo {author} {\bibfnamefont {M.}~\bibnamefont {Lisowski}}, \bibinfo {author} {\bibfnamefont {U.}~\bibnamefont {Bovensiepen}}, \bibinfo {author} {\bibfnamefont {H.}~\bibnamefont {Eisaki}}, \ and\ \bibinfo {author} {\bibfnamefont {M.}~\bibnamefont {Wolf}},\ }\href {\doibase 10.1103/PhysRevLett.99.197001} {\bibfield  {journal} {\bibinfo  {journal} {Phys. Rev. Lett.}\ }\textbf {\bibinfo {volume} {99}},\ \bibinfo {pages} {197001} (\bibinfo {year} {2007})}\BibitemShut {NoStop}%
\bibitem [{\citenamefont {Kr{\"{o}}mker}\ \emph {et~al.}(2008)\citenamefont {Kr{\"{o}}mker}, \citenamefont {Escher}, \citenamefont {Funnemann}, \citenamefont {Hartung}, \citenamefont {Engelhard},\ and\ \citenamefont {Kirschner}}]{Kromker2008}%
  \BibitemOpen
  \bibfield  {author} {\bibinfo {author} {\bibfnamefont {B.}~\bibnamefont {Kr{\"{o}}mker}}, \bibinfo {author} {\bibfnamefont {M.}~\bibnamefont {Escher}}, \bibinfo {author} {\bibfnamefont {D.}~\bibnamefont {Funnemann}}, \bibinfo {author} {\bibfnamefont {D.}~\bibnamefont {Hartung}}, \bibinfo {author} {\bibfnamefont {H.}~\bibnamefont {Engelhard}}, \ and\ \bibinfo {author} {\bibfnamefont {J.}~\bibnamefont {Kirschner}},\ }\href@noop {} {\bibfield  {journal} {\bibinfo  {journal} {Rev. Sci. Instr.}\ }\textbf {\bibinfo {volume} {79}} (\bibinfo {year} {2008})}\BibitemShut {NoStop}%
\bibitem [{\citenamefont {Stern}\ \emph {et~al.}(2017)\citenamefont {Stern}, \citenamefont {de~Cotret}, \citenamefont {Otto}, \citenamefont {Chatelain}, \citenamefont {Boisvert}, \citenamefont {Sutton},\ and\ \citenamefont {Siwick}}]{Stern2017}%
  \BibitemOpen
  \bibfield  {author} {\bibinfo {author} {\bibfnamefont {M.~J.}\ \bibnamefont {Stern}}, \bibinfo {author} {\bibfnamefont {L.~P.~R.}\ \bibnamefont {de~Cotret}}, \bibinfo {author} {\bibfnamefont {M.~R.}\ \bibnamefont {Otto}}, \bibinfo {author} {\bibfnamefont {R.~P.}\ \bibnamefont {Chatelain}}, \bibinfo {author} {\bibfnamefont {J.-P.}\ \bibnamefont {Boisvert}}, \bibinfo {author} {\bibfnamefont {M.}~\bibnamefont {Sutton}}, \ and\ \bibinfo {author} {\bibfnamefont {B.~J.}\ \bibnamefont {Siwick}},\ }\href {\doibase 10.1103/PhysRevB.97.165416} {\bibfield  {journal} {\bibinfo  {journal} {Phys. Rev. B}\ }\textbf {\bibinfo {volume} {79}},\ \bibinfo {pages} {165416} (\bibinfo {year} {2017})}\BibitemShut {NoStop}%
\bibitem [{\citenamefont {Seiler}\ \emph {et~al.}(2021)\citenamefont {Seiler}, \citenamefont {Zahn}, \citenamefont {Zacharias}, \citenamefont {Hildebrandt}, \citenamefont {Vasileiadis}, \citenamefont {Windsor}, \citenamefont {Qi}, \citenamefont {Carbogno}, \citenamefont {Draxl}, \citenamefont {Ernstorfer},\ and\ \citenamefont {Caruso}}]{Seiler2021}%
  \BibitemOpen
  \bibfield  {author} {\bibinfo {author} {\bibfnamefont {H.}~\bibnamefont {Seiler}}, \bibinfo {author} {\bibfnamefont {D.}~\bibnamefont {Zahn}}, \bibinfo {author} {\bibfnamefont {M.}~\bibnamefont {Zacharias}}, \bibinfo {author} {\bibfnamefont {P.~N.}\ \bibnamefont {Hildebrandt}}, \bibinfo {author} {\bibfnamefont {T.}~\bibnamefont {Vasileiadis}}, \bibinfo {author} {\bibfnamefont {Y.~W.}\ \bibnamefont {Windsor}}, \bibinfo {author} {\bibfnamefont {Y.}~\bibnamefont {Qi}}, \bibinfo {author} {\bibfnamefont {C.}~\bibnamefont {Carbogno}}, \bibinfo {author} {\bibfnamefont {C.}~\bibnamefont {Draxl}}, \bibinfo {author} {\bibfnamefont {R.}~\bibnamefont {Ernstorfer}}, \ and\ \bibinfo {author} {\bibfnamefont {F.}~\bibnamefont {Caruso}},\ }\href {\doibase 10.1021/acs.nanolett.1c01786} {\bibfield  {journal} {\bibinfo  {journal} {Nano Lett.}\ }\textbf {\bibinfo {volume} {21}},\ \bibinfo {pages} {6171} (\bibinfo {year} {2021})}\BibitemShut {NoStop}%
\bibitem [{\citenamefont {Rohwer}\ \emph {et~al.}(2011)\citenamefont {Rohwer}, \citenamefont {Hellmann}, \citenamefont {Wiesenmayer}, \citenamefont {Sohrt}, \citenamefont {Stange}, \citenamefont {Slomski}, \citenamefont {Carr}, \citenamefont {Liu}, \citenamefont {Avila}, \citenamefont {Kall{\"{a}}signne}, \citenamefont {Mathias}, \citenamefont {Kipp}, \citenamefont {Rossnagel},\ and\ \citenamefont {Bauer}}]{Rohwer2011}%
  \BibitemOpen
  \bibfield  {author} {\bibinfo {author} {\bibfnamefont {T.}~\bibnamefont {Rohwer}}, \bibinfo {author} {\bibfnamefont {S.}~\bibnamefont {Hellmann}}, \bibinfo {author} {\bibfnamefont {M.}~\bibnamefont {Wiesenmayer}}, \bibinfo {author} {\bibfnamefont {C.}~\bibnamefont {Sohrt}}, \bibinfo {author} {\bibfnamefont {A.}~\bibnamefont {Stange}}, \bibinfo {author} {\bibfnamefont {B.}~\bibnamefont {Slomski}}, \bibinfo {author} {\bibfnamefont {A.}~\bibnamefont {Carr}}, \bibinfo {author} {\bibfnamefont {Y.}~\bibnamefont {Liu}}, \bibinfo {author} {\bibfnamefont {L.~M.}\ \bibnamefont {Avila}}, \bibinfo {author} {\bibfnamefont {M.}~\bibnamefont {Kall{\"{a}}signne}}, \bibinfo {author} {\bibfnamefont {S.}~\bibnamefont {Mathias}}, \bibinfo {author} {\bibfnamefont {L.}~\bibnamefont {Kipp}}, \bibinfo {author} {\bibfnamefont {K.}~\bibnamefont {Rossnagel}}, \ and\ \bibinfo {author} {\bibfnamefont {M.}~\bibnamefont {Bauer}},\ }\href {\doibase 10.1038/nature09829} {\bibfield  {journal} {\bibinfo  {journal} {Nature}\ }\textbf
  {\bibinfo {volume} {471}},\ \bibinfo {pages} {490} (\bibinfo {year} {2011})}\BibitemShut {NoStop}%
\bibitem [{\citenamefont {Hellmann}\ \emph {et~al.}(2010)\citenamefont {Hellmann}, \citenamefont {Beye}, \citenamefont {Sohrt}, \citenamefont {Rohwer}, \citenamefont {Sorgenfrei}, \citenamefont {Redlin}, \citenamefont {Kall{\"{a}}ne}, \citenamefont {Marczynski-B{\"{u}}hlow}, \citenamefont {Hennies}, \citenamefont {Bauer}, \citenamefont {F{\"{o}}hlisch}, \citenamefont {Kipp}, \citenamefont {Wurth},\ and\ \citenamefont {Rossnagel}}]{Hellmann2010}%
  \BibitemOpen
  \bibfield  {author} {\bibinfo {author} {\bibfnamefont {S.}~\bibnamefont {Hellmann}}, \bibinfo {author} {\bibfnamefont {M.}~\bibnamefont {Beye}}, \bibinfo {author} {\bibfnamefont {C.}~\bibnamefont {Sohrt}}, \bibinfo {author} {\bibfnamefont {T.}~\bibnamefont {Rohwer}}, \bibinfo {author} {\bibfnamefont {F.}~\bibnamefont {Sorgenfrei}}, \bibinfo {author} {\bibfnamefont {H.}~\bibnamefont {Redlin}}, \bibinfo {author} {\bibfnamefont {M.}~\bibnamefont {Kall{\"{a}}ne}}, \bibinfo {author} {\bibfnamefont {M.}~\bibnamefont {Marczynski-B{\"{u}}hlow}}, \bibinfo {author} {\bibfnamefont {F.}~\bibnamefont {Hennies}}, \bibinfo {author} {\bibfnamefont {M.}~\bibnamefont {Bauer}}, \bibinfo {author} {\bibfnamefont {A.}~\bibnamefont {F{\"{o}}hlisch}}, \bibinfo {author} {\bibfnamefont {L.}~\bibnamefont {Kipp}}, \bibinfo {author} {\bibfnamefont {W.}~\bibnamefont {Wurth}}, \ and\ \bibinfo {author} {\bibfnamefont {K.}~\bibnamefont {Rossnagel}},\ }\href {\doibase 10.1103/PhysRevLett.105.187401} {\bibfield  {journal} {\bibinfo
  {journal} {Phys. Rev. Lett.}\ }\textbf {\bibinfo {volume} {105}},\ \bibinfo {pages} {187401} (\bibinfo {year} {2010})}\BibitemShut {NoStop}%
\bibitem [{\citenamefont {Mikitik}\ and\ \citenamefont {Sharlai}(2006)}]{Mikitik2006}%
  \BibitemOpen
  \bibfield  {author} {\bibinfo {author} {\bibfnamefont {G.~P.}\ \bibnamefont {Mikitik}}\ and\ \bibinfo {author} {\bibfnamefont {Y.~V.}\ \bibnamefont {Sharlai}},\ }\href {\doibase 10.1103/PhysRevB.73.235112} {\bibfield  {journal} {\bibinfo  {journal} {Phys. Rev. B}\ }\textbf {\bibinfo {volume} {73}},\ \bibinfo {pages} {235112} (\bibinfo {year} {2006})}\BibitemShut {NoStop}%
\bibitem [{\citenamefont {Orlita}\ \emph {et~al.}(2008)\citenamefont {Orlita}, \citenamefont {Faugeras}, \citenamefont {Martinez}, \citenamefont {Maude}, \citenamefont {Sadowski},\ and\ \citenamefont {Potemski}}]{Orlita2008}%
  \BibitemOpen
  \bibfield  {author} {\bibinfo {author} {\bibfnamefont {M.}~\bibnamefont {Orlita}}, \bibinfo {author} {\bibfnamefont {C.}~\bibnamefont {Faugeras}}, \bibinfo {author} {\bibfnamefont {G.}~\bibnamefont {Martinez}}, \bibinfo {author} {\bibfnamefont {D.~K.}\ \bibnamefont {Maude}}, \bibinfo {author} {\bibfnamefont {M.~L.}\ \bibnamefont {Sadowski}}, \ and\ \bibinfo {author} {\bibfnamefont {M.}~\bibnamefont {Potemski}},\ }\href {\doibase 10.1103/PhysRevLett.100.136403} {\bibfield  {journal} {\bibinfo  {journal} {Phys. Rev. Lett.}\ }\textbf {\bibinfo {volume} {100}},\ \bibinfo {pages} {4} (\bibinfo {year} {2008})}\BibitemShut {NoStop}%
\bibitem [{\citenamefont {Na}\ \emph {et~al.}(2019)\citenamefont {Na}, \citenamefont {Mills}, \citenamefont {Boschini}, \citenamefont {Michiardi}, \citenamefont {Nosarzewski}, \citenamefont {Day}, \citenamefont {Razzoli}, \citenamefont {Sheyerman}, \citenamefont {Schneider}, \citenamefont {Levy}, \citenamefont {Zhdanovich}, \citenamefont {Devereaux}, \citenamefont {Kemper}, \citenamefont {Jones},\ and\ \citenamefont {Damascelli}}]{Na2019}%
  \BibitemOpen
  \bibfield  {author} {\bibinfo {author} {\bibfnamefont {M.~X.}\ \bibnamefont {Na}}, \bibinfo {author} {\bibfnamefont {A.~K.}\ \bibnamefont {Mills}}, \bibinfo {author} {\bibfnamefont {F.}~\bibnamefont {Boschini}}, \bibinfo {author} {\bibfnamefont {M.}~\bibnamefont {Michiardi}}, \bibinfo {author} {\bibfnamefont {B.}~\bibnamefont {Nosarzewski}}, \bibinfo {author} {\bibfnamefont {R.~P.}\ \bibnamefont {Day}}, \bibinfo {author} {\bibfnamefont {E.}~\bibnamefont {Razzoli}}, \bibinfo {author} {\bibfnamefont {A.}~\bibnamefont {Sheyerman}}, \bibinfo {author} {\bibfnamefont {M.}~\bibnamefont {Schneider}}, \bibinfo {author} {\bibfnamefont {G.}~\bibnamefont {Levy}}, \bibinfo {author} {\bibfnamefont {S.}~\bibnamefont {Zhdanovich}}, \bibinfo {author} {\bibfnamefont {T.~P.}\ \bibnamefont {Devereaux}}, \bibinfo {author} {\bibfnamefont {A.~F.}\ \bibnamefont {Kemper}}, \bibinfo {author} {\bibfnamefont {D.~J.}\ \bibnamefont {Jones}}, \ and\ \bibinfo {author} {\bibfnamefont {A.}~\bibnamefont {Damascelli}},\ }\href {\doibase
  10.1126/science.aaw1662} {\bibfield  {journal} {\bibinfo  {journal} {Science}\ }\textbf {\bibinfo {volume} {366}},\ \bibinfo {pages} {1231} (\bibinfo {year} {2019})}\BibitemShut {NoStop}%
\end{thebibliography}%

\end{document}

\title{Supplemental Material for\texorpdfstring{\\}{ }``Ultrafast low-energy photoelectron diffraction for the study of surface-adsorbate interactions with 100 femtosecond temporal resolution''}

%%%%%%%%%%%%%%%%%%%%%%%%%%%%%%%%%%%%%%%%%%%%%%
\author{H.~Erk}
\email{erk@physik.uni-kiel.de}
\affiliation
{Institute of Experimental and Applied Physics, Kiel University, 24098 Kiel, Germany
}
\author{C. E.~Jensen}
\affiliation{Institute of Experimental and Applied Physics, Kiel University, 24098 Kiel, Germany
}
%\author{N.~N}
%\affiliation{Institute of Experimental and Applied Physics, Kiel University, 24098 Kiel, Germany
%}
\author{S.~Jauernik}
\affiliation{Institute of Experimental and Applied Physics, Kiel University, 24098 Kiel, Germany
}
\author{M.~Bauer}
\homepage{https://www.physik.uni-kiel.de/de/institute/ag-bauer}
\affiliation{Institute of Experimental and Applied Physics, Kiel University, 24098 Kiel, Germany
}
\affiliation{
Kiel Nano, Surface and Interface Science KiNSIS, Kiel University, 24118 Kiel, Germany
}

%%%%%%%%%%%%%%%%%%%%%%%%%%%%%%%%%%%%%%%%%%%%%%

\date{\today}

\maketitle

\section*{S.\,I Comparison of LEPD data with simulations\quad }
\begin{figure}
        \includegraphics[width=0.4\linewidth]{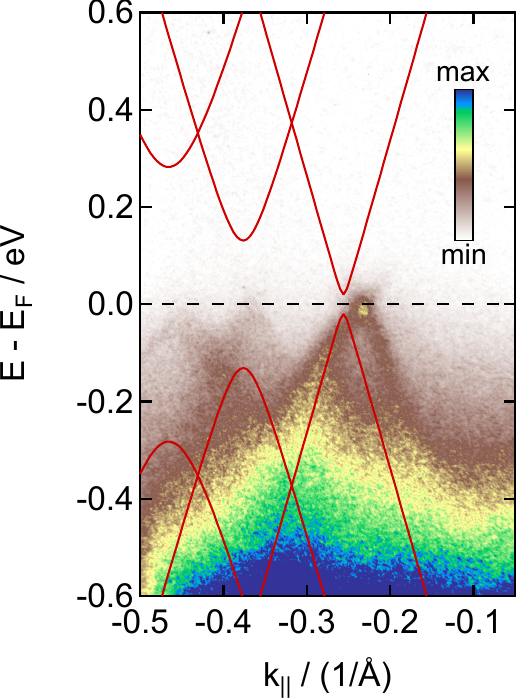}
        \caption{Experimental LEPD intensity map in comparison to simulation results for the LEPD signal for SnPc/SCG.} 
    	\label{fig:FigS1}
\end{figure}

\noindent
Figure \ref{fig:FigS1} shows the LEPD intensity map in Fig. 2(a) of the manuscript, overlaid with simulation results for the LEPD signal of graphite near $E_\text{F}$ under consideration of the SnPc overlayer epitaxial matrix. Here, the band structure of pristine graphite was approximated by a two-dimensional isotropic Dirac cone at at \Kbar, so that the simulated data represent hyperbolic slices of Dirac cones. As in the simulated diffraction pattern shown in the top panel of Fig. 1(b) of the main manuscript, we have considered diffraction orders up to the fourth order. The simulations show a good agreement with the LEPD data with respect to the momentum distribution and qualitative reproduce the main cone in the data as well as the two additional side cones at larger momenta. We attribute the relative energy shifts between simulations and experiments to the simplified band structure used in the model. Further details are given in Ref. \cite{Erk2023}.

\section*{S.\,II Contribution of the Fermi-Dirac distribution to the temperature dependence of the LEPD intensity\quad }
\begin{figure}
        \includegraphics[width=0.6\linewidth]{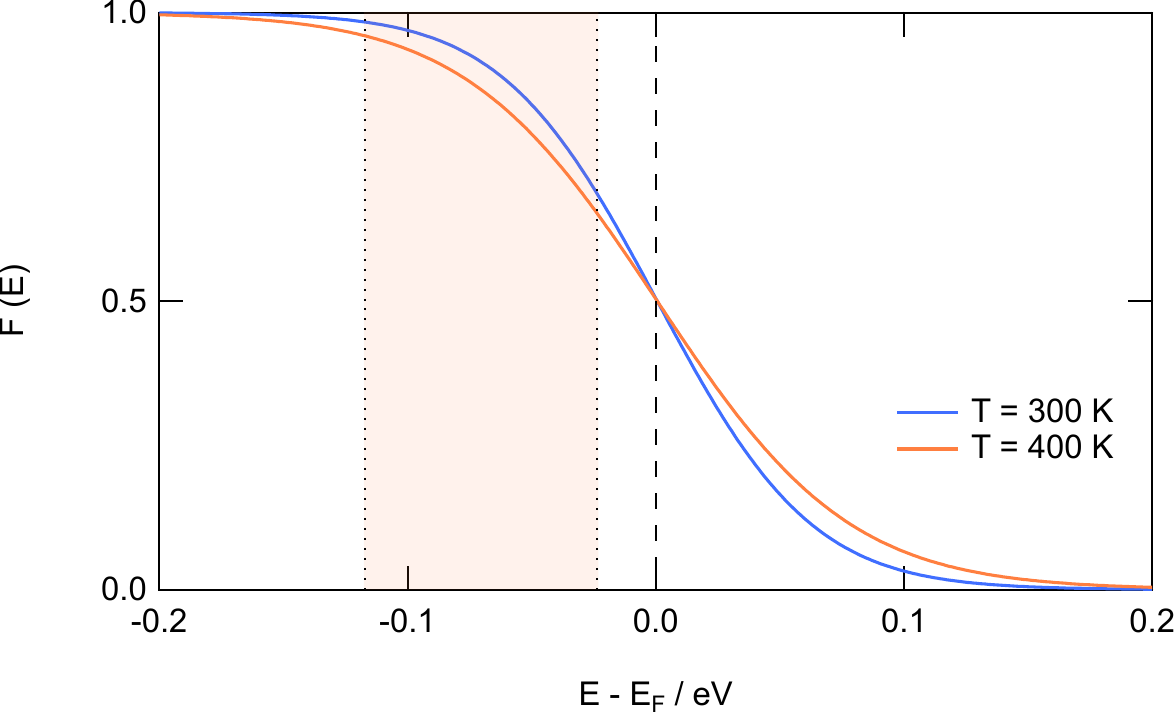}
        \caption{Calculated Fermi-Dirac distribution functions $F(E)$ to estimate its maximum contribution to the observed changes in the LEPD intensity as a function of temperature. To account for the finite energy resolution of the experiment, the distributions were convolved with a Gaussian with a FWHM of \SI{60}{\milli eV}.} 
    	\label{fig:FigS2}
\end{figure}

\noindent
Figure \ref{fig:FigS2} shows calculated Fermi-Dirac distribution functions $F(E)$ for $T=\SI{300}{K}$ and $T=\SI{400}{K}$, which corresponds to initial and final temperature of the static LEED and LEPD measurements on the temperature dependence of the diffraction intensities shown in Fig. 3(a). Both curves were convolved with a Gaussian to account for the energy resolution of \SI{60}{\milli eV} of the experiment. The two dotted lines mark the energy interval used to evaluate the temperature dependence of the LEPD signal. The comparison of the integrals of the two Fermi-Dirac distribution functions in this energy interval shows that the heating of the electron gas can account for a maximum relative intensity change of the LEPD signal in the temperature range of $\approx \SI{4.7}{\%}$. This value reduces to $\approx \SI{4.6}{\%}$ if the density of state of graphite is taken into account \cite{Ooi2006}. The value is significantly lower than the relative change in LEPD intensity of $\approx$~40\% observed in the experiment.

\section*{S.\,III Calculation of the temperature increase of the SCG substrate due to absorption of the pump pulse\quad } 

%Give details on how the final temperature has been estimated
The increase in temperature of the SCG substrate due to the pump pulse excitation and after complete thermalization of all electronic and lattice degrees of freedom has been calculated from the absorbed fluence by

\begin{equation}
    \Delta T = \frac{0.06 \cdot E_{\mathrm{pulse}} \cdot \cos(\Theta)}{V_{\mathrm{exc}} \cdot c_{\mathrm{graphite}} \cdot \rho_{\mathrm{graphite}}} \approx 16~\mathrm{K}
\end{equation}.

Here $E_{\mathrm{pulse}} = 0.045~\upmu \mathrm{J}$ is the net energy deposited in the SCG substrate due to absorption calculated using the Fresnel equations. $\Theta = 33^\circ$ is the incidence angle of the pump pulse, $c_{\mathrm{graphite}} = 706~\mathrm{J~kg^{-1}K^{-1}}$ \cite{Picard2007} and $\rho_{\mathrm{graphite}} = 2.1~\mathrm{g~cm^{-3}}$ are the specific heat and density of graphite, respectively. The factor 0.06 accounts for the probe pulse diameter ($d_{\mathrm{probe}} = 50~\mathrm{\upmu m}$) being much smaller than the ($d_{\mathrm{pump}} = 185~\mathrm{\upmu m}$), which guarantees that only the center of the excitation region is probed. The probed excitation volume $V_\mathrm{exc} = 98~\mathrm{\upmu m^3}$ is given by the probe pulse diameter  and the pump pulse penetration depth of $50~\mathrm{nm}$. The later value has been calculated from optical constants of graphite \cite{Djurišić1999}.

\section*{S.\,IV Comparison of the ULEPD transients from $\pi$ and the $\pi^*$ band\quad } 
\begin{figure}
        \includegraphics[width=0.7\linewidth]{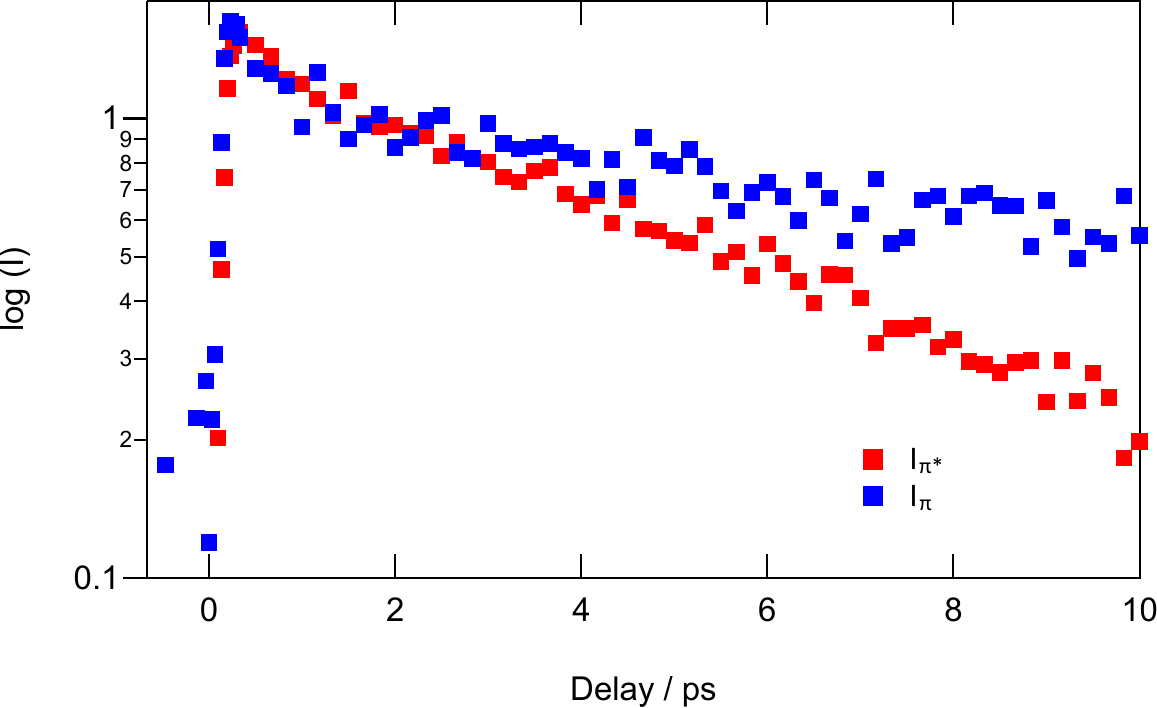}
        \caption{$\pi$ and the $\pi^*$ band transients in Fig. 2(d) of the main text in semi-logarithmic representation. The $\pi$-band transient was multiplied by a factor $(-1)$} 
    	\label{fig:FigS3}
\end{figure}

\noindent
Figure \ref{fig:FigS3} compares the ULEPD transients from the $\pi$ and the $\pi^*$ band on a semi-logarithmic scale, with the transient from the $\pi$ band inverted. In this representation of the data, it can be seen that for time delays between \SI{1}{\pico s} and \SI{2}{\pico s} the slopes of the two transients are virtually the same. We conclude that in this time window the the temporal evolution of both signals is governed by the same process, namely the decay of the hot carrier distribution. We used the integrated signals in this time interval to normalize the two transients in order to compensate for the differences in the matrix element for photoemission from the $\pi$ and the $\pi^*$ band \cite{Gruneis2008}. By summing, we were able to separate quantitative information about the structural dynamics in the SnPc top layer from the carrier dynamics in the SCG substrate in a further step (see section S. V). Note that on longer time scales the $\pi$-band transient starts to deviate from the $\pi^*$-band transient as the structural dynamics in the adsorbate layer becomes more important.

\section*{S.\,V Summation of $\pi$-band and $\pi^*$-band transients\quad } 
\noindent
The photoexcitation temporarily reduces the electron population in the $\pi$ band, so that the corresponding ULEPD signal intensity $I_{\pi}(\Delta t)$ contains information about the subsequent hole relaxation dynamics. Similarly, the ULEPD signal intensity $I_{\pi^*}(\Delta t)$ appears in the ULEPD data as soon as the $\pi^*$ band is occupied by photoexcitation. Due to charge conservation and neglecting differences in the photoemission matrix elements, $I_{\pi}(\Delta t)$ can be written as $I_{0\pi}-I_{\pi^*}(\Delta t)$, where $I_{0\pi}$ is the ULEPD signal from the $\pi$ band before photoexcitation. At the same time, both the transient ULEPD signals from the $\pi$ band and from the $\pi^*$ band are reduced by the (time-delay dependent) Debye-Waller factor DWF$(\Delta t)$ as soon as the SnPc layer is heated. The summation of the two signals, taking DWF$(\Delta t)$ into account, results in
\begin{align}
     [I_{0\pi}-I_{\pi^*}(\Delta t)]\cdot \text{DWF}(\Delta t)+I_{\pi^*}(\Delta t)\cdot \text{DWF}(\Delta t)=I_{0\pi}\cdot \text{DWF}(\Delta t)
\end{align}
so that a time-delay dependent signal remains that only governed by the DWF, i.e., the  vibrational disorder in the adsorbate layer. To account for the differences in the photoemission matrix elements \cite{Gruneis2008, Stange2015}, the transients were normalized before summation, as described in section S. IV.    

\section*{S.\,VI Assignment of the temperature scale to the adsorbate transient in Fig. 3(b)\quad } 
\noindent
\begin{figure}
        \includegraphics[width=0.7\linewidth]{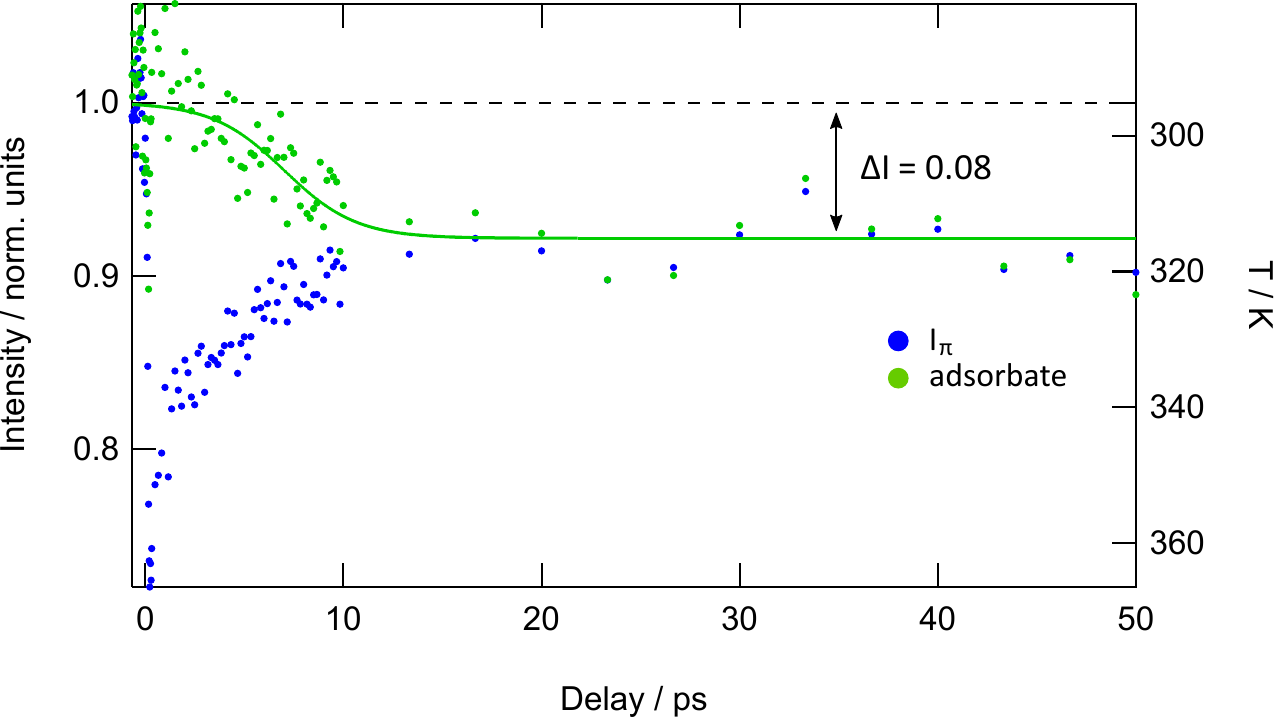}
        \caption{Total intensity changes in the ULEPD signal of the $\pi$ band overlaid with the ULEPD intensity changes associated with the excitation of the adsorbate layer (green line and data points). The intensity level before time zero was set to 1. For large time delays ($\Delta t > \SI{20}{\pico s}$) the adsorbate layer transient was scaled to match the intensity level of the ULEPD signal of the $\pi$ band.} 
    	\label{fig:FigS4}
\end{figure}

All data in Fig. 2(d) show difference intensities with respect to the ULEPD signal before excitation with the data being normalized to the average difference intensity in the time window between \SI{1}{\pico s} and \SI{2}{\pico s} (see section S.III). In order to relate the time-resolved data to the temperature scale derived from the analysis of the temperature dependent LEED and LEPD data in Fig. 3(a), we have to determine the relative intensity changes associated with the vibrational excitation of the adsorbate [green data points in Fig. 2(d)] with respect to the ULEPD signal from the $\pi$ band before excitation. The procedure is illustrates in Fig. \ref{fig:FigS4}. From the raw ULEPD data we extracted the total transient intensity change of the $\pi$ band signal and normalized this data set with respect to the intensity level before time zero (blue data points in Fig. \ref{fig:FigS4}). For large time delays, i.e., at time scales where the changes in the diffraction signal are given by the DWF, i.e., the vibrational excitation of the adsorbate layer, this data set yields a relative intensity change in the ULEPD signal of $\Delta I \approx \SI{8}{\%}$. We then overlaid this data with the adsorbate lattice transient derived from the difference intensity data. Here the zero level of the data (which corresponds to the initial adsorbate state before excitation) is set to 1, the intensity level at large delays is re-scaled to match the intensity level of the raw data. The relative intensity changes can now be referred to the temperature dependent LEPD data in Fig. 3(a) yielding a scale for the time evolution of the temperature in the adsorbate layer.

\bibliography{bibliography}